\documentclass[fleqn,useAMS,usenatbib]{mnras}

\usepackage{amssymb}

\usepackage{newtxtext,newtxmath}
\usepackage[T1]{fontenc}

\DeclareRobustCommand{\VAN}[3]{#2}
\let\VANthebibliography\thebibliography
\def\thebibliography{\DeclareRobustCommand{\VAN}[3]{##3}\VANthebibliography}

\usepackage{graphicx,url}
\usepackage{amsmath}
\usepackage{verbatim}
\usepackage{xspace}
\usepackage{booktabs}
\usepackage{textcomp} % for the TM symbol

\usepackage{enumitem}   
\hypersetup{draft}

\def \deg  {$^{\circ}$\xspace}
\def \degC {$^{\circ}$C\xspace}
\newcommand\ddfrac[2]{\ensuremath{\frac{\displaystyle #1}{\displaystyle #2}}}  % for nicer fractions with more space

\usepackage{color}
\definecolor{amethyst}{rgb}{0.6, 0.4, 0.8}
\definecolor{green}{rgb}{0.55, 0.71, 0.0}
\definecolor{apricot}{rgb}{0.98, 0.81, 0.69}
\definecolor{auburn}{rgb}{0.43, 0.21, 0.1}
\definecolor{babyblueeyes}{rgb}{0.63, 0.79, 0.95}
\definecolor{bittersweet}{rgb}{1.0, 0.44, 0.37}
\definecolor{blue(munsell)}{rgb}{0.0, 0.5, 0.69}
\newcommand{\mg}[1]{\textcolor{babyblueeyes}{Markus Gaug: #1}}  % Markus's comments
\usepackage[normalem]{ulem} % for strikethrough
\newcommand{\cfsout}{\bgroup\markoverwith{\textcolor{red}{\rule[0.5ex]{2pt}{0.4pt}}}\ULon}

\newcommand{\ccol}[1]{\textcolor{black}{#1}}  % highlight changes made for the referee

\newcommand{\orcid}[1]{{\href{http://orcid.org/#1}{% 
\openin1 images/orcid-ID.png \ifeof1
\else%
\hskip2pt\includegraphics[width=9pt]{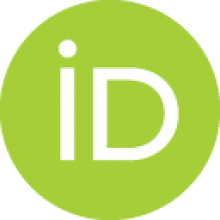}\fi}}{}}%

\title[20 Years of MAGIC Weather Station Data]{Detailed Analysis of Local Climate at the CTAO-North Site \ccol{on La Palma} from 20 Years of MAGIC Weather Station Data}
\author[M. Gaug et al.]{Markus Gaug$^{1,2}$\thanks{E-mail:markus.gaug@uab.cat}\orcid{0000-0001-8442-7877},  
Alessandro Longo$^{3,4}$\orcid{0000-0003-4254-8579},   
Stefano Bianchi$^5$\orcid{0000-0001-6729-8655},
Llu\'is Font$^{1,2}$\orcid{0000-0003-2109-5961},\newauthor
Sofia Almirante$^1$\orcid{0009-0009-3961-4193},
Harald Kornmayer$^6$\orcid{0000-0002-8878-7623},
Michele Doro$^7$\orcid{0000-0001-9104-3214},
Alexander Hahn$^8$\orcid{0000-0003-0827-5642},\newauthor
Oscar Blanch$^9$\orcid{0000-0002-8380-1633},
Wolfango Plastino$^{5,10}$\orcid{0000-0002-5737-6346},
Daniela Dorner$^{11}$\orcid{0000-0001-8823-479X}
\\
$^1$Departament de F\'isica, Universitat Aut\`onoma de Barcelona, 08193 Bellaterra, Spain.\\
$^2$CERES-IEEC, Universitat Aut\`onoma de Barcelona, 08193 Bellaterra, Spain.\\
$^3$Universit\`a degli Studi di Urbino ``Carlo Bo'', I-61029 Urbino, Italy.\\
$^4$INFN, Sezione di Firenze, I-50019 Sesto Fiorentino, Firenze, Italy.\\
$^5$Department of Industrial, Electronics and Mechanical Engineering, Roma Tre University,  I-00146 Rome, Italy.\\
$^6$Duale Hochschule Baden-W\"urttemberg, Mannheim, Germany.\\
$^7$University of Padova \& Istituto Nazionale di Fisica Nucleare, Sez. di Padova, Italy.\\
$^8$Max Planck Institute for Physics (Werner-Heisenberg-Institut), Boltzmannstr. 8, 85748 Garching, Germany. \\
$^9$Insitut de Fisica d'Altes Energies, Bellaterra E-08153, Spain.\\
$^{10}$INFN Sezione Tor Vergata, Via della Ricerca Scientifica, 1, I-00133 Rome, Italy.\\
$^{11}$Julius-Maximilians-Universit\"at W\"urzburg, Fakult\"at f\"ur Physik und Astronomie, Institut f\"ur Theoretische Physik und Astrophysik, \\ Lehrstuhl f\"ur Astronomie, Emil-Fischer-Str. 31, D-97074 W\"urzburg, Germany.}

\date{Accepted XXX. Received YYY; in original form ZZZ}

\pubyear{2024}

\begin{document}
\label{firstpage}
\pagerange{\pageref{firstpage}--\pageref{lastpage}}
\maketitle

\begin{abstract}
%Meteorology plays an important role in the process of astronomical data  taking. Optimal weather conditions are needed for a good interpretation of the  data, because the light transmission through the atmosphere and its subsequent  detection is affected by the current atmospheric properties. In addition, atmospheric conditions are monitored to operate the telescopes within the operational safety limits.
The Observatorio del Roque de los Muchachos will host the northern site of the Cherenkov Telescope Array Observatory (CTAO), in an area about 200~m below the mountain rim, where the optical telescopes are located. The site currently hosts the MAGIC Telescopes, which have gathered a unique series of 20 years of weather data.
%To characterize the environmental conditions at the MAGIC Telescopes site, their implications on observation duty cycle  and long-term weather evolution, as a proxy for CTAO-North.} 
We use advanced profile likelihood methods to determine seasonal cycles, the occurrence of weather extremes, weather downtime, and long-term trends correctly taking into account data gaps. The fractality of the weather data is investigated by means of multifractal detrended fluctuation analysis.  The data are published according to the \ccol{Findable, Accessible, Interoperable, and Reusable (FAIR)} principles.
We find that the behaviour of wind and relative humidity show significant differences compared to the mountain rim. We observe an increase in temperature of $0.55\pm0.07\mathrm{(stat.)}\pm0.07\mathrm{(syst.)}$\degC/decade, the diurnal temperature range of 
$0.13\pm0.04\mathrm{(stat.)}\pm0.02\mathrm{(syst.)}$\degC/decade (accompanied by an increase of seasonal oscillation amplitude of $\Delta C_m=0.29\pm0.10\mathrm{(stat.)}\pm0.04\mathrm{(syst.)}$\degC/decade)  
and relative humidity of $4.0\pm0.4\mathrm{(stat.)}\pm1.1\mathrm{(syst.)}$\%/decade, and a decrease in trade wind speeds
of $0.85\pm0.12\mathrm{(stat.)}\pm0.07\mathrm{(syst.)}$(km/h)/decade.
 The occurrence of extreme weather, such as tropical storms and long rains, remains constant over time. We find a significant correlation of temperature with the North Atlantic Oscillation Index and multifractal behaviour of the data.
The site shows a weather-related downtime of 18.5\%-20.5\%, depending on the wind gust limits employed. No hints are found of a degradation of weather downtime under the assumption of a linear evolution of environmental parameters over time.
%  We cannot find any hints of a degradation of weather downtime under the assumption of a linear evolution of environmental parameters with time.
\end{abstract}


\section{Introduction}

%Meteorology plays an important role in the process of astronomical
%data taking. Optimal weather conditions are needed for a good
%interpretation of the data, 

%The complementary is called the
%\emph{downtime} of the telescope. It is composed of technical
%failures, accidental problems, telescope preparation and repositioning
%time  and adverse weather conditions. Sometimes also human errors
%cause an increase in downtime.

Ground meteorology of the Observatorio del Roque de los Muchachos (ORM, 28$^\circ$N, 18$^\circ$W,) on the island of La Palma,
Canary Islands, Spain, has been extensively characterized over the past 50~years
\citep{Kiepenheuer:1972,mcinnes1974,murdin1985,Mahoney:1998,lombardi2006,lombardi2007,lombardi2008,Jabiri:2000,munoztunon:2015,Castro-Almazan:2018,Hidalgo:2021}. 
% Castro-Almazan:2009,
A peak of site-testing activity can
be identified in coincidence with the site selection campaigns carried out for the Extremely Large Telescope~\citep[ELT,][]{Varela:2009,lombardi2009,Vernin:2011,Varela:2014}, for which the ORM remained the
only candidate site in the northern hemisphere on the shortlist, and the Thirty Meter Telescopes~\citep[TMT,][]{Teran:2018,Vogiatzis:2018}, for which the ORM is treated as alternative site. 
 Lately, another large infrastructure, the Cherenkov Telescope Array Observatory\footnote{\url{www.cta-observatory.org}} (CTAO),
has chosen the ORM for its northern
array. The prototype of the first CTAO Large Size Telescope (LST1), a project for a 23-m dish Cherenkov
 telescope~\citep{Mazin:2021}, is already operating, and 
 the next three telescopes LST2-4 are being constructed. 
 Lately, the impact of climate change has raised concerns about the future viability and quality of the few world-class astronomical observatories, among them the ORM~\citep{Haslebacher:2022}. 

\begin{figure}
  \centering
\includegraphics[width=0.99\linewidth,clip, trim={5.8cm 0 4.2cm 0}]{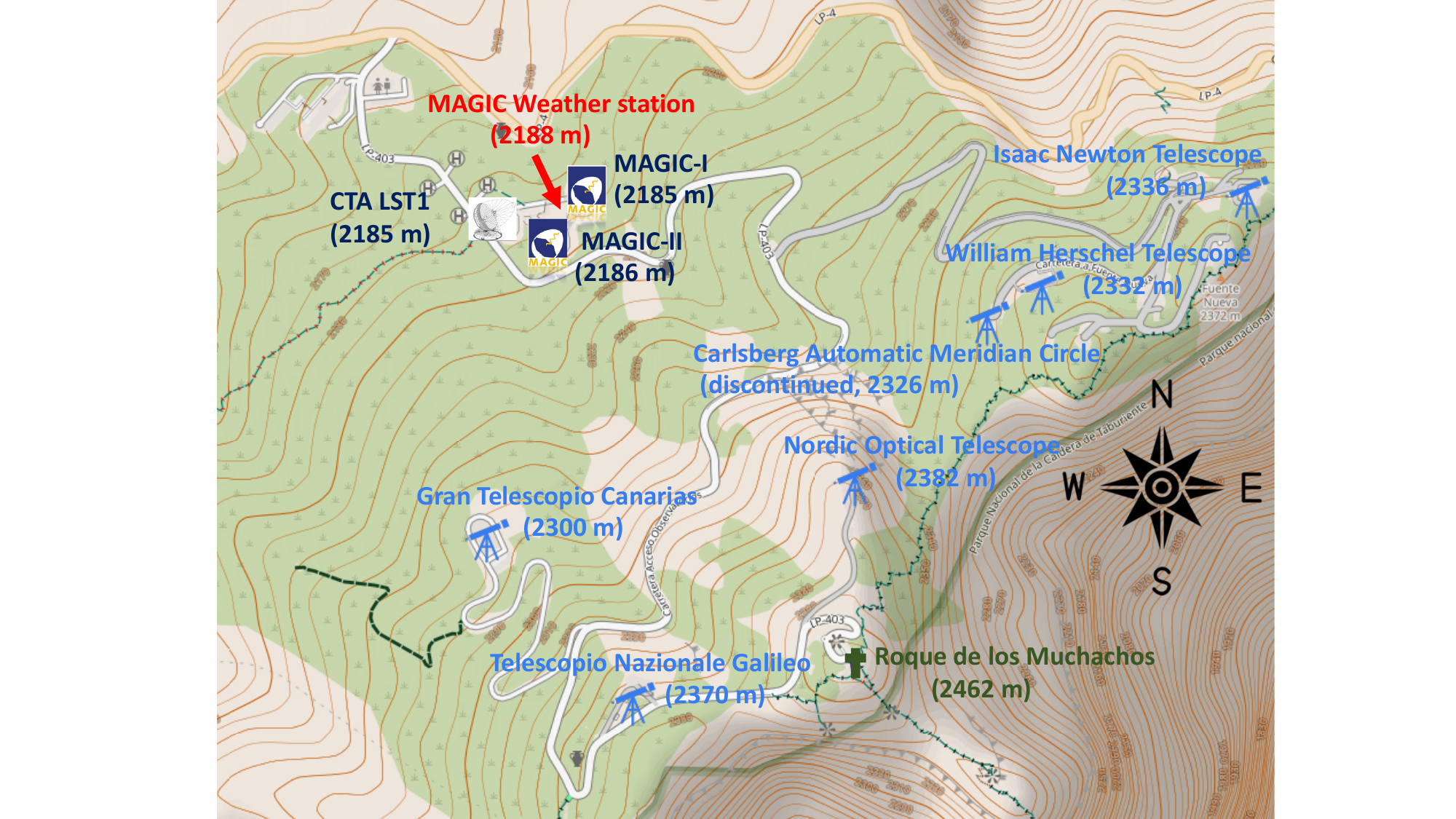}
  \caption[Location of CAMC, TNG, NOT and MAGIC telescopes at ORM.]{
    \label{fig:ORMsite}
    Location of GTC, TNG, NOT, WHT, INT and MAGIC telescopes at ORM (Source:
%    \mg{\url{https://osm.org/go/bkp0PhfH?m=} 
\textsuperscript{\textcopyright}OpenStreetMap contributors, see license \url{https://www.openstreetmap.org/copyright})
  }
\end{figure}

The ORM is located on the upper part of the \textit{Roque de los Muchachos} caldera rim, with several optical and infrared telescopes placed at the top, at altitudes around 2400\,m a.s.l.\ (see Fig.~\ref{fig:ORMsite}). The observatory extends, however, to lower altitudes reaching 2100\,m a.s.l., with the MAGIC Telescopes~\citep{magicperformance1,magicperformance2}
located at $\sim$2200\,m a.s.l.\ That lower part generally shows a bit milder climate and is better
protected against strong winds blowing from Southern directions, due to the orography of the
mountain slope. It is also here
that the array of CTAO telescopes will be placed. This article describes the weather conditions at this particular part of the ORM, hitherto neglected in site characterization studies. 
%Beside describing the atmospheric conditions at La Palma, one of the goals of
%this chapter is to determine the fraction of MAGIC downtime caused by weather
%since it started operations back in 2004.

The atmosphere in the subtropical region of the Canary Islands is characterized by great stability throughout the year. Due to the combination of large-scale atmospheric circulation of the descending branch of the \emph{Hadley cell}\footnote{\ccol{a global-scale tropical atmospheric circulation that features air rising near the equator, flowing poleward near the tropopause at a height of 12--15~km above the Earth's surface, cooling and descending in the subtropics and then returning equatorward near the surface.}} around 30$^\circ$N~\citep{Longo:2019} and the Trade Winds coming from the high area of the Azores, a
quasi-permanent temperature inversion layer appears around 1300\,m a.s.l. on average~\citep{font-tullot,palmen1969,Carrillo_2015}, which can usually be well identified by the \textit{sea of stratocumulus} on the northern coasts of the islands. This layer separates two
well-defined regimes: below it, the moist marine boundary layer, and above it the dry free troposphere. The inversion layer is present 78\% of the time throughout the year~\citep{Carrillo_2015}. Its altitude and thickness have a seasonal dependence, being higher and thinner during winter (when it is found between 1350\,m a.s.l.\ and 1850\,m a.s.l., being only 350\,m thick) and lower and thicker during summer (between 750\,m a.s.l.\ and 1400\,m a.s.l., being approximately 550\,m thick)~\citep{torres2002}.  Furthermore, \citet{Carrillo_2015} have found two separate inversion layers around 20\% of the time on Tenerife, where the second is located \textit{above} the main one, at $\sim$800\,hPa ($\sim$2000\,m a.s.l.) during winter and $\sim$820-850\,hPa ($\sim$1500-1800\,m a.s.l.) during summer. The authors associate the lower main inversion layer with the top of the convective marine boundary layer, while the second upper inversion is related to synoptic-scale subsidence and sharp changes in the $u$ component of the trade wind shear. 
The MAGIC Telescopes site is usually located above the (second) 
inversion layer, in the free troposphere, at pressure levels below 795\,hPa~\citep{Gaug:2017site}, and is therefore characterized by excellent observing conditions, with dry and clean air. 
 
For several years, various telescope sites at the ORM have been continuously monitored by automatic weather stations, among other reasons due to the long-lasting lack of professional measurement of weather conditions by the Spanish Meteorological Weather Service (AEMET) only available since 2022\footnote{see \url{https://www.aemet.es/es/eltiempo/observacion/ultimosdatos?k=coo&l=C101A} (AEMET local identifier \textit{C101A}) and \url{https://www.boe.es/boe/dias/2022/04/07/pdfs/BOE-A-2022-5672.pdf}}.
%which provide measurements of local meteorological parameters. 
All these instantaneous and long-term records of the meteorological data are important tools not only for meteorological or climatological studies but also to assess the environmental conditions to which new installations, like CTAO, will be exposed. 
%for the safety and operation of the telescopes.

 This article presents an analysis of the 20~years meteorological data base
obtained with the MAGIC weather station (28.7619\deg~N, 17.8907\deg~W, 2188\,m
a.s.l.), and compares its results with the known conditions at the higher-altitude
sites of the ORM, particularly the Telescopio Nazionale Galileo (TNG), the Carlsberg Automatic Meridian Circle (CAMC) telescope~\citep{lombardi2006,lombardi2007,lombardi2008}, the Gran Telescopio de Canarias (GTC) and Extremely Large Telescope (ELT) site testing campaigns~\citep{Mahoney:1998,Varela:2014}, and the Northern Optical Telescope (NOT)~\citep{Haslebacher:2022}. 
Moreover, we compare the behavior of meteorological
parameters with the telescope downtime for the MAGIC site, for the hypothetical case of no
downtime due to hardware failures and upgrades. 
%\md{To avoid having to  pass through CTA SAPO, I use a milder version like "extend our  results to'' We apply then the same procedure to estimate the AAOT for the CTA, using the corresponding foreseen operation limits.}
%Combining the weather data and results available from other telescopes at the
%ORM, this study covers more than 25 years of meteorological measurements. A
%goal of this article is to better understand the influence of local
%meteorological conditions for astronomical observations at the ORM, using a
%statistical approach.

In Sect.~\ref{sec:orm}, the local climatological conditions are discussed, and an overview is given of the MAGIC telescopes, its weather monitoring system and the data collection process. 
After that, Sect.~\ref{sec:results} discusses the results of the analysis: temperature, relative humidity, pressure, wind direction, wind speed 
and derived parameters. 
%: diurnal temperature range, temperature and humidity gradients, occurrence of rain, snow and storms. 
We show average distributions, long-term behavior, seasonal cycles, and interrelations
between the parameters. 
A statistical assessment is made to test hypothetic linear increases or decreases of all weather parameters over time, and the occurrence of long precipitation periods and storms over time. 
Section~\ref{sec:dutycycle} evaluates the impact of the meteorological parameters found on the duty cycle of the MAGIC telescopes. In Sect.~\ref{sec:fractal}, we present a fractal analysis of the measured meteorological parameters by means of Detrended Fluctuation Analysis (DFA) and multifractal DFA (MFDFA).
%~\citep{ihlen2012introduction}. 
This allowed us to quantify spectral width and persistency of the meteorological time series. 
Fairification of the weather station data is described in Sect.~\ref{sec:fairification}. 
Finally, in Sect.~\ref{sec:discussion} , the relevance of the results and systematic uncertainties are discussed. Section~\ref{sec:conclusions} summarizes the most relevant results and presents the conclusions of the article.
%Since 1970, the Observatorio del Roque de Los Muchachos hosts several European astronomical telescopes. 

%All telescopes are located along the northern rim of
%the Caldera de Taburiente, at the north-western side of La Palma island. The
%irregular shapes produce a complex orography and the crowdedness at the top, due
%to the presence of the astronomical observatories, suggests a possible
%modification of the local micro climate, making difficult to foresee in advance
%the precise local meteorological parameters. Therefore, 

\section{The MAGIC telescopes and its weather monitoring system}
\label{sec:orm}
%\md{<--Responsible}

MAGIC (Major Atmospheric Gamma-ray Imaging
Cherenkov)\footnote{\url{http://magic.mpp.mpg.de}} is an instrument
constituted by a pair of 17\,m diameter telescopes that investigate
the nonthermal emission from astrophysical objects in the energy band between
about 10\,GeV ($2.5\times10^{24}$\,Hz) and 10\,TeV
($2.5\times10^{27}$\,Hz). 
%For cost reasons, 
Both telescopes stand freely without a protective dome. All elements, like structure, the tessellated mirror, the camera, etc. are hence continuously exposed to the weather conditions. 

Whereas the other (optical) telescopes are located along the mountain rim, %La Caldera de Taburiente, 
the MAGIC Telescopes are found 200\,m further downhill, as shown in Fig.~\ref{fig:ORMsite}. The observatory area is not flat, so differences
in height and the complex orography have an effect on local weather conditions.
It is, for instance, possible that clouds move up the mountain hill and cover the MAGIC Telescopes, while the telescopes on the mountain rim are unaffected. Conversely, clouds sometimes move up the southern "Caldera de Taburiente" and spill over the rim, which affects the highest situated telescopes, but not the MAGIC Telescopes. 
Finally, the local orography with its ravines and shoulders strongly affects local wind fields~\citep{Mahoney:1998}.

\subsection{The MAGIC Weather Station}
\label{sec:analysis}
%{\bf TBD Martin Will and add location, history (Lluis Font) and sketch}

%\mg{More on quality assurance}

\begin{figure}
  \centering
 \includegraphics[width=0.75\linewidth]{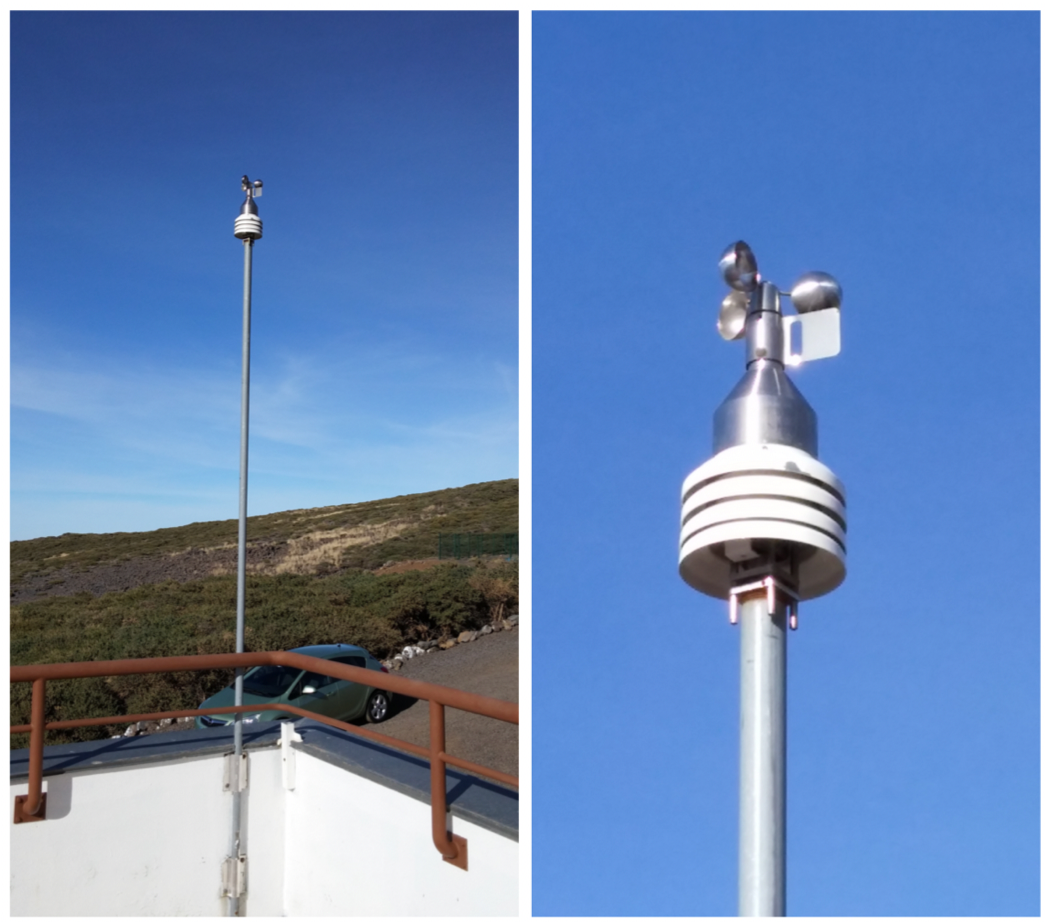}
  \caption{ \label{fig:weatherstation}
    Weather station model \texttt{MWS-5MV} from \textit{Reinhardt System- und
    Messelektronik GmbH} installed at MAGIC telescope site.
 %   \mw{This is a promotional image from the company website, probably
 %     better to replace it with an image of our own.}
      %\md{Do we have one taken at close distance from the roof of MAGIC?} 
  }
\end{figure}

A weather station (hereafter \textit{WS}) was first installed on 27/09/2003 on the lightning rod of the first MAGIC Telescope. On 01/03/2004, it was relocated to its final position on the roof of the MAGIC control building at 2188\,m height (see Fig.~\ref{fig:weatherstation}). 
Initially, the model Reinhardt MWS5
%\footnote{The supplier's model initials \textit{MWS} stand for \textit{Microprocessor Weather Station}.} 
was employed and upgraded to an MWS~5MW in 2007. 
Currently, an MWS-55V\footnote{Reinhardt System- und Messelektronik GmbH \url{https://www.reinhardt-testsystem.de/english/climate_sensors/weatherstations/weather_station_mws_55v.php}} is used, alternating with an MWS~5MW\footnote{\url{https://www.reinhardt-testsystem.de/_pdf/english/MWS5MV-10_MWS55MV_e.pdf}}. One of either is used as spare, whenever the other is getting repaired. The WS is located at a height of 5\,m above ground and a distance of approximately 50\,m from the telescopes.
Table~\ref{tab:WSreplacements} lists  all replacements and 
repairs carried out during the past 20~years. 
%on the MWS 

The WS  
measures the following parameters:
\begin{list}{--}{\itemsep=0pt}
\item temperature,
\item relative humidity,
\item barometric pressure,
\item wind speed (incl.\ wind speed peak and average wind),
\item wind direction (incl.\ prevalent wind direction).
\end{list}
%
%\item dew point, \mw{sure this is measured?}
%The station is the \texttt{MWS-5MV} %model from German manufacturer {\it
%Reinhardt System- und Messelektronik GmbH}~\citep{Reinhardt}, and it is shown in
%Fig. \ref{fig:weatherstation}.

%Average values 

The temperature measurement is based on a precision temperature sensor
\textit{PT100}. 
%This platinum resistance thermometer offers excellent accuracy
%over a wide temperature range. The principle of operation is to measure the
%resistance of a platinum element, which has a resistance of 100\,$\Omega$ at
%0\degC and 138.4\,$\Omega$ at 100\degC. 
The measured value is linearized by the software. By default, the temperature sensor is mounted on the lower side of the
weather station. A white lacquered pagoda protects it from direct radiation
and prevents the accumulation of heat. The measurable temperatures range from
$-$40\degC to +60\degC with an accuracy of $\pm$~0.3\degC.

%The {\bf relative humidity} is the fraction of water vapor in the air with
%respect to the amount necessary to reach condensation at the same temperature.
The humidity sensor is a fast responding capacitive sensor (monolithic) based on
a dielectric polymer. This polymer absorbs or releases water proportional to the relative humidity of the environment and thus changes capacitance, which is measured
by an on-board electronic circuit. The sensor is also mounted on the lower side
of the weather station. A Gore-Tex\textsuperscript{\textregistered}  
 cover protects it from pollution or
destruction by dust or insects. The sensor can be used in a temperature
range between $-$40\degC and +60\degC and
is capable of measuring relative humidity between 0\% and 100\%, with a precision of about 2\%. This precision worsens to 3\% when the humidity drops to less than 10\% and the temperature is lower than 15\degC at the same time.

Long-term exposure to conditions outside the "normal" range (i.e., relative humidities above 80\%), can introduce an error in the relative humidity signal, increasing by 3\% after 60~hours, according to the manufacturer. However, we have noticed that erroneous measurements are possible after at least two days of continuous rain, resulting in erroneous displays of 100\% humidity for up to a few hours after the actual humidity has dropped to values below 50\%. 

%Its range extends from 10\% to 100\%,
%with a measuring accuracy of $\pm$~2\%. 

%In addition, the {\bf dew point} in
%\degC is provided. The dew point it the temperature to which the air needs to be
%cooled in order to reach saturation. If the relative humidity reaches 100\%, the
%dew point and outside temperatures are identical.
The pressure sensor is a monolithic laser-trimmed sensor for absolute
pressure, which is linearized to 5~hPa for the whole temperature range.
Another temperature linearization reduces the deviation to less than 2~hPa over the temperature range of interest. The measuring signal is elaborated by an instrumental amplifier. 
%The sensor can be used across the entire above-mentioned temperature range. 
 The measurement limits for
this sensor are 600 and 1100\,hPa with $\pm$0.8~hPa accuracy. This sensor can
be used at altitudes from 50\,m below zero to altitudes of 3000\,m. In our case, it was calibrated for an altitude of 2200\,m a.s.l.

The wind speed sensor is made of a three-cup anemometer with magnetic scanning.  
Wind speed is measured without touch using a Hall sensor. 
%, a transducer that varies its output voltage in response to a magnetic field. 
%Using a Hall effect sensor and a wheel with 12 magnets, as the wind blows the device generates a series of pulses at a speed that increases with the speed of the wind and is read by a processor. 
Given that the increase in the speed of rotation (and therefore the frequency of the electric pulses) is not
proportional to the wind speed, the device firmware incorporates linearization. The device provides three different measurements for the wind speed, these being the current speed, the average speed (averaged over ten-minute intervals) and the wind gusts, the maximum of an average of three seconds obtained during the last two minutes.
The sensor range covers speeds between 0 and 157.3~km/h with a precision of 2.5~km/h (by MWS~55) or 2~km/h (for MWS~5MW and MWS-55V) for small and medium speeds. 
However, at wind speeds greater than 100~km/h, the effect of the drag force of the cups gradually worsens the accuracy, which reaches only 9~km/h for the highest speed that the anemometer is capable of measuring. 
For this reason, the manufacturer has artificially imposed nine discrete intervals in this regime following approximately:

\begin{align}
 \textit{Wind~Speed~Bin}_{\,i} &= \left(  101 + 4.2 + 4 \cdot i + 0.3\cdot i^2 \right) ~ \mathrm{km/h} \nonumber\\
 {} & \mathrm{for}~i \in [0,8]~.
\end{align}

%\begin{equation}
% \textit{Wind~Speed~Bin}_{\,i} = \left(  101 + 4.2 + 4 \cdot i + 0.3\cdot i^2 \right) ~ \mathrm{km/h} \quad \mathrm{for}~i \in [0,8]~.
%\end{equation}

At temperatures below 0\degC the MWS5 sensor could freeze, which would result in erroneous measurements of no wind for the three parameters mentioned. The issue has been solved with the more recent MWS~5MV and MWS-55V models, which incorporate internal heating.
%We must also take into account the phenomenon of overspeeding that we find in this type of device due to its geometry, where a faster response to the acceleration of the air flow is obtained than the obtained after a deceleration of the wind flow [3, 17, 16].

Due to the difficult orography of the place, the wind sensor could not be installed according to all recommendations of the World Meteorological Organization~\citep{WMO_recommendations:2018}, requiring 10\,m height above ground and no obstacles on the horizon. Instead, the wind sensor is located 5~m above ground, and the MAGIC LIDAR~\citep{MAGICLIDAR:2022} dome is found at a distance of 6\,m in SSW direction, surpassing the height of the weather station by about a meter and creating a shadow with a full width of about 28$^\circ$. For winds blowing from northern directions, where the terrain slopes downward, the conditions are closer to those recommended. The two MAGIC Telescopes are found at 49\,m distance  towards E and 50\,m towards S, respectively. Since 2016, the LST1 telescope 
is located at 42~m distance towards WSW.

%The wind sensor is made up of an anemometer with magnetic scanning.
%Hall sensors are used for proximity switching, positioning and speed detection. 
%A peak detector finds every wind peak and hands them down to the measuring software. 
%An average value is determined within the respective memory intervals. 
%The range is from 0 to 150\,km/h with $\pm$~2\,km/h measuring accuracy. 

A weather vane with a precision
magnetic encoder and a rotation angle of 360\deg measures the direction of the wind.
%The direction is given in degrees (\deg), with 0\deg being North, and 90\deg being East. 
The measuring accuracy is 5\deg, with a
starting speed less than 2~km/h, and a hysteresis of less than 8\deg.
The device provides two different measurements for the wind direction, these being the current wind direction and the moving average for the prevailing wind direction. 

On 25 June 2018, a variable resistor rain drop counter and a condensation detector, controlled by an Arduino microcontroller, were added to the system and mounted close to the weather station. Drops and condensation measurements were found to behave similarly when summed over 30~s. This led to the definition of
a rain parameter as the maximum of either drops or condensation
in 30\,s time bins, normalized to a range from 0 to 100. The system showed $\lesssim$4\% of false positives and a similar amount of false negatives. 

The recorded data are transmitted to the serial port of a server machine in the MAGIC Counting House. Every two seconds, the station is read out, and a data record is received, which starts with time and date. 
%Separated by commas, 
The single measured values with sensor identification are written in a text file.
The MAGIC Telescope Central Control and the Camera Control programs read this file and acts in
case a safety limit is surpassed.  Moreover, the weather information is displayed online\footnote{\url{http://www.magic.iac.es/site/weather/}} and publicly accessible.
A data storage script reads the single
string file every two minutes and generates a daily file in which the data are continuously stored.

%Weather information can be seen online at
%\texttt{http://www.magic.iac.es/site/weather/}, as shown in
%figure~\ref{fig:WebWS}. When safety limits are surpassed, an alarm is visible
%on the website and the color of the plots change to yellow in case of a warning
%or red in case of alarm. This weather monitoring system was created and is
%maintained by the Grup de Física de les Radiacions of the Universitat Aut\`onoma
%de Barcelona.

%\begin{figure}[h!tb]
%  \begin{center}
%    %\includegraphics[bb = 0 0 920 793,width=0.6\linewidth]{images/webpage_v2.jpg}
%  \end{center}
%  \caption[Sketch of the webpage for Atmospheric Monitoring of the
%    MAGIC telescope.]{Sketch of the webpage for Atmospheric Monitoring of the
%    MAGIC telescope. The webpage has open access and can be found at
%    \texttt{http://www.magic.iac.es/site/weather/}.}
%  \label{fig:WebWS}
%\end{figure}

\begin{table}
\centering
\begin{tabular}{lp{6.8cm}}
\toprule
Date  &  Comments \\
\midrule
2003-09-27 & Installation of model MWS5 on lightning rod \\
2004-03-26 & Relocation to final site \\
2005-03-01 & Exchange to a new unit, old was kept as spare.\\
2007-03-28 & Wind vane bent, replacement by new MWS 5MV\\
2009-03-12 & Short circuit on board, damaged wind cups, replacement by MWS 5MV\\
2010-12-19 & Replacement by repaired and calibrated MWS5 \\
2011-01-26 & Replacement by repaired and calibrated MWS 5MV   \\
2011-06-27 & Replacement by repaired and calibrated MWS 5MV   \\
2012-12-05 & Humidity drift detected, replacement by repaired and calibrated MWS 5MV \\
2015-01-27 & Humidity sensor broken, replacement by calibrated spare MWS 5MV \\
2017-04-10 & Replacement by new MWS 55V   \\
2019-07-20 & Defective memory card, replacement by repaired and calibrated MWS 5MV \\
%2020-12-11 & \mg{Llu\'is, t'en recordes qu\`e va passar llavors?}\\
2023-01-16 & After a gradual increase of lowest humidities had been observed, the station was sent for inspection to the provider and replaced by a repaired and calibrated MWS 55V. \\
\bottomrule
\end{tabular}
\caption{\label{tab:WSreplacements} Replacement dates of the WS. Each time, the weather station had been replaced by a new or a repaired one. In both cases, the station came with a new calibration certificate. }
\end{table}

\subsection{Data Reduction}
\label{sec:datareduction}

%\md{This section was badly written. I reviewed it. However, more work  to do.}

The WS has been operational since
September 2003. 
Due to maintenance, upgrades, power outages, and malfunctioning of sensors, several time intervals have remained completely without data: 2004/03/01-26, 2004/10/21-2005/03/01, 2007/03/19-28, 2009/02/05-03/12, 2011/06/14-28, 2012/11/01-12/05,  2019/06/29-07/20 and 2020/11/28-12/11.
Furthermore, on two occasions the MAGIC Telescopes reflected sunlight onto the weather station during installation or replacement of mirrors (and hence the telescopes pointing upward instead of northward in their usual parking position). This caused unusual and sudden temperature increases to more than 40\degC in the mornings of 2010/10/11 and 2020/03/24. 
Further consistency checks on parameter gradients have revealed software-induced dating errors after early software updates, which could be corrected offline. 
The pressure measurements taken before the relocation of the station to its final place in 03/2004 have been corrected for the altitude difference of 12\,m, assuming a barometric scale height of 8.4\,km. 
Finally, the humidity sensor experienced an apparent drift starting from 2020 seemingly over-predicting particularly very low relative humidity. However, inspection of the sensor by the provider and further checks and cross-correlation with the publicly available NOT weather data\footnote{\url{https://www.not.iac.es/weather/}} revealed that the sensor was not affected by drifts, and the observed increase of low relative humidities must have been due to a change in the local atmosphere.

\begin{figure*}
    \centering
    \includegraphics[width=\textwidth]{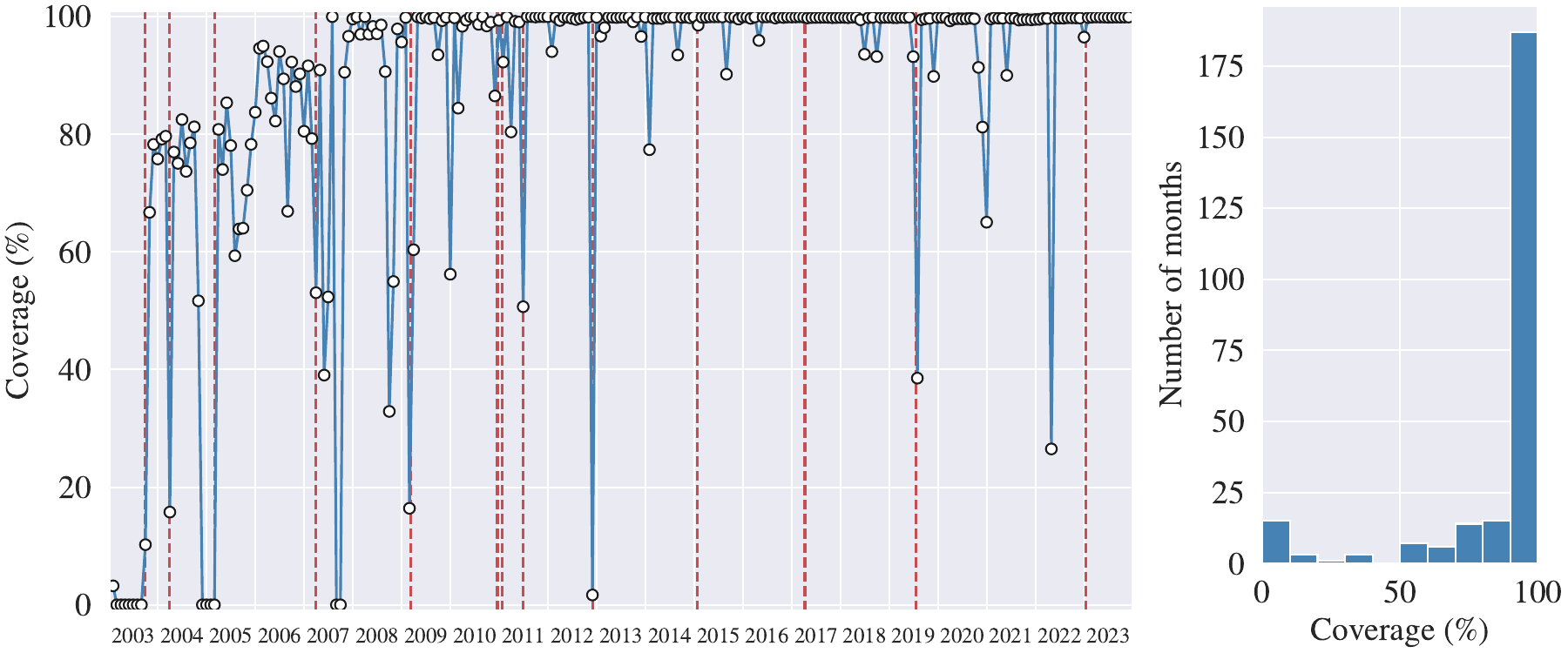}
    \caption{Monthly coverage of WS data by months (left) and the distribution of monthly coverage (right). The red dashed lines denote the times when a WS was exchanged. \label{fig:coverage}}
\end{figure*}

%Unfortunately, that drift started to be very small and only gradually increased over time until it was discovered at the beginning of 2023. 
%Therefore, we have discarded those three years of data for the analysis of relative humidity.  
%Data registered by the weather station are acquired every two seconds and directly sent 
%to the MAGIC central control for online checks whether the safety limits for telescope operation are exceeded. Additionally, the data are archived every two minutes. 
Fig.~\ref{fig:coverage} shows the available uptime of the weather data archive.  One can immediately notice the enormous improvement of reliability and availability of the WS data after the upgrade to the MWS~5MW in 2007.

%\mg{To be created} 

%On the global data sample, we had to perform several consistency checks. The retrieval of monthly and
%yearly averages are hereafter discussed. The result of the consistency control is shown in Tabble~\ref{table:Qcuts}. 

From each raw
data series,  medians and median standard deviations are computed following the 
recipes outlined in~\citet{Castro-Almazan:2018}\footnote{Note that \citet{Castro-Almazan:2018} present a climatological study for a former CTA candidate site at the neighbouring island of Tenerife, not the final one surrounding the MAGIC site on the island of La Palma.}. 

For studies involving fits to possible long-term changes in weather behavior, we required a monthly data coverage of at least 80\%, whereas for studies involving daily medians, 
we required daily data coverage of at least 85\%. 

Only for the overall twenty-years statistics, we followed the method outlined in~\citet{MAGICLIDAR:2022}, namely: 
 \begin{enumerate}
     \item Months with less than 50\% coverage were discarded. Counting from October 2003 on, this affects 15 months in total.
     \item The month-wise parameter distributions of the rest were normalized and then added according to the number of days that each month counts (28.25 days were assumed for February).
     \item The summed distribution was normalized again.
 \end{enumerate}
 For the overall minima and maxima of each parameter, the entire data set was used, in order not to miss any extreme value.

\subsection{\ccol{Robustness of results and systematic errors}}
\label{sec:discussion}
%{\it Problems of SITE characterization, need updates. Daniela Resp., start in 2nd half of November}

\ccol{The Observatorio del Roque de los Muchachos has long been deprived of professional weather recording by the Spanish Meteorological Weather Service (AEMET), established only in 2022. For this reason, individual telescope installations have operated and maintained their own automated weather stations~\citep{Mahoney:1998,garciagil2010,Jabiri:2000,lombardi2006,Varela:2009}, without being included in a network of professional meteorologists. As a consequence of this, the installation and location of the weather stations did not always follow the recommendations of the WMO~\citep{WMO_recommendations:2018} (anyhow a difficult task at that site given the complicated local orography) and obey the necessary calibration intervals.  The MAGIC weather station is no exception to that practice. At the time of installation, twenty years ago, it was designed to provide some guidance on telescope operation, without planning a professional characterization of the local climate, let alone the effects of climate change on the measured parameters. Only after two decades of almost uninterrupted operation has the added value of the acquired data become apparent. }

\ccol{Bearing in mind these caveats, we have dedicated additional efforts to data selection and quality assurance, the search and detection of systematic biases in combination with a wide range of systematic tests, and the assessment of systematic uncertainties. Contrary to what one might have naively expected, the choice of relatively cheap weather station models, which resulted in frequent repairs, was a benefit for the quality of our data. Due to the frequent weather station replacements, on time scales of few years typically, by repaired and freshly calibrated models, we have been able to control and limit possible long-term sensor drifts. }

\ccol{Among the 
%additional 
tests for systematic errors carried out are: }
\ccol{%
\begin{itemize}
\item Possible systematic calibration offsets specific for the three employed weather station models. This test was carried out by including an additional offset parameter between the MWS-5MV and the MWS-55V model in the list of nuisance parameters for the profile likelihoods (see Sect.~\ref{sec:results}) and excluding the first years when the MWS5 was operative. No significant shift of any of the tested long-term evolution parameters 
%$b$ 
could be detected beyond statistical expectations, except for atmospheric pressure, which became compatible with no evolution after allowing an offset between the two WS models. 
\item Individual miscalibration or drifting periods. This test was carried out by asking each coauthor to provide two periods of suspicious drift or offset behavior from visual inspection of the normalized difference of our WS parameters and the corresponding ones obtained with the publicly available NOT weather data. Then these two periods (different for each coauthor) were excluded from the likelihood fits and its effect on the profile likelihoods of the long-term evolution parameter 
%$b$ 
tested.  No significant shifts could be detected, and in particular the possibility of no long-term evolution could not be recovered with this test. 
For the study of relative humidity, the deviating period from 2020 through 2022 was additionally excluded for tests, with the same results. 
\item We contacted the WS provider and explicitly asked for known long-term sensor drifts of one or several of our purchased models. Apart from the general maximally possible sensor drifts specified in the data sheets, there was no further known drift to our concrete models, all of which had been repaired and recalibrated at least once. 
\end{itemize}
}

\section{Results of Long-term Data Analysis}
\label{sec:results}

In this section, we present a statistical analysis of the WS data set and compare the results with those previously published from higher altitude sites at the ORM, particularly the TNG, the CAMC telescopes~\citep{lombardi2006,lombardi2007,lombardi2008}, the GTC and ELT site testing campaigns~\citep{Mahoney:1998,Varela:2014}, and the NOT~\citep{Haslebacher:2022}.
Furthermore, the long-term behavior of all original and derived parameters is studied and put in context with expectations from climate change.

\subsection{Long-term Temperature Behavior}
\label{sec:temperature}

\begin{figure*}
\centering
\includegraphics[width=0.99\textwidth, clip, trim={0cm 0 0cm 0}]{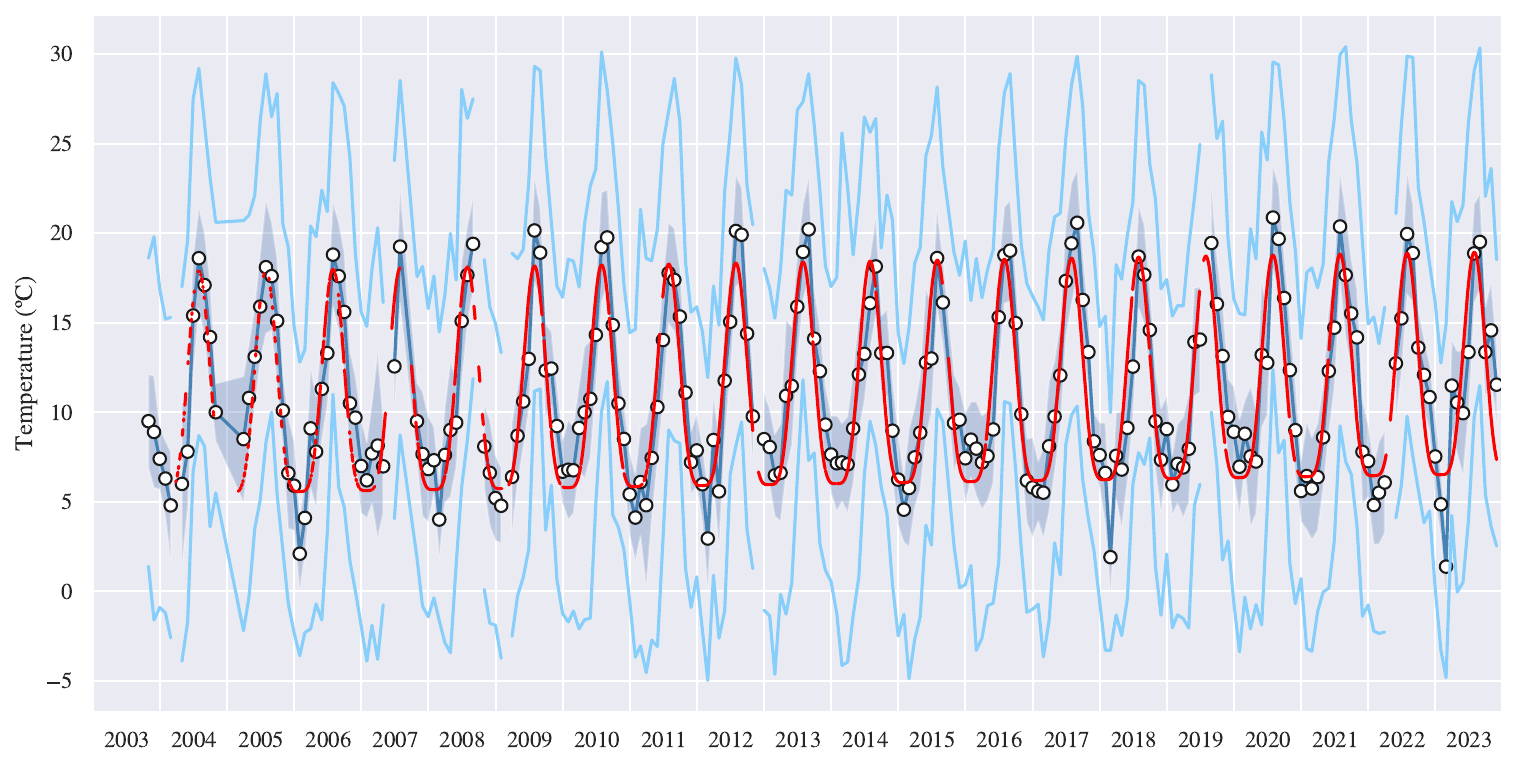}
\caption{
  \label{fig:annualT}
  Temperature times series observed with the MAGIC WS: white circles show the monthly medians, the blue shadow fills the IQR, and the light blue lines show the observed monthly maxima and minima. The red lines show the result of the maximum likelihood analysis, Eq.~\protect\ref{eq:mui}, for those daily medians that were used for the fit. The additional gaps visible here are due to the daily completeness requirement of 85\%.} 
\end{figure*}

\begin{figure}
\centering
\includegraphics[width=0.99\linewidth, clip, trim={0cm 0 0cm 0}]{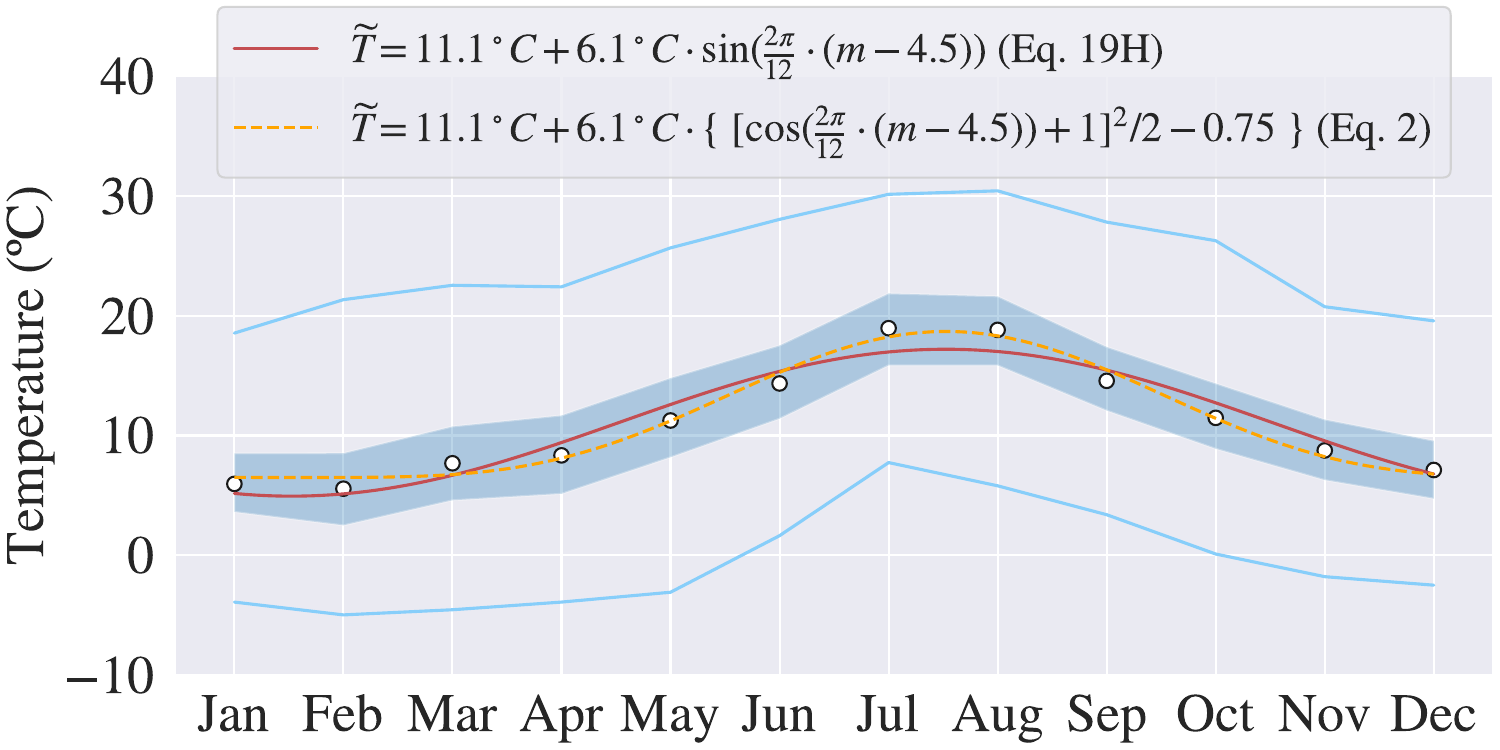}
\caption{
  \label{fig:monthT}
  Seasonal temperature cycle at MAGIC: white circles display the month-wise medians, the blue shadow fills the IQR, and the light blue lines show the observed month-wise maxima and minima. The two fits have been obtained using Eq.~19H and Eq.~\ref{eq:mui}, respectively, under and assumption of no temperature increase over time ($b$=0).}
\end{figure}

%In this section the historic behavior of temperatures is presented. 
Figure~\ref{fig:annualT} shows the temperature time series of the data set, in terms of month-wise statistics: median, the interquartile range (IQR) from 25\% to 75\% percentile of the data, and month-wise temperature extremes. Seasonal variations are clearly visible, reaching temperatures around $\sim$17$^\circ$C--22$^\circ$C on average during the summer months and $\sim$4$^\circ$C--8$^\circ$C on average during winter. Harsher winters were observed at the beginnings of 2006, 2012 and 2018, whereas hotter summers occurred in 2009, 2012, 2013, 2017, 2020, 2022 and 2023. 
%\dd{how are 'harsher' and 'hotter' defined? certain deviation from the 20y average? For the winters, it seems to be the years going below 3-4 deg. Then also 2023 should be listed. Btw 2006 means 2005/06 right? I mean winter starts typically in Nov/Dec already. For the summers, the selection of the years is less clear to me, especially 2022.} 
The lowest temperatures reaching -5$^\circ$C were found in 2012, 2015 and 2023, whereas the high temperature extremes reaching $\geq$30\degC are found in 2010, 2012, 2017, 2021, 2022 and 2023. 
%Note that hotter summers on average do not correlate well with high temperature extremes, similarly to cold winters and low temperature extremes. 
\begin{figure}
\centering
\includegraphics[width=0.99\linewidth, clip, trim={0cm 0 0cm 0}]{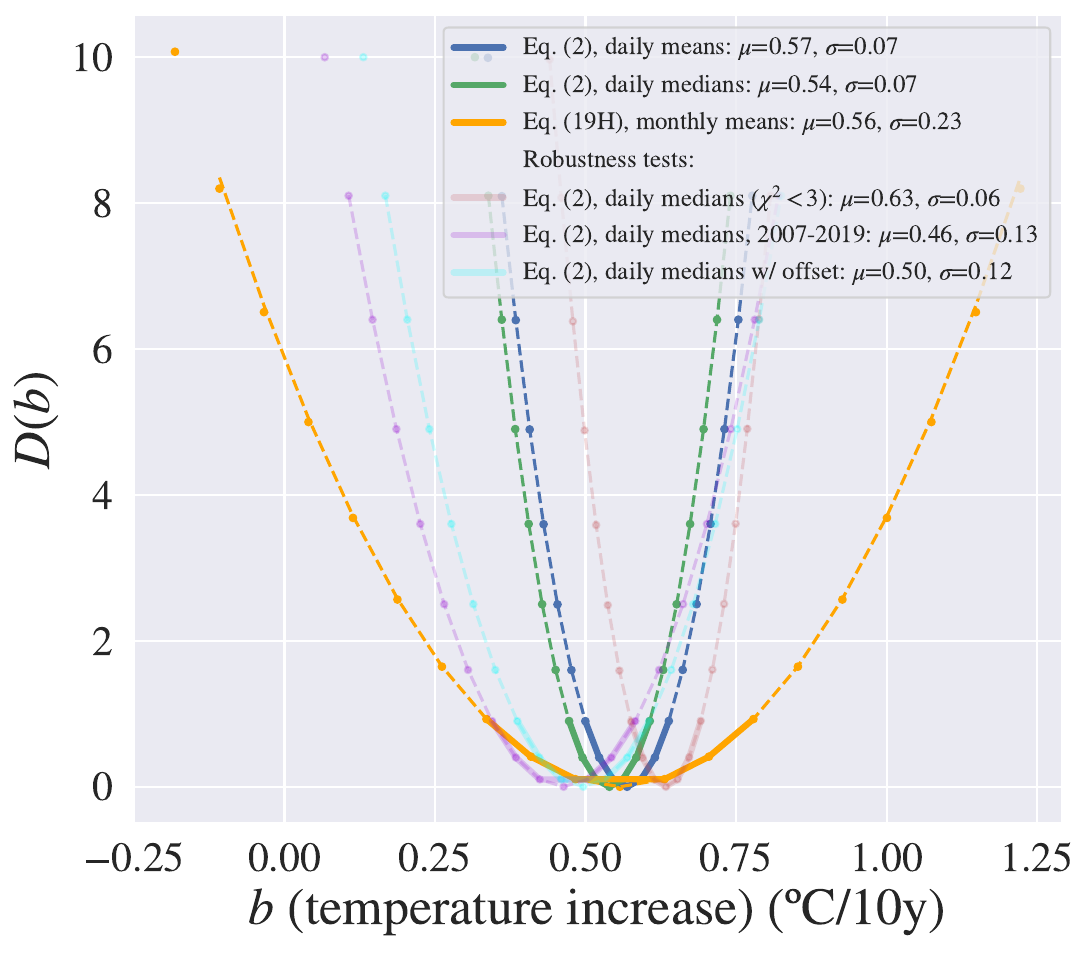}
\caption{
  \label{fig:tempL}
  Profile likelihood test statistic $D(b)$ (see Eq.~\protect\ref{eq.lr1}) as a function of the temperature increase parameter $b$. Shown are the profile likelihoods obtained from daily mean temperatures (blue), daily median temperatures (green), obtained with the fit function Eq.~\ref{eq:mui}, and from monthly means (yellow), fitted with Eq.~19H.  For the latter, a monthly coverage cut of 80\% had been applied. Furthermore, $D(b)$ from three robustness tests are shown: a robust version of the previous fit to daily medians after removing outliers $>$3$\sigma$ (pink), from a reduced period of time that contains only data from 2007-2019 (violet) and after including a possible systematic offset between the WS models MWS~5MV and MWS~55V.  Daily means or medians required a daily coverage of at least 85\%.
  All profile likelihoods have been fitted to a parabolic function $(b-\mu)^2/\sigma^2$ around the minimum, the fit parameters obtained are shown in the legend.}
\end{figure}

\citet{lombardi2006} had been the first to claim an increase in temperatures over time at the ORM, in their case 1.0\degC/decade through a linear fit to annual temperature means obtained from 1985-2004 Carlsberg Meridian Telescope (CAMC) weather data, without providing a fit uncertainty. However, this result was criticized~\citep{Castro-Almazan:2009}. Later, \citet{Graham:2007} found $\sim$0.3\degC/decade for data ranging from 1971 to 2000, however again without calculating statistical uncertainties.
\citet{Pepin:2015} predicted elevation dependent warming (EDW) all over the planet and highlighted the causes of it. This finding was confirmed for the Canary Islands~\citep{Exposito:2015}.
\citet{Martin:2012} predict increasing temperature averages due to climate change in mountainous Canarian areas, confirmed in very long-term data from 36 AEMET-operated weather stations at the neighboring island of Tenerife, where a temperature increase of $(0.31\pm 0.12)^\circ$C/decade (1970-2010) was found for mountainous regions above the stratocumulus layer. \citet{Haslebacher:2022} find past temperature increases  
% Table A.8
%0.11-0.04/1.645
% 0.19+0.09/1.645
ranging from 0.09\degC/decade  to 0.24\degC/decade from historical (until 2014) ERA5 atmospheric reanalysis data, and PRIMAVERA GCM simulations (1950-2014) coupled to local NOT telescope meteo data taken between 2004 and 2019. Their simulation predictions for the future (2015-2050) range from 0.3\degC/decade to 0.5\degC/decade. 
%0.37-0.1/1.645
%0.44+0.11/1.645

\begin{table*}
\centering
\begin{tabular}{l|ccccc|c}
\toprule
Modality  &  $\hat{a}$ & $\hat{b}$ & $\hat{C}$ & $\hat{\phi}$ & $\hat{\sigma}_0$ & $\chi^{2}/\mathrm{NDF}$\\ 
          &  (\degC)  & (\degC/10y) & (\degC) & (month)  & (\degC)  & (1) \\
\midrule
Daily means, Eq.~\ref{eq:mui}    & 10.53 &  0.57 &  6.5 &  7.0 &  3. & 1.00 \\
Daily medians, Eq.~\ref{eq:mui} & 10.06 & 0.54 & 6.2 & 7.0 &  3.0 & 1.00 \\
%Daily medians, Eq.~\ref{eq:mui} (no outliers) &  10.11 & 0.65 & 6.1  & 7.0  & 2.5 & 1.00  \\
%Daily medians, Eq.~\ref{eq:mui} (2007-2019)  & 10.11 & 0.49 & 6.2 & 7.0 & 2.9 & 1.00 \\
Monthly means, Eq.~19H &
                                10.53 & 0.56 &  6.2 & 4.0 & 1.7 & 1.03 \\
\bottomrule
\end{tabular}
\caption{\label{tab:lresults} Parameters that maximize the likelihood, Eq.~\protect\ref{eq:ltemp}, using  Eq.~\ref{eq:mui} for the seasonal cycle, from daily mean and median temperatures and using Eq.~19H from monthly means: $a$ corresponds to the average temperature as of  01/01/2003, $b$ the decadal temperature increase, $C$ the seasonal oscillation amplitude, $\phi$ the location of the seasonal maximum and $\sigma_0$ the average Gaussian spread of temperatures around the fitted seasonal cycle. 
\ccol{The column $\chi^{2}/\mathrm{NDF}$ provides the sum of squared residuals of all data points with respect to the function that maximizes the likelihood, divided by the number of degrees of freedom of the fit.}
}
\end{table*}

In an attempt to compare these findings with our own temperature data set, we performed a maximum likelihood (ML) analysis with a model similar to Eq.~19 of \citet{Haslebacher:2022} (hereafter called \textit{Eq.~19H}). We used, however, a slightly adapted seasonal model for the temperatures, instead of the simple sine function employed by \citet{Haslebacher:2022}, namely a function that allows for a certain adjustment of the kurtosis (see, e.g.~\citet{Soldevilla:2016}):

\begin{align}
\mu_i &= a + \ddfrac{b}{120}\cdot x_i + {} \nonumber\\ 
      &  {} ~ + \ddfrac{C}{2}\cdot \left(\left(\cos\left(\ddfrac{2\pi}{12}\cdot (x_i-\phi)\right)+1\right)^2 - \ddfrac{3}{2} \right)  \label{eq:mui} \quad.
\end{align}
\noindent
Here, $\mu_i$ denotes the mean or median, daily or monthly, temperature, $a$ the average temperature expectation for 01/01/2003, $b$ the linear temperature increase per decade (120~months), $C$ the yearly oscillation amplitude and $\phi$ the oscillation phase. Data points $x_i$ contain the corresponding day or month, always in units of months, counted since 01/01/2003.

The seasonal cycle of temperature is shown in Fig.~\ref{fig:monthT}, in which also two fits with functions Eq.~19H and our Eq.~\ref{eq:mui}, respectively, are shown. The summer temperature medians lie at $\sim$18$^\circ$C, whereas winter temperatures reach on average $\sim$6--7$^\circ$C. As expected, temperature minima are found in February, whereas the maxima are reached in July and August. The observed maximum temperature in February is somewhat surprising because it exceeds the seasonal trend. 

\begin{comment}
\begin{table}
\centering
\begin{tabular}{l|ccc}
\toprule
Modality  &  68\% & 95\% & 99\% \\ 
&  (\degC/10y) &  (\degC/10y)&  (\degC/10y)\\
\midrule
From daily means    & (0.51, 0.65) & (0.43, 0.73) & (0.36,0.80) \\
From daily medians &  (0.48, 0.62) & (0.41, 0.69) & (0.34,0.76) \\
\bottomrule
\end{tabular}
\caption{\label{tab:CLb} 1$\sigma$, 2$\sigma$ and 3$\sigma$ CIs for the temperature increase parameter $b$, obtained from the maximum likelihood ratio test.  }
\end{table}
\end{comment}

\begin{figure}
\centering
\includegraphics[width=0.99\linewidth, clip, trim={0cm 0 0cm 0}]{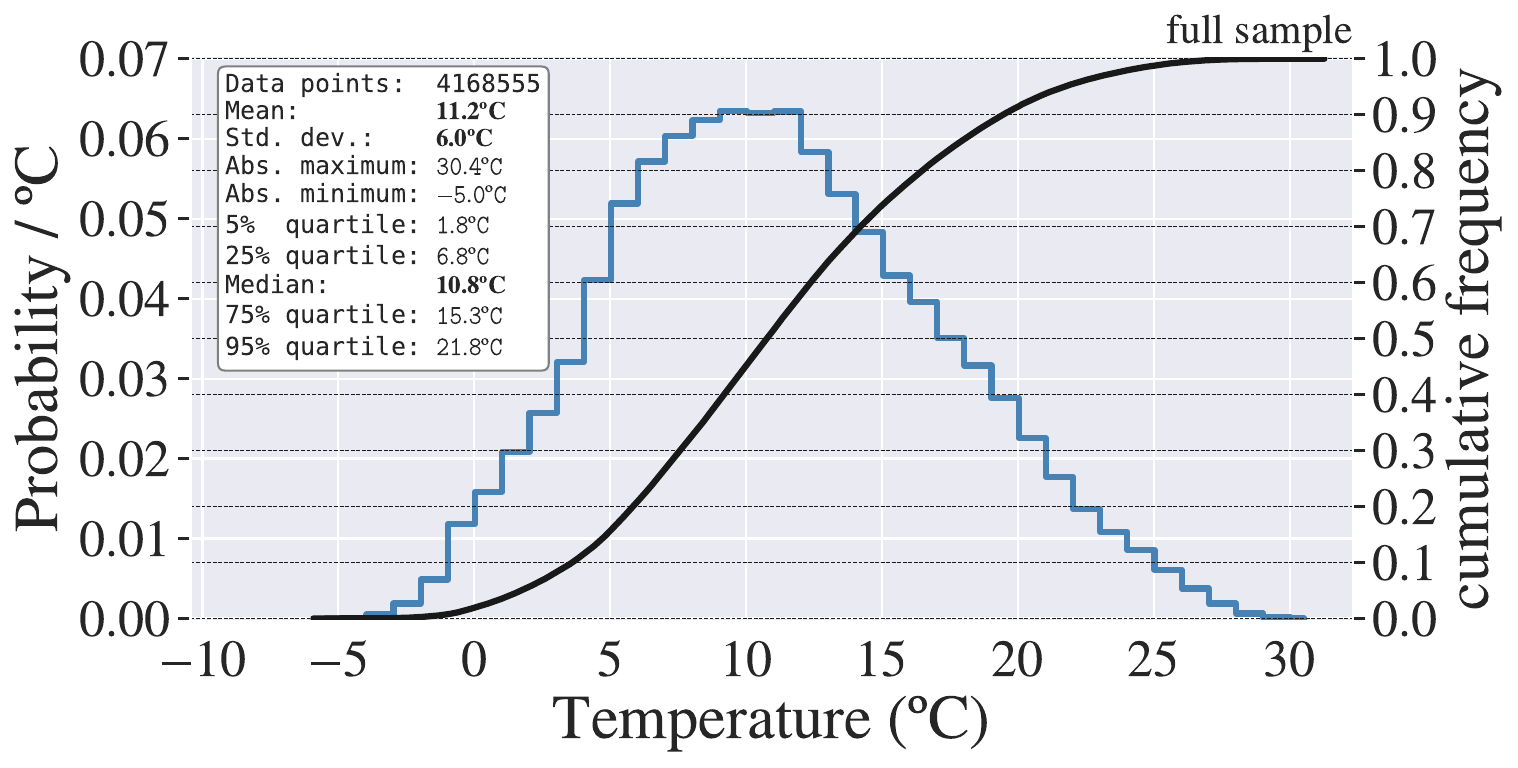}
\includegraphics[width=0.99\linewidth, clip, trim={0cm 0 0cm 0}]{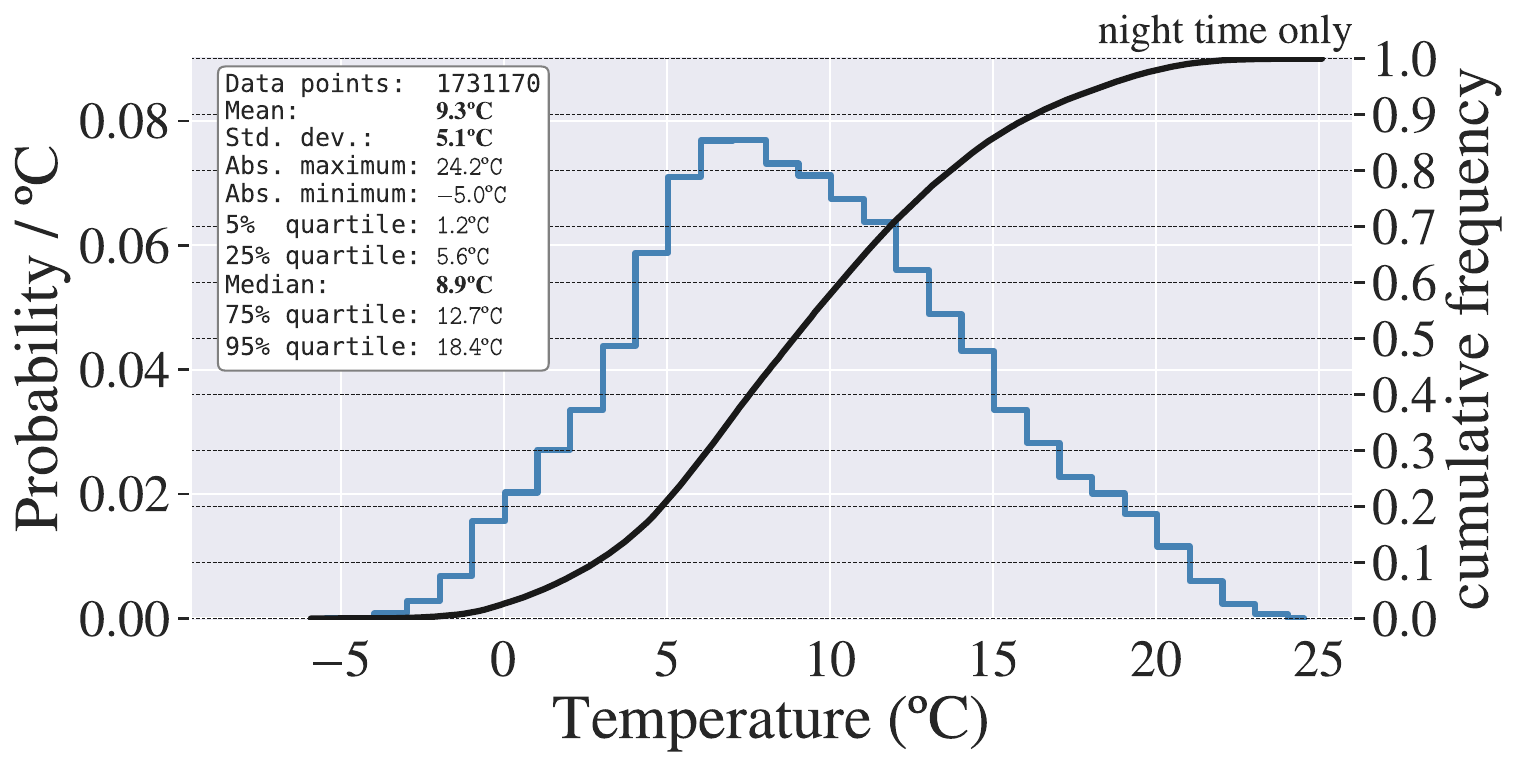}
\caption{
  \label{fig:histT}
  Distribution and statistical parameters of temperatures at the MAGIC site.}
\end{figure}

We finally used the profile likelihood ratio test method (see Eq.~\ref{eq.lr1} and further discussion of the method in Appendix~\ref{sec:likelihood}) to determine a possible temperature increase through the parameter of interest, $b$. First, we tested that the distribution of daily mean or median temperatures follows a normal distribution $\mathcal{N}(\mu_i,\sigma_i)$ around the expectation Eq.~\ref{eq:mui}. We found reasonable Gaussian behavior and a maximum change of less than 10\% for the seasonal cycle of $\sigma_i$. With this result, we decided to fit our data to a single global spread $\sigma_0$ and constructed a likelihood of the form: 
\begin{align}
\mathcal{L} &= \prod_{i=0}^{N} \mathcal{N}(t_i,\mu_i(b,\vec{\nu};x_i),\sigma_0) \quad,
\label{eq:ltemp}
\end{align}
\noindent
where $t_i$ are the (daily or monthly) temperature means or medians, the parameters $a$, $C$ and $\phi$ have been subsumed under the nuisance parameter vector $\vec{\nu}$. The Gaussian spread $\sigma_0$ is also treated as a nuisance parameter. 
For the numerical likelihood maximization, we initialized all parameters to values that fit the seasonal cycle (Fig.~\ref{fig:monthT}) and $b$ to zero.
Table~\ref{tab:lresults} shows the results of the parameters that maximize the likelihood, Eq.~\ref{eq:ltemp}, for the different cases studied: daily mean and median temperatures, fitted to Eq.~\ref{eq:mui}, and, for a direct comparison with \citet{Haslebacher:2022}'s method, the monthly means fitted to their Eq.~19H.
In addition, a robust analysis has been carried out that eliminates outliers from the previous case of daily medians, a reduced data set from a time period of only 2007 to 2019, when only one WS model was used, and an analysis that includes a possible systematic offset between the two WS models MWS~5MV and MWS~55V to the list of nuisance parameters. The latter yielded only 0.05\degC as the best-fit systematic offset, well below the sensor’s specified accuracy.  
All scenarios obtain a considerably larger temperature increase of $\sim$0.55\degC/decade, about twice as high as \citet{Haslebacher:2022}'s historical simulations and even slightly higher than the upper range obtained from their simulation predictions for the period 2015-2050. 
In addition to that, means yield 0.5\degC higher temperature averages than medians, pointing to an asymmetric distribution of the diurnal temperatures. 
%The spread between both methods gets slightly reduced after removal of outliers, as expected. 
All methods yield an excellent fit quality, shown in the column $\chi^{2}/\mathrm{NDF}$, with Eq.~19H being only marginally worse.

Since the interdecadal temperature increase is somewhat higher than expected, we profiled this parameter with the likelihood ratio test (Eq.~\ref{eq.lr1}). The results are shown in Fig.~\ref{fig:tempL}. One can see that all methods produce compatible results, with improved precision obtained by daily data points over monthly ones, as expected. 
Considering that the other methods can only be considered robustness tests, we obtain $b=0.55\pm0.07\mathrm{(stat.)}\pm0.07\mathrm{(syst.)}$\degC/decade. 
%show 1$\sigma$, 2$\sigma$ and 3$\sigma$ confidence intervals (CI) for the parameter $b$, obtained for daily means and daily medians in Table~\ref{tab:CLb}. 
A systematic uncertainty of about 0.07\degC/decade can be obtained from the spread of the results between the different methods and data samples. Note that the fact that the uncertainty is smaller than the accuracy of the temperature sensor of 0.3\degC is not a contradiction, since the frequent exchange of (calibrated) weather stations must cancel out most of the calibration uncertainties if we assume that these are normally distributed. Even in the unlikely case of systematic absolute scale differences between the three WS models employed, conspiring in a way to mimic a temperature increase in the range of 0.3\degC calibration accuracy, we have to take into account that the three different models were exchanged back and forth intermittently, hence largely reducing such an effect on the retrieved long-term temperature increase. In any case, we explicitly tested a possible systematic calibration offset hypothesis between the two models MWS~5MW and MWS~55V, and included such an offset in the list of nuisance parameters. The likelihood got maximized with an offset of only 0.05\degC and an according shift of $\hat{b}$ by the same amount. Figure~\ref{fig:tempL} includes the profiled test statistic on $b$ for this test. We incorporated the small displacement found in the systematic uncertainty of $b$. 
% 0.58 +- 0.07
% 0.62 +- 0.07
% 0.39 +- 0.13
% 0.49 +- 0.08 
% --> weighted total: 0.55 +- 0.04
\ccol{The test of possible individual miscalibration or drifting periods (see Sect.~\ref{sec:discussion}) yielded compatible results within the stated systematic uncertainty.}

\begin{figure}
\centering
\includegraphics[width=0.99\linewidth, clip, trim={0cm 0 0cm 0}]{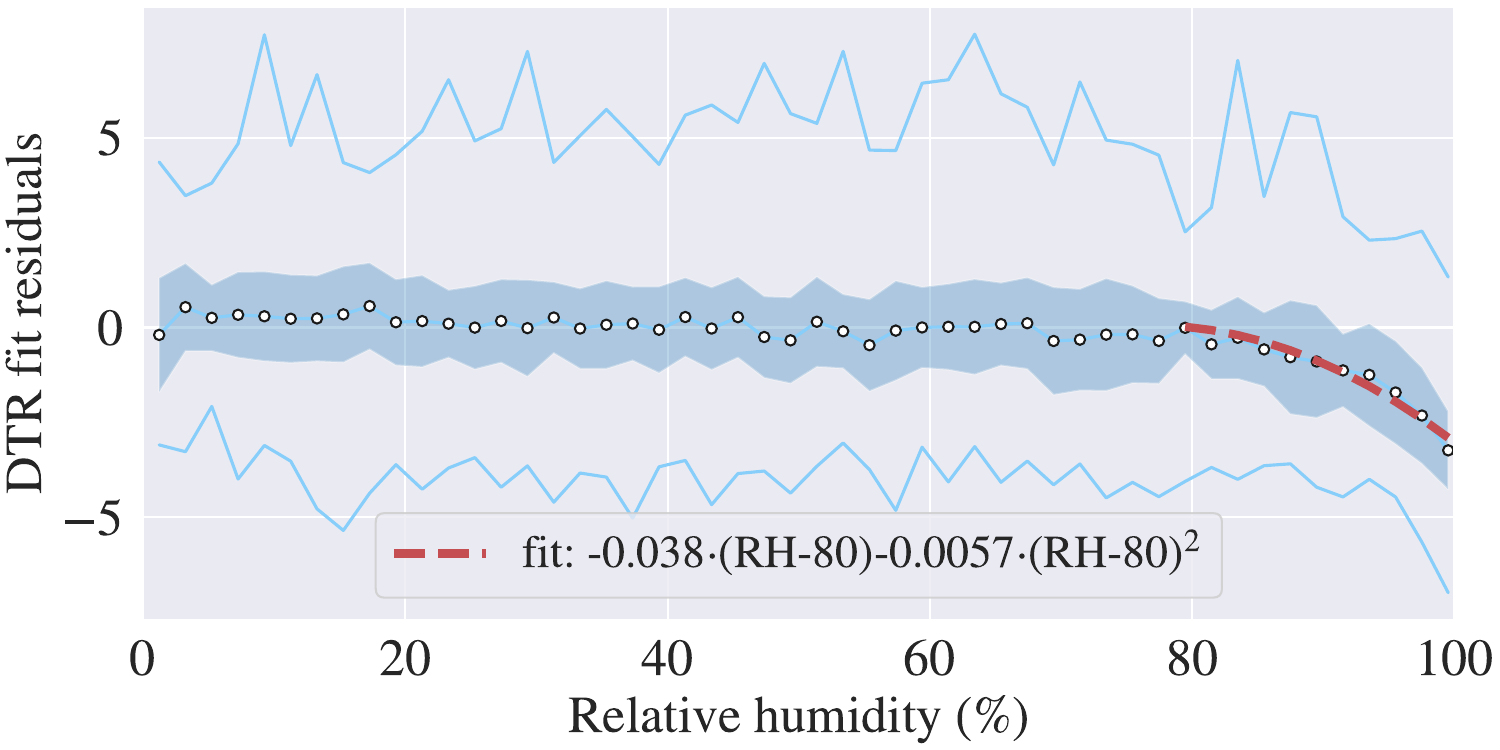}
\caption{
  \label{fig:diurnal_corr_RH} Diurnal Temperature Ranges (DTR) fit residuals as a function of mean relative humidity (RH) of the corresponding day. 
   White circles display the observed medians, the blue shadow fills the IQR, and the light blue lines display the maxima and minima found in each RH bin. The diurnal temperature range medians found for average RHs larger than 80\% have been fitted to a quadratic function (red dashed line).}
\end{figure}

\begin{figure*}
\centering
\includegraphics[width=0.99\textwidth, clip, trim={0cm 0 0cm 0}]{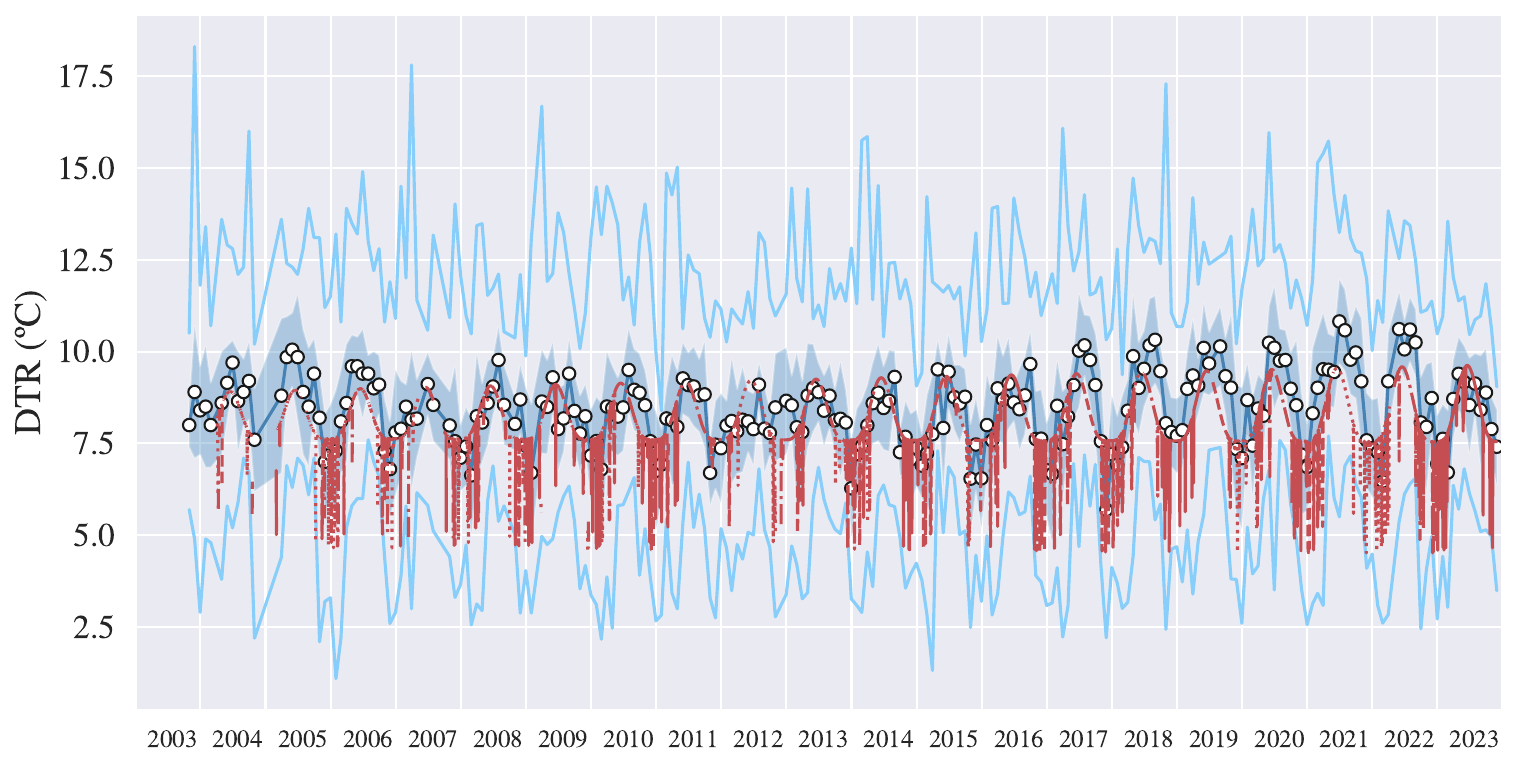}
\caption{
  \label{fig:diurnalT}
  DTR times series: the white circles show the monthly medians, the blue shadow fills the IQR, and the light blue lines show the observed monthly maxima and minima. The red lines show the result of the rain-bias-corrected maximum likelihood analysis for those daily medians that were used for the fit. The additional gaps visible here are due to the (higher) daily completeness requirement of 95\%.} 
\end{figure*}

\begin{table*}
\centering
\begin{tabular}{l|cccccc|c}
\toprule
Modality  &  $\hat{a}$ & $\hat{b}$ & $\hat{C}$ & $\hat{\phi}$ & $\widehat{\Delta C}_m$ & $\hat{\sigma}_0$ & $\chi^{2}/\mathrm{NDF}$\\ 
          &  (\degC)  & (\degC/10y) & (\degC) & (month)  & (\degC/10y) & (\degC)  & (1) \\
\midrule
Eq.~\ref{eq:mui}     & 7.94 & 0.14 & 1.0 & 5.7 & n.a. & 1.8 & 1.00 \\
Eq.~\ref{eq:muiCx}   & 7.96 & 0.13 & 0.8 & 5.7 & 0.25 & 1.8 & 1.00 \\
Eq.~\ref{eq:muiCx} 
(with high RH corr.) & 8.14 & 0.13 & 0.6 & 5.5 & 0.33 & 1.7 & 1.00 \\
\bottomrule
\end{tabular}
\caption{\label{tab:lresultsDTR}  Parameters that maximize the likelihood, Eq.~\protect\ref{eq:ltemp} for DTRs using Eq.~\ref{eq:mui}, Eq.~\ref{eq:muiCx} and Eq.~\ref{eq:muiCx} with a correction applied for \textit{RH}$>$80\%. The parameter $a$ corresponds to the average DTR as of 01/01/2003, $b$ the decadal increase in DTR, $C$ the seasonal oscillation amplitude, $\phi$ the location of the seasonal maximum, $\Delta C_x$ the decadal increase in the oscillation amplitude, and $\sigma_0$ the average Gaussian spread of DTRs around the fitted seasonal cycle. 
\ccol{The column $\chi^{2}/\mathrm{NDF}$ provides the sum of squared residuals of all data points with respect to the function that maximizes the likelihood, divided by the number of degrees of freedom of the fit.}}
\end{table*}

\begin{figure}
\centering
\includegraphics[width=0.99\linewidth, clip, trim={0cm 0 0cm 0}]{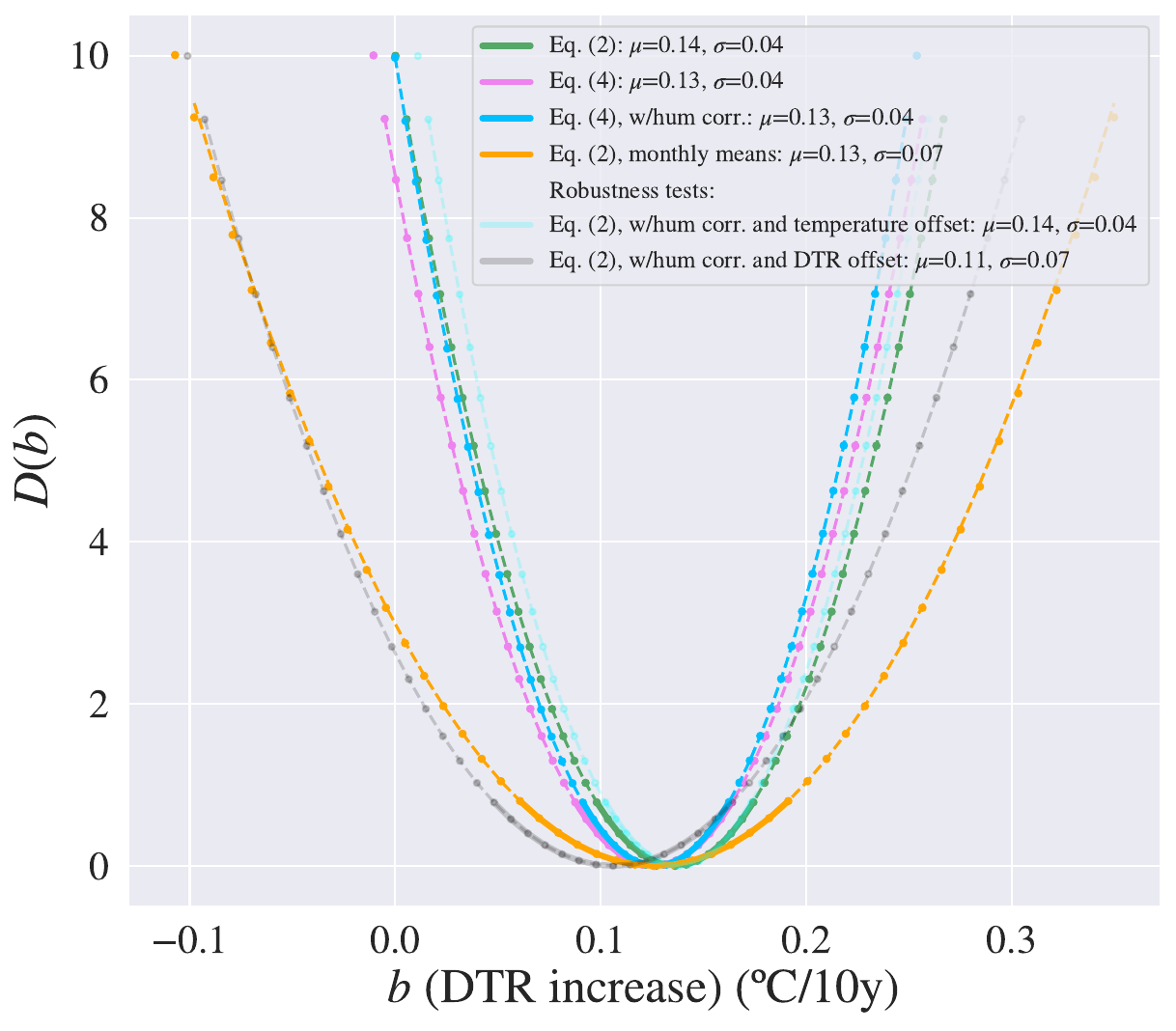}
\caption{
  \label{fig:DTRL}
  Profile likelihood test statistic $D(b)$ (see Eq.~\protect\ref{eq.lr1}) as a function of the DTR increase parameter $b$. Shown are the profile likelihoods obtained  with the fit function Eq.~\ref{eq:mui} (green), with Eq.~\ref{eq:muiCx} (pink), with a correction for $\textit{RH}>80\%$ (light blue), and from monthly means (yellow).  For the latter, a monthly coverage cut of 80\% had been applied. Furthermore, $D(b)$ from two robustness tests are shown: with the systematic temperature offset between the WS models MWS~5MV and MWS~55V, obtained from the robustness test carried out in Sect.~\ref{sec:temperature} (cyan), and with a possible systematic DTR offset between both models added to the list of nuisance parameters (gray). Daily means or medians required a daily coverage of at least 95\%.
  All profile likelihoods have been fitted to a parabolic function $(b-\mu)^2/\sigma^2$ around the minimum, the obtained fit parameters are shown in the legend.}
\end{figure}

\begin{figure}
\centering
\includegraphics[width=0.85\linewidth, clip, trim={0cm 0 0cm 0}]{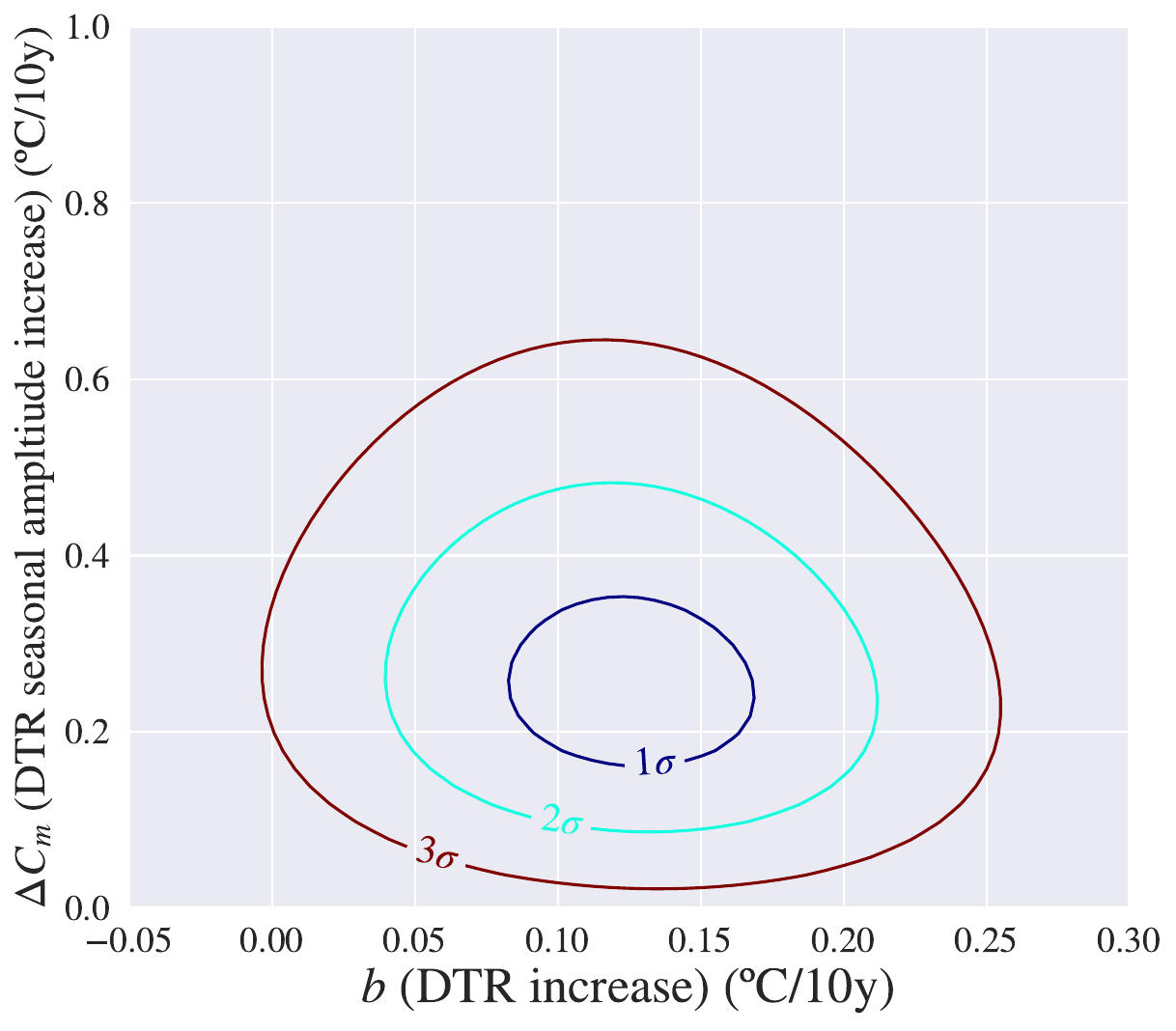}
\includegraphics[width=0.85\linewidth, clip, trim={0cm 0 0cm 0}]{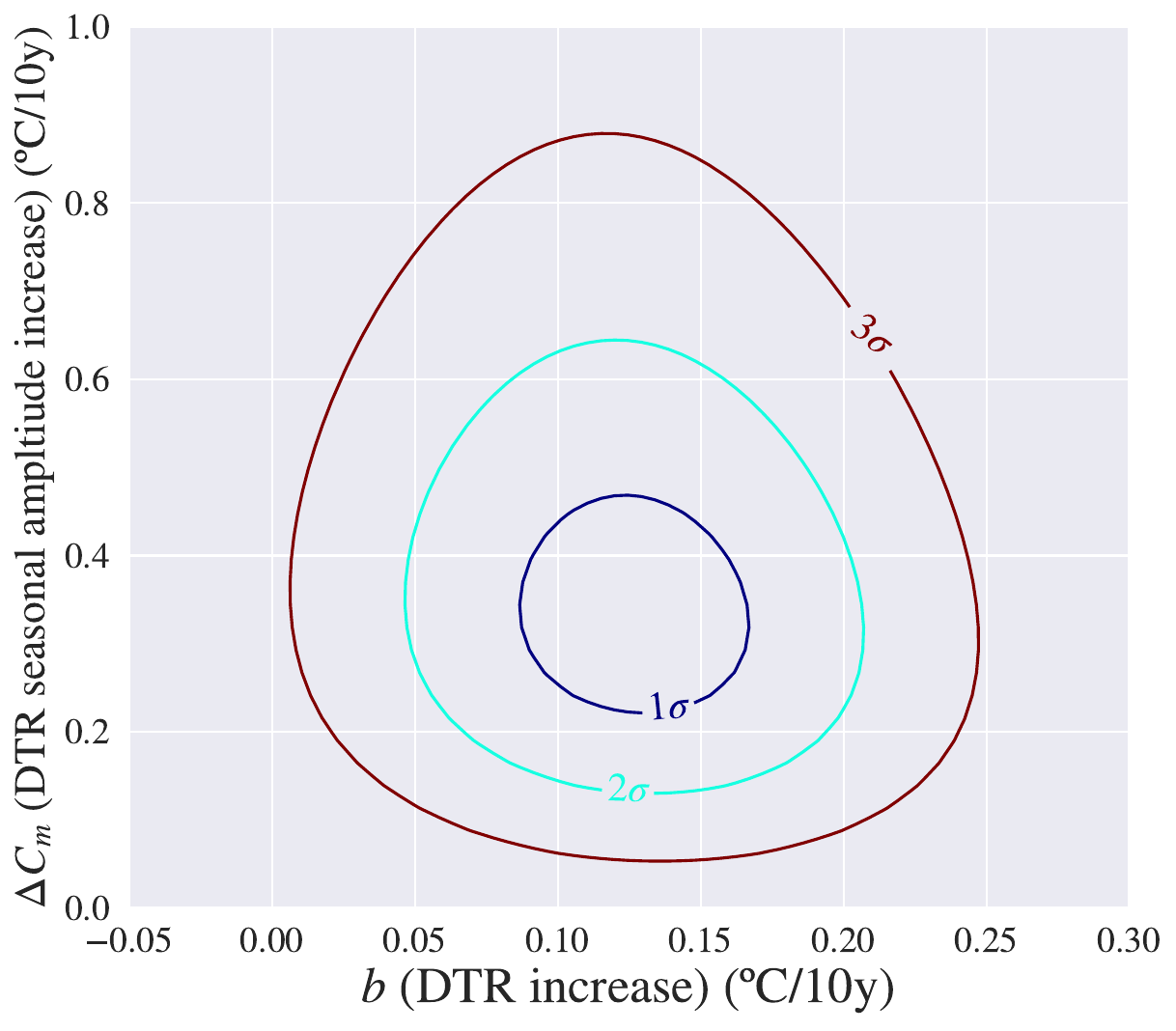}
\caption{
  \label{fig:diurnal_contour} Profile likelihood uncertainty ellipses for the two parameters of interest: a linear increase of the DTR over time (horizontal axis) and a linear increase of seasonal oscillation amplitude of the DTR with time (vertical axis), see Eq.~\protect\ref{eq:muiCx}. The upper plot shows the case of no correction for high humidity, the  lower one with a correction applied for rains (see Fig.~\protect\ref{fig:diurnal_corr_RH}).}
\end{figure}

\begin{figure*}
\centering
\includegraphics[width=0.99\textwidth, clip, trim={0cm 0 0cm 0}]{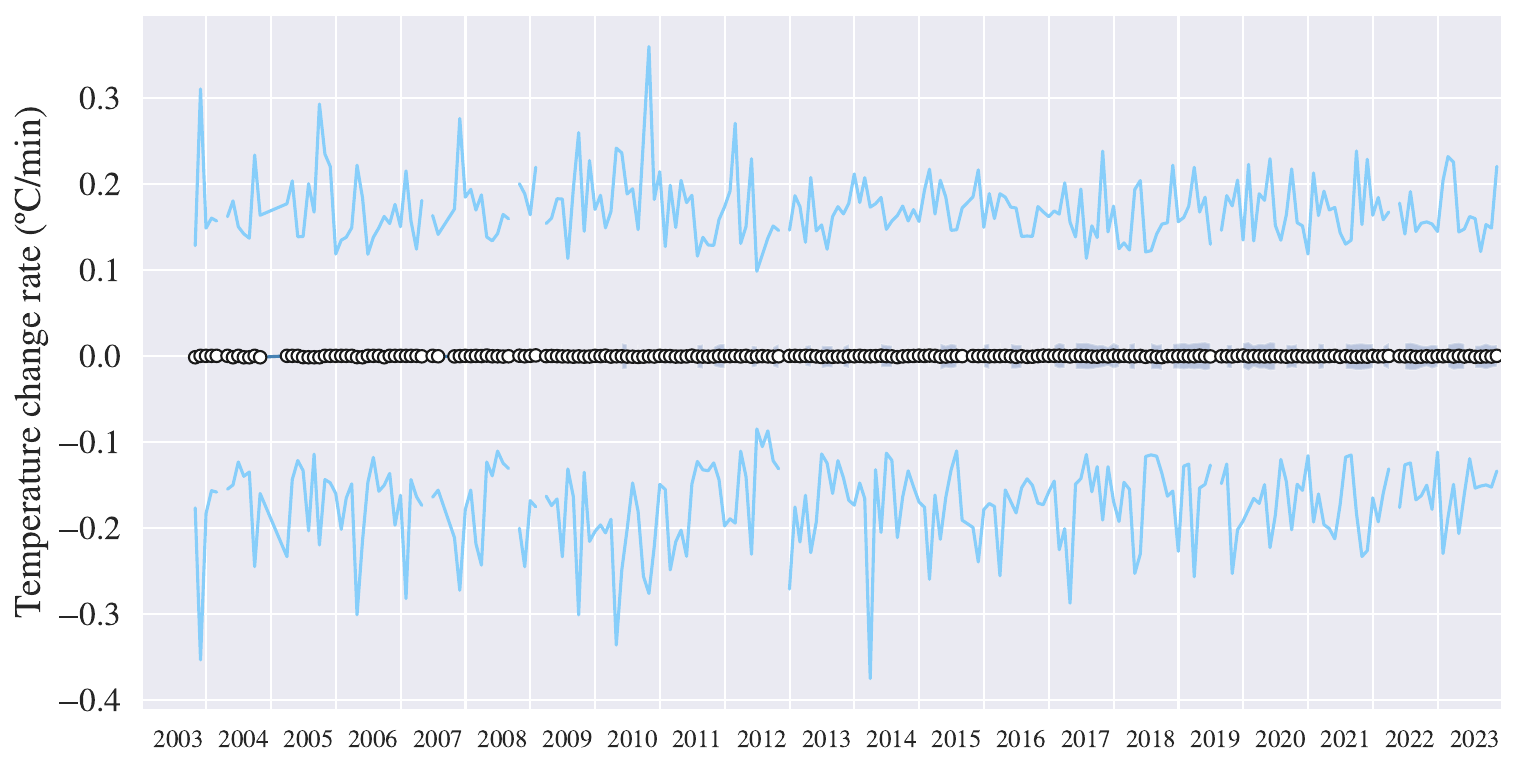}
\caption{
  \label{fig:tcrsencer}
   Times series of temperature change rates for 20 minutes: white circles show the monthly medians, the blue shadow fills the IQR, and the light blue lines show the observed monthly maxima and minima. } 
\end{figure*}

\begin{figure}
\centering
\includegraphics[width=0.99\linewidth, clip, trim={0cm 0 0cm 0}]{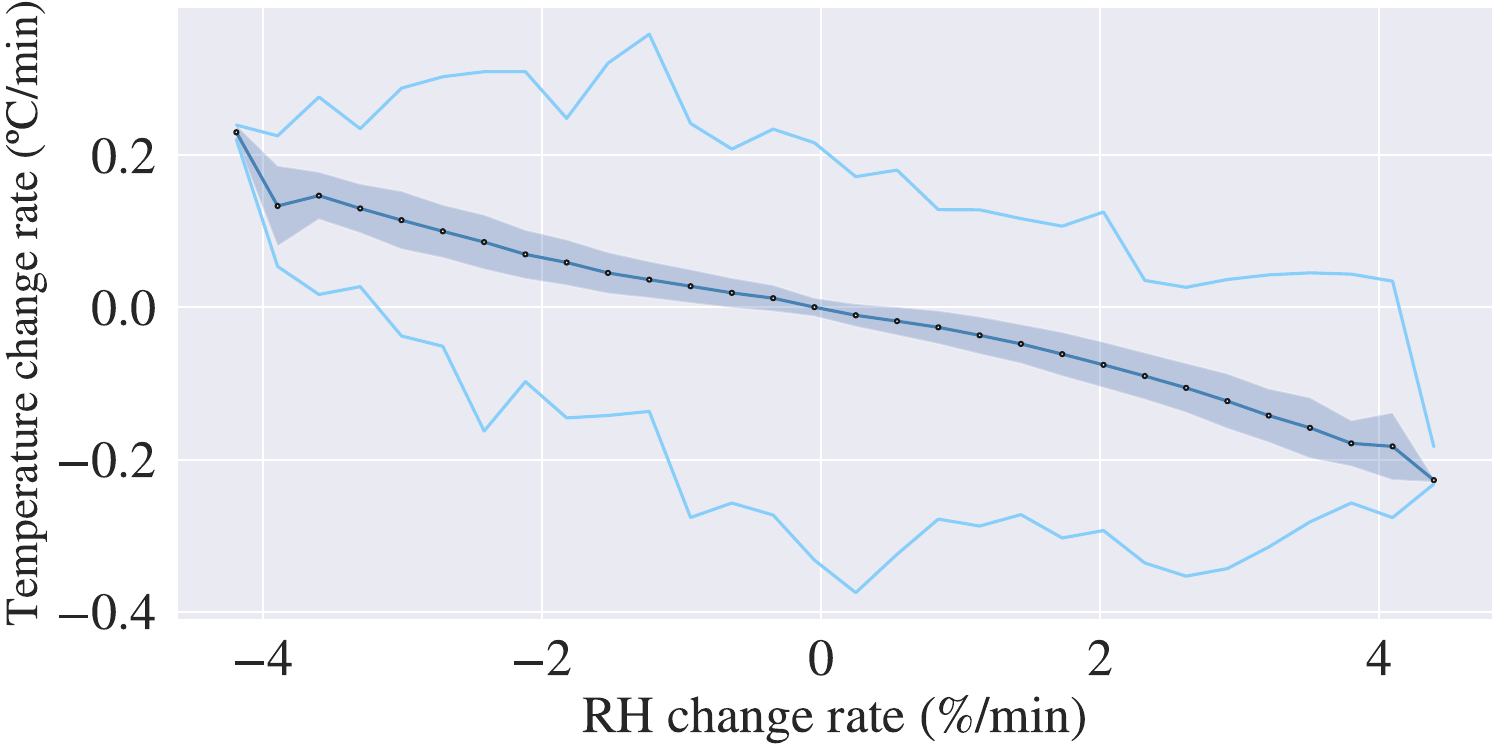}
\caption{
  \label{fig:TgradientvsRgradient}
   Profile of temperature change rates during 20 minutes vs. relative humidity change rates during 20 min: white circles show the medians of each bin, the blue shadow fills the IQR, and the light blue lines the observed maxima and minima. 
   } 
\end{figure}

As a last systematic check, we tested possible seasonal phase shifts or amplitude increases over the years but found no significant effect, particularly none that would alter the result on the global temperature increase.

%Finally, 
Figure~\ref{fig:histT} shows the complete distribution and statistical analysis of the temperatures, separately for the entire sample and only at night (Sun elevation $<$~-12$^\circ$). For this analysis, the procedure described in Sect.~\ref{sec:datareduction} was used.

% ORM: 9.5 = 24h of Table 3 (Varela: 2346 m)
% TNG: 9.8 = average of Table 2 (Lombardi, 2387+10 = 2397 m)
% CAMC: 8.8 = average of Table 2 (Lombardi, 2326+10 = 2336 m)
An average temperature of 11.2$^\circ$C for the full sample is found. This is 1.4$^\circ$ (compared to TNG) and 2.4$^\circ$C (compared to CAMC) higher than those obtained on the mountain rim~\citep{lombardi2006, Varela:2014}, compatible with the expectation of adiabatic lapse rate for the comparison with TNG, but $\sim$1.4$^\circ$ higher than expectations for the CAMC comparison. The latter were obtained from
earlier years (1985-2004); hence, the difference may be partially explained by the global temperature increase over time.
% (2346-2188)*0.0065 = 1.03 degC vs. 1.7 deg C
% (2402-2188)*0.0065 = 1.4 degC vs.  1.4 deg.C
% (2336-2188)*0.0065 = 0.96 degC vs. 2.4 deg C

\subsubsection{Diurnal Temperature Ranges}

We have calculated diurnal temperature ranges (DTR), that is, the maximum temperature found during a 24-hour time span minus its minimum. For this analysis, days were defined as starting and ending at 8\,h a.m. This definition ensures that the maxima (normally found after noon) are associated with their subsequent nightly minima (normally found before sunset). In order to minimize possible undesired effects from varying sensor precision over time, we calculated the daily maxima from rolling temperature averages over 10 minutes (i.e., five data recordings around the absolute daily temperature maxima) and accordingly for the daily minima. We tested that our results are not sensitive to the exact size of the averaging window chosen and even the single daily maximum or minimum temperature entries. 

It is generally known that a
decrease in DTR can be associated with an increase in cloud
coverage because clouds cause stronger decreases in daily temperature maxima than in minima.  Cloudy days do not allow the sun to heat the
atmosphere as strongly as during sunny days. Similarly, DTRs tend to be smaller during rain. 
Other reasons for changes in DTR are related to aerosols~\citep{Sanroma:2010}.
Moreover, there is a seasonal cycle of DTRs, associated with the change of length of a solar day and the maximum altitude of the Sun. In a recent study, \citet{Wang:2023} found significant changes in DTR over the past 20 years in North America, Africa, and Antarctica, all with a negative index, which could be attributed to changes in land use~\citep{Kalnay:2003}.  \citet{Sun:2019} found a reversal of that trend, around 2014, on a global scale, but with different magnitudes for the northern and southern hemispheres.  A trend of DTR for the mountainous area of Tenerife~\citep{Sanroma:2010,Martin:2012}
could not be confirmed with data from 1970 to 2010, however, a strong positive trend of DTR is predicted using detailed \textit{Weather Research and Forecasting model} (WRF)\footnote{\url{https://www.mmm.ucar.edu/models/wrf}} simulations for the period 2045-2054~\citep{Exposito:2015}, caused by a decrease in soil moisture and a slight reduction in cloud cover.

We fitted the seasonal cycle of the DTRs to Eq.~\ref{eq:mui} by maximizing the likelihood of Eq.~\ref{eq:ltemp}. Figure~\ref{fig:diurnal_corr_RH} shows the residuals of the actual DTRs minus the fit expectation, as a function of the daily relative humidity (RH) averages. One can clearly observe a systematic trend towards negative residuals when the RH reaches 80\%, as expected.  That trend can be fitted to a quadratic function, as shown in Fig.~\ref{fig:diurnal_corr_RH}. 
In a next step, we remove the bias introduced by rain and subtract the quadratic fit result from Eq.~\ref{eq:mui}, whenever the daily RH averages reach values higher than 80\%. As in the case of the overall temperatures, a profile likelihood analysis reveals a significant increase in the DTRs over time. Moreover, visual inspection of the data makes an increase in the seasonal oscillation amplitude with time plausible. For that reason, we tested a modified seasonal cycle hypothesis of the form: 
\begin{align}
\mu_i &= a + \ddfrac{b}{120}\cdot x_i + {} \nonumber\\ 
      &  {} ~ + \ddfrac{C}{2}\cdot \left(1+ \ddfrac{\Delta C_m}{120}\cdot x_i\right) \cdot \left(\left(\cos\left(\ddfrac{2\pi}{12}\cdot (x_i-\phi)\right)+1\right)^2 - \ddfrac{3}{2} \right)  \label{eq:muiCx} \quad,
\end{align}
\noindent
with or without humidity correction. The parameters that maximize the likelihood are shown in Table~\ref{tab:lresultsDTR}, showing a positive trend, both for the overall DTR over time (parameter $b$) and for the amplitude of the seasonal oscillation (parameter $\Delta C_m$). Figure~\ref{fig:diurnalT} shows the monthly DTRs, together with the fit result to Eq.~\ref{eq:muiCx}, with a high-humidity correction applied. 

We present the test statistic $D(b)$ of the profiled likelihood for the decadal DTR increase in Fig.~\ref{fig:DTRL}, and a two-dimensional profile likelihood for the parameters $b$ and $\Delta C_m$ in Fig.~\ref{fig:diurnal_contour}, for both cases with and without high humidity correction. We observe that both parameters are largely uncorrelated and that the removal of high humidity does not alter the location of $\hat{b}$ and $\widehat{\Delta C}_m$. However, it improves the precision on $\widehat{\Delta C}_m$. The increase in DTR over time is robust against possible calibration offsets of the temperature sensor and possible offsets of the DTR itself, shown as robustness tests in Fig.~\ref{fig:DTRL}. 

The found increase in DTR of $b=0.13\pm0.04\mathrm{(stat.)}\pm0.02\mathrm{(syst.)}$ \degC/decade,
accompanied by an increase in the amplitude of the seasonal oscillation of $\Delta C_m=0.29\pm0.10\mathrm{(stat.)}\pm0.04\mathrm{(syst.)}$ \degC/decade, 
confirms the predictions for the mountainous areas of the Canary Islands, that is, we have stronger increases in DTR during summer than during winter.

\subsubsection{Temperature Change Rates}

Temperature change rates or temperature gradients cause materials to age faster due to thermal fatigue~\citep{Buschow:2001a}. For this reason, the CTAO has elaborated maximum temperature change rate requirements for their sites, e.g., that damage shall not occur
due to air temperature gradients of 0.5\degC/min for 20 minutes.  

We have calculated temperature change rates by first replacing the temperatures by the respective mean of a rolling window of 7~minutes (i.e.\ averaging over three consecutive samples), in order to reduce small fluctuations of the sensor. These averaged temperatures were then subtracted from the previous temperature of ten samples (corresponding to time differences of $\sim$20~min), and divided by the actual time difference. Data gaps were taken into account in this procedure. 

The resulting temperature change rates are shown in Fig.~\ref{fig:tcrsencer}. 
One can see that change rates above 0.4\degC/min or below -0.4\degC/min are not observed, except for the few exceptional cases due to reflected sunlight and possible lightnings explained in Sect.~\ref{sec:qualitychecks}, which had been removed from this analysis. Moreover, temperature change rates are typically very small, and lie only very occasionally above 0.01\degC/min or below -0.01\degC/min. 

The temperature change rates are strongly anti-correlated with the change rates of relative humidity, as shown in Fig.~\ref{fig:TgradientvsRgradient}. We interpret this finding with the frequent occurrence of clouds moving into and out of the site. These clouds cover the MAGIC counting house during a given time interval (often less than half an hour) and move away again. Such a situation causes relative humidity to increase immediately and the temperature to drop. When the cloud moves away, the opposite happens.  Other, less important causes of sudden temperature changes are higher clouds that catch sunlight.

\begin{figure*}
\centering
\includegraphics[width=0.99\textwidth, clip, trim={0cm 0 0cm 0}]{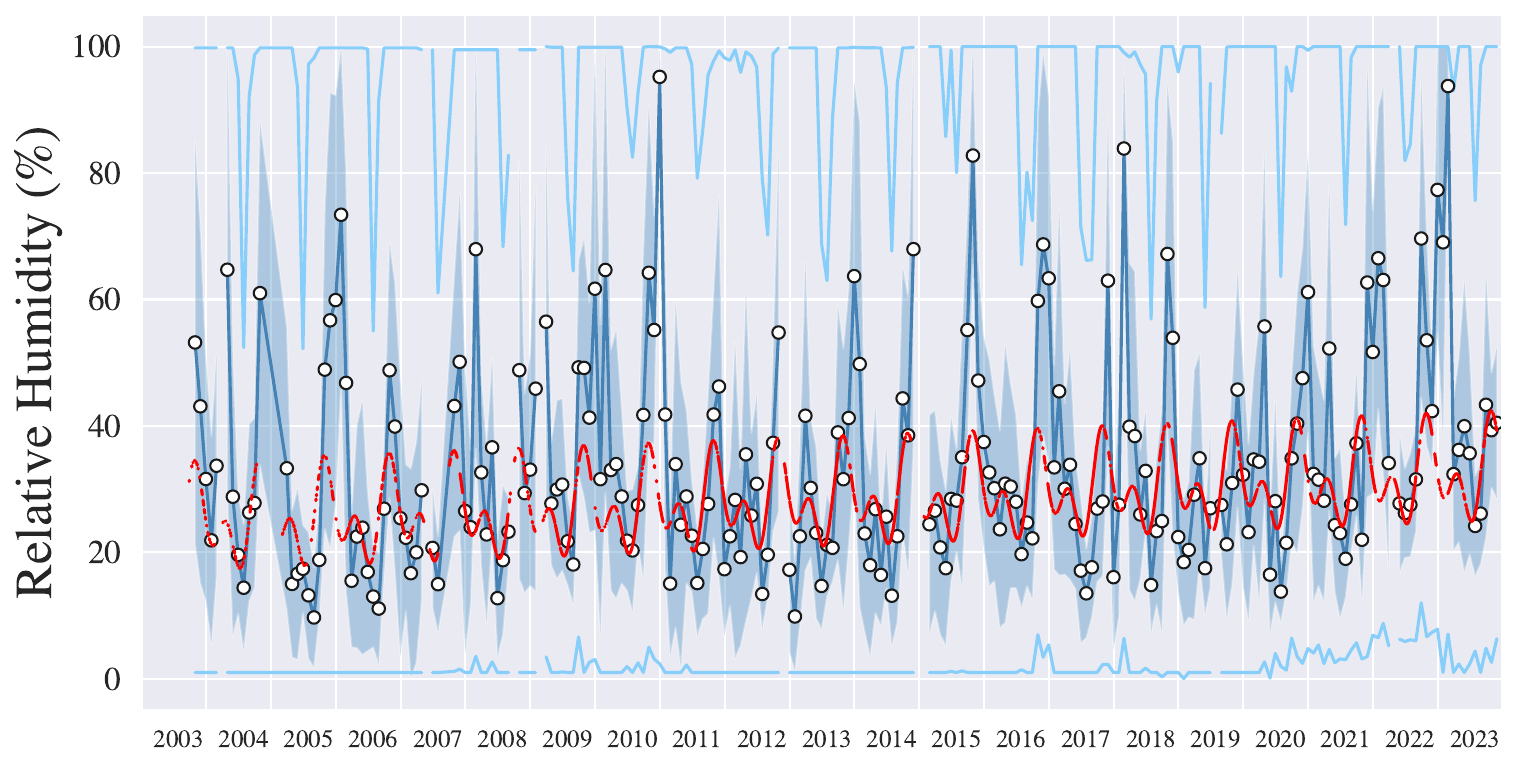}
\caption{
  \label{fig:annualRH}
 Relative humidity times series observed with the MAGIC weather station: white circles show the monthly medians, the blue shadow fills the IQR, and the light blue lines show the observed monthly maxima and minima. The red lines show the result of the fit Eq.~\protect\ref{eq:murh} to the parts of the data with no precipitation (i.e.\ RH$<$90\%). 
 Additional gaps visible in the fit are due to the (higher) daily data completeness requirement of 95\%.}
\end{figure*}

\begin{figure}
\centering
\includegraphics[width=0.99\linewidth, clip, trim={0cm 0 0cm 0}]{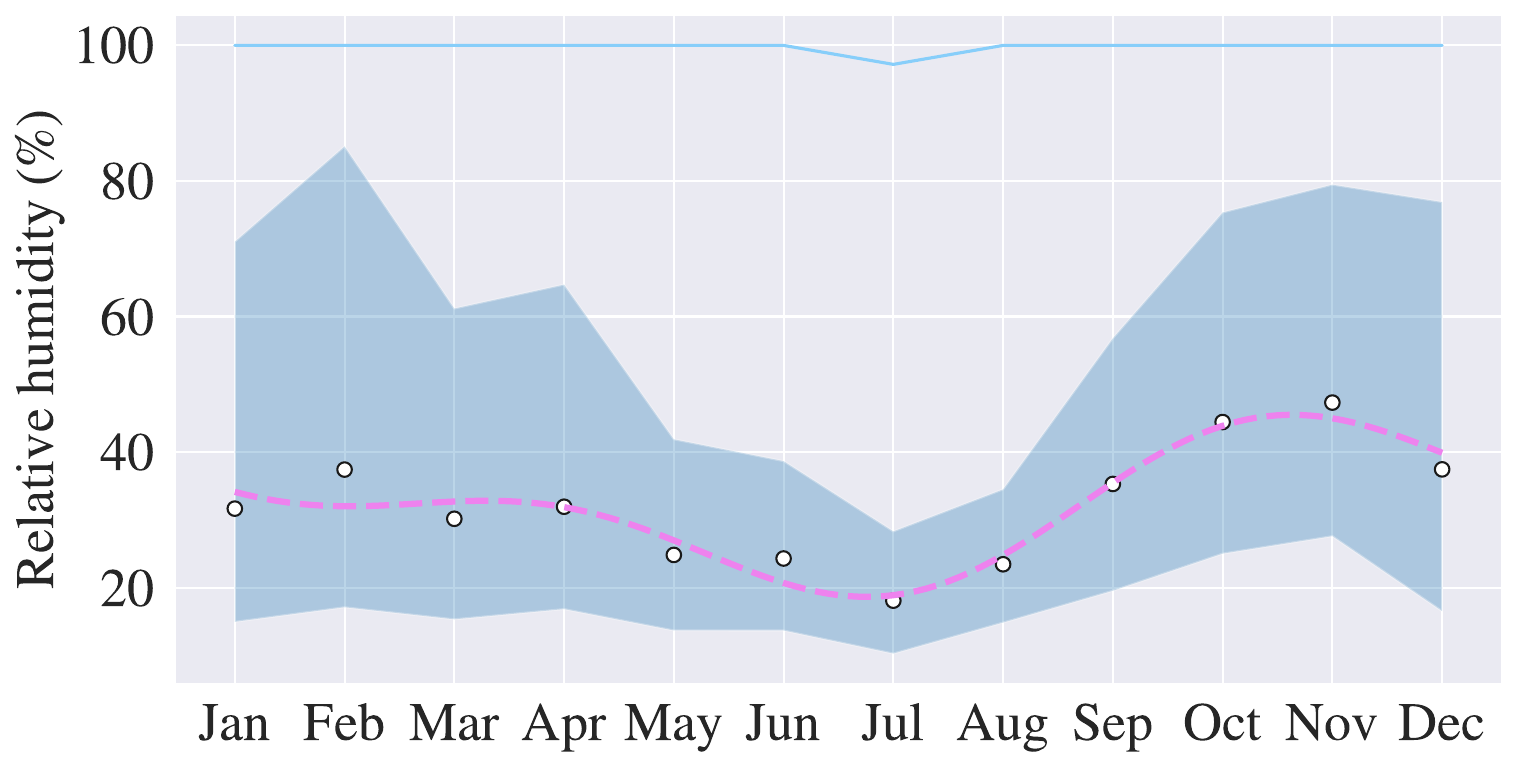}
\caption{
  \label{fig:monthRH}
    Seasonal cycle of relative humidity at the MAGIC site: white circles display the month-wise medians, the blue shadow fills the IQR, and the light blue lines display the month-wise maxima and minima observed. The magenta dashed line shows a fit to the medians obtained with Eq.~\protect\ref{eq:murh} under the assumption of no temperature increase over time ($b$=0).}
\end{figure}

%From table~\ref{table:annualT}, the average temperature
%difference for the two years of overlap between MAGIC and TNG is 1.6\degC. This
%value is therefore subtracted from the MAGIC data. In Fig.~\ref{fig:annualT},
%the behavior of the average annual temperatures of table~\ref{table:annualT} are
%shown. An increasing trend over the CAMC and TNG data was already reported
%in~\citet{lombardi2006}.

%\subsubsection*{Temperature relation with the telescope down-time}

%The MAGIC Telescopes operate at any temperature within the absolute minimum and maximum shown in Fig.~\ref{fig:histT}. Hence, temperature alone is not a reason for telescope downtime. 

%If, however, materials were used rated for positive temperatures only, then the temperature-induced night downtime would amount to 2.5\%.  

\begin{figure}
\centering
\includegraphics[width=0.99\linewidth, clip, trim={0cm 0 0cm 0}]{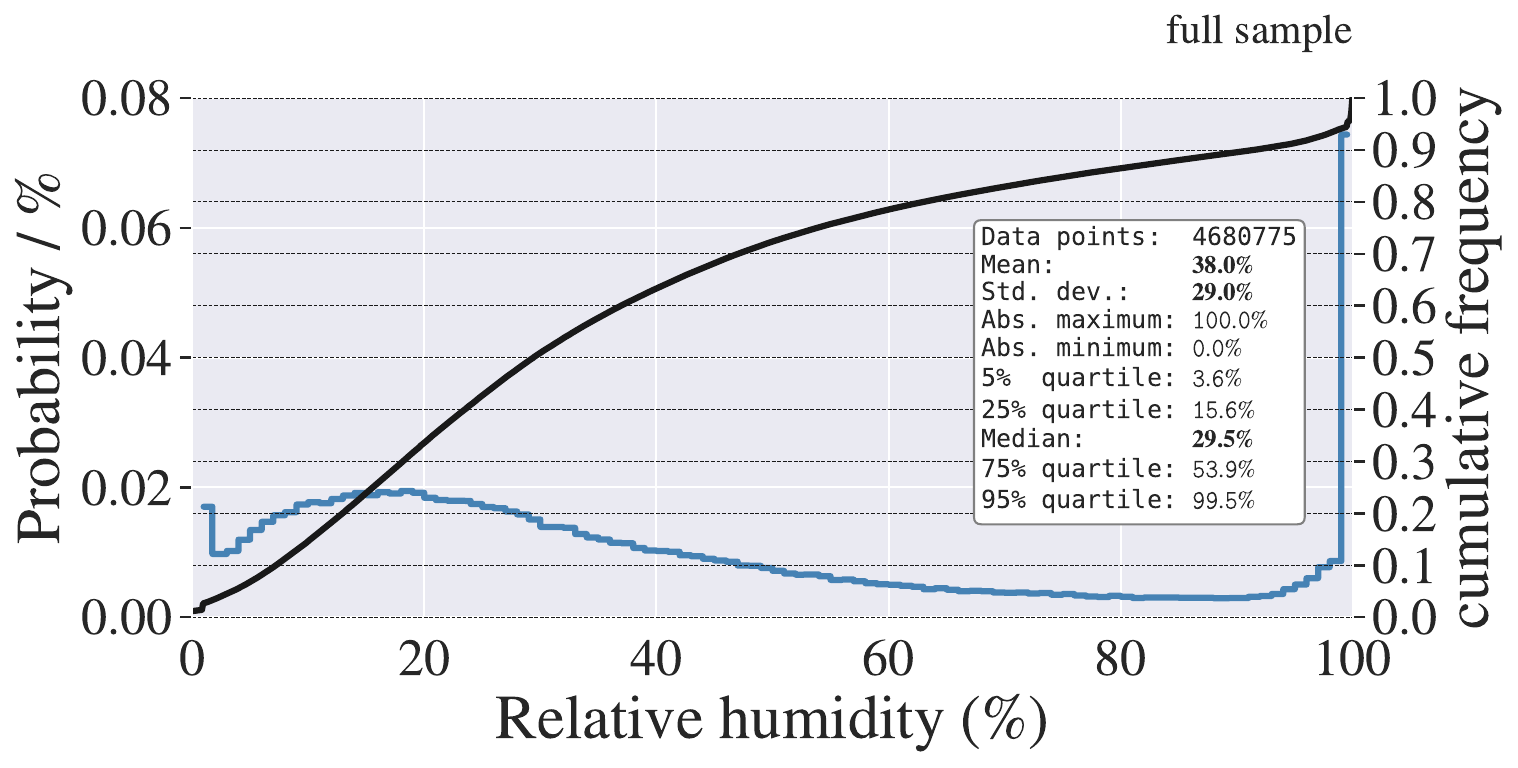}
\includegraphics[width=0.99\linewidth, clip, trim={0cm 0 0cm 0}]{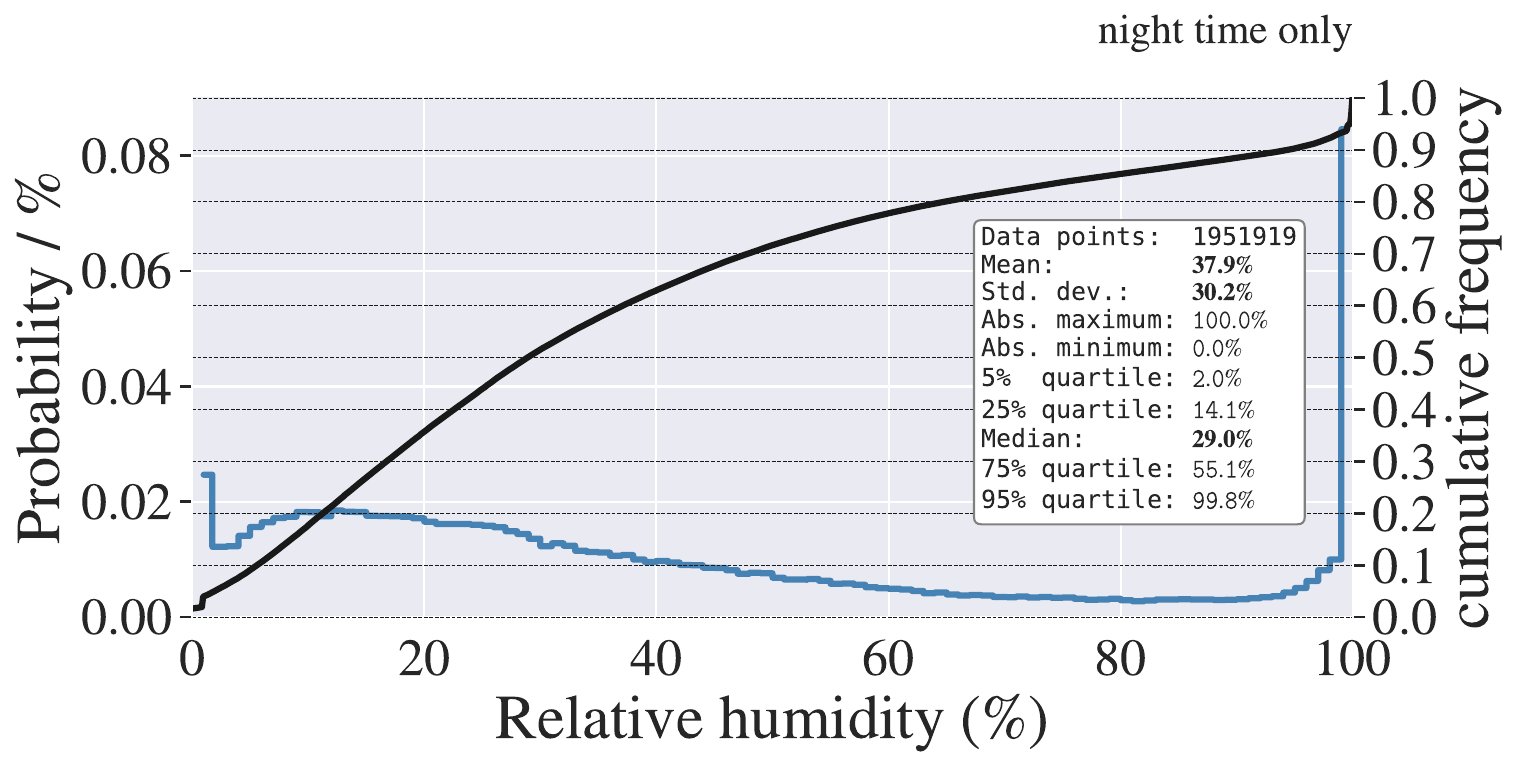}
\caption{
  \label{fig:histRH}
  Distribution and statistical parameters of relative humidity at the MAGIC site.}
\end{figure}

\begin{figure}
\centering
\includegraphics[width=0.99\linewidth, clip, trim={0cm 0 0cm 0}]{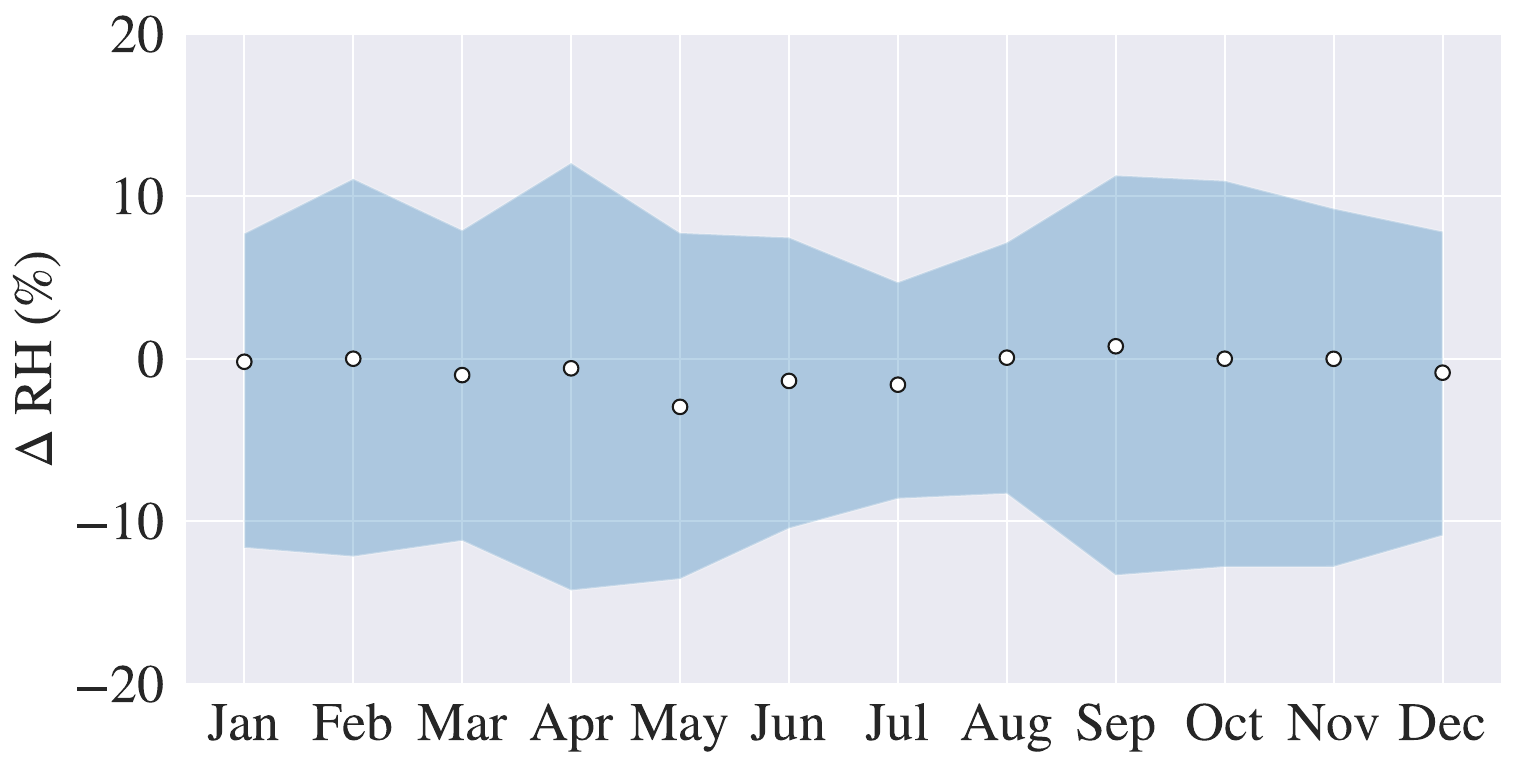}
\caption{
  \label{fig:RHdiurnal}
  Seasonal cycle of average night-time \textit{RH} ((22:00-04:00) local time, as defined in~\citet{lombardi2007}) minus average day-time \textit{RH} ((10:00-16:00) local time, as defined in~\citet{lombardi2007}). White circles display the month-wise medians, the blue shadow fills the IQR.}
\end{figure}

\subsection{Long-term Relative Humidity Behavior}

Figure~\ref{fig:annualRH} shows the time series of relative humidity (\textit{RH}) measurements flagged as reliable, in terms of month-wise statistics: median, IQR, and month-wise humidity extremes. Seasonal variations are clearly visible, although less pronounced than in the case of temperature. On average, the summer and spring months show lower humidity %around $\sim$ 20\% and
than winters. % show a higher humidity from $\sim$ 45\% to $>$ 80\% on average. 
Particularly humid winter months were observed during the winters of 2005/2006, 2010/2011, 2015/2016, 2016/2017, 2017/2018, 2018/2019 and beginning of 2023.  Most of these years are also accompanied by lower \textit{RH} extremes that do not reach the minimum sensor sensitivity at 2\%. 
%Note that the data set shows a 3-year gap for the years 2020 to 2022, due to the drifting humidity sensor being considered unreliable. 
Starting in 2020, a notable absence of the very lowest \textit{RH}s is observed accompanied by an overall slight increase of median \textit{RH}. When this effect was discovered in early 2022, the WS was exchanged and sent for repair to the provider, who confirmed, however, that the humidity sensor had \textit{not} suffered any drift or damage. In fact, the drop in low \textit{RH} extreme observed at the beginning of 2023 occurred \textit{before} the WS was exchanged. 

The seasonal cycle of \textit{RH} is shown in Fig.~\ref{fig:monthRH}. Summer \textit{RH} medians lie around $\sim$ 20\%, while \textit{RH} in October and November averages $\sim$45--50\%. The remaining months are found to lie in-between these limits. There is only one month, that is, July, that never reaches 100\% \textit{RH}. 
The months with the highest variations in \textit{RH} are February, followed by October and November, while June, July and August show the smallest variations. 

%Finally, 
Figure~\ref{fig:histRH} shows the complete distribution and statistical analysis of \textit{RH}, separately for the entire data sample and only during astronomical night (Sun elevation $<$-12$^\circ$).  Note the strong asymmetry of the distribution and, therefore, the difference between the sample mean and the median. Such differences have already been observed earlier~\citep{Varela:2014}. 
The median \textit{RH} is found about 7\% 
% Varela: 24h, table 4: 
% MED: 22%, MEAN: 35%
% Hidalgo: MED: 22.5% 
higher than those obtained at the very mountain rim~\citep{Varela:2014,Hidalgo:2021}, whereas mean RH is only 3\% higher~\citep[see][]{Varela:2014} and the CAMC results of~\citep[][]{lombardi2007}. Using only data taken before 2022 (not shown in Fig.~\ref{fig:histRH}), a 0.7\% lower median is obtained, whereas the mean is almost unaffected. 

Other locations, such as those of the TNG and NOT telescopes, located just below the caldera rim, have shown even slightly higher annual \textit{RH} means of 41--43\% in the past~\citep{lombardi2007}. 
%  Lombardi 2007 CAMC data (years 1985-2004):
% 43.6 42.4 47.9 42.5 45.6 47.4 42.9 35.9 34.1 28.0 35.2 33.7 24.8 14.1 28.7 33.6 35.4 31.4 32.1 38.6 
% --> mean: 35.9%
%  TNG (years 1998-2005)
%  38.1 39.8 37.0 36.8 37.0 33.5 52.7 49.3 
% --> mean: 40.5%
%  NOT (years 1998-2005)
%  39.4 43.7 37.1 38.6 44.2 40.9 50.3 46.3
% --> mean: 42.6%
This finding can be interpreted in terms of a higher probability that the clouds move uphill than downhill on the northern slope of the Roque de los Muchachos. 
% Varela: night, table 4: 
% MED: 21%, MEAN: 34%
%
%  Lombardi 2007 NIGHT CAMC data (years 1985-2004):
% 43.6 42.4 47.9 42.5 45.6 47.4 42.9 35.9 34.1 28.0 35.2 33.7 24.8 14.1 28.7 33.6 35.4 31.4 32.1 38.6 
% --> mean: 35.9%
%  TNG (years 1998-2005)
%  38.1 39.8 37.0 36.8 37.0 33.5 52.7 49.3 
% --> mean: 40.5%
%  NOT (years 1998-2005)
%  39.4 43.7 37.1 38.6 44.2 40.9 50.3 46.3
% --> mean: 42.6%
It is interesting to note that \citet{lombardi2007} find significant differences of 5--6\% between night-time and day-time \textit{RH} on average for CAMC data, which is not reflected in our data. In \citet{Varela:2014}, such differences are much smaller, of the order of 1-2\%, and reversed (that is, the daytime humidity is higher on average), consistent with our findings. The difference may be actually attributed to different definitions of night: while \citet{lombardi2007} define night time as (22:00-04:00) and day time as (10:00-16:00) local time, \citet{Varela:2014} 
use the time of sunset/sunrise to determine the beginning and end of the night or the start of the day. Adopting exactly the same definition of daytime and nighttime as in \citet{lombardi2007}, we have produced the seasonal cycle of diurnal \textit{RH} differences (see Fig.~\ref{fig:RHdiurnal}). However, we cannot yet reproduce the findings of \citet{lombardi2007} for our site: \textit{RH}s at daytime tend to be even slightly higher than those at night.

In the following, we will try to understand possible trends of \textit{RH} over the past two decades. \citet{Haslebacher:2022} made the interesting observation that relative and specific humidity have \textit{decreased} over the past 40 years for all major astronomical observatories, except for the island observatories on La~Palma and Mauna~Kea, for which no significant trend in relative humidity was observed, but a slight, but significant \textit{increase} of \textit{specific} humidity. 

Due to the different behavior of precipitation, defined here as \textit{RH}$>$90\%, and humidity of the remaining time periods, we have separated periods of precipitation from the rest of the data sample and studied both separately.

\subsubsection{Long-term behavior of relative humidity under  conditions without precipitation}

\begin{figure}
\centering
\includegraphics[width=0.485\linewidth]{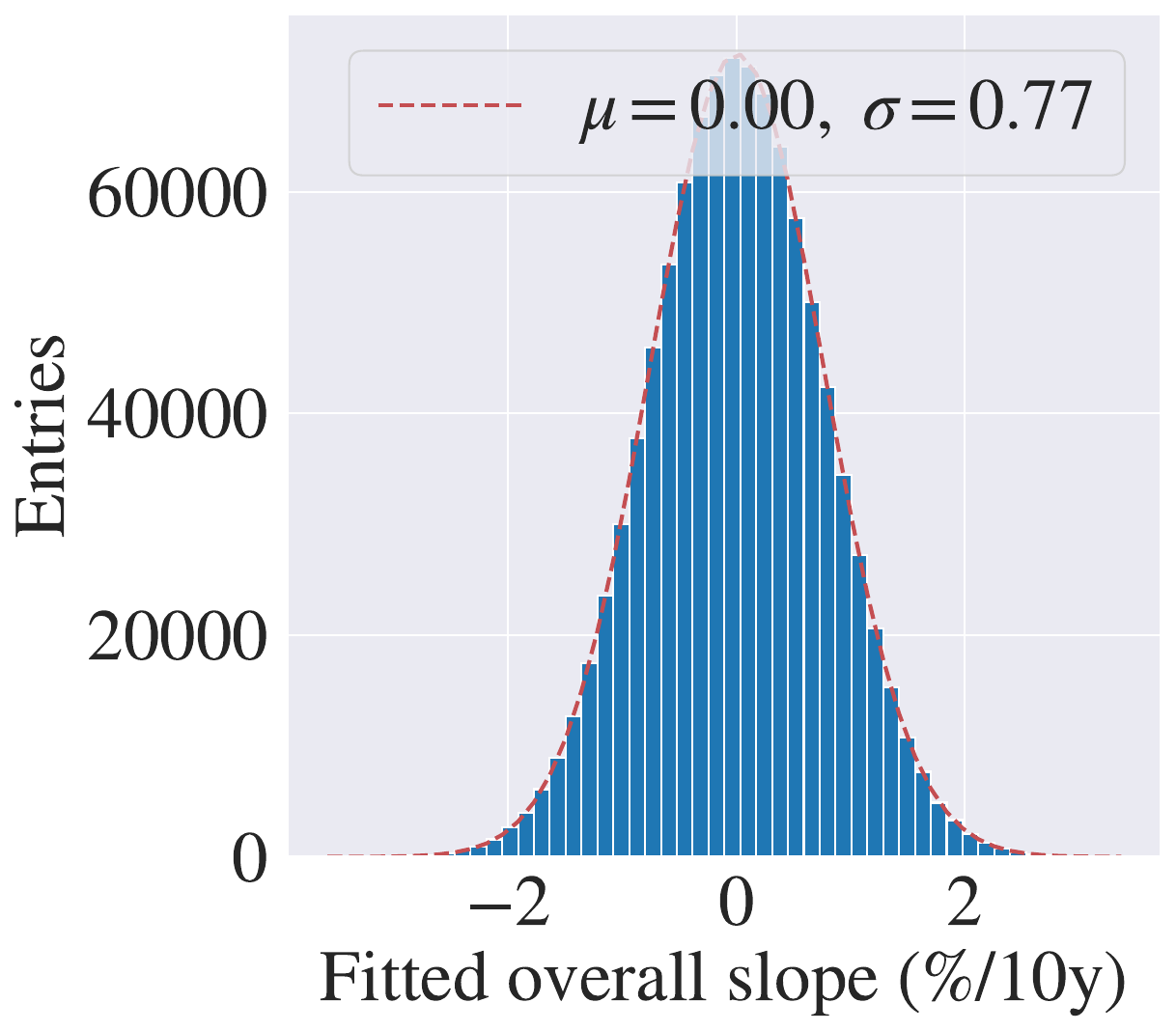}
\includegraphics[width=0.485\linewidth]{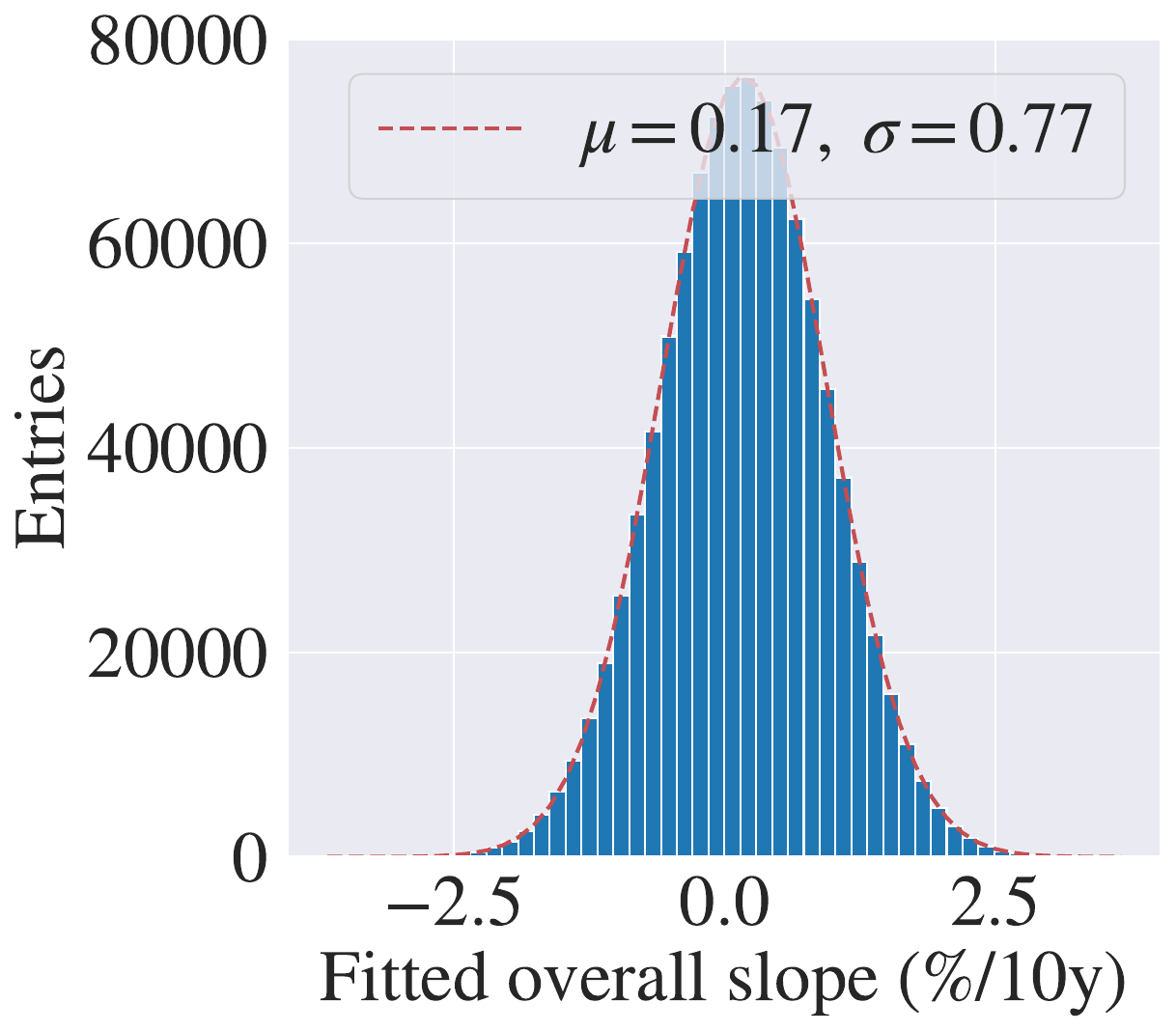}
\caption{\label{fig:sensordrift}
Distribution of fitted slopes to a toy data set spanning 20 years with simulated random sensor drifts of zero (left) and 1\%/year (right) on average and 1.5\%/year spread for each time period between weather station replacements.}
\end{figure}

We studied long-term changes of \textit{RH}$<$90\% with the use of a likelihood, %(Eq.~\ref{eq:ltemp}), 
which incorporates a seasonal cycle of relative humidity modeled by the function:
\begin{align}
\mu_i &= a + \ddfrac{b}{120}\cdot x_i + {} \nonumber\\ 
      &  {} ~ + \ddfrac{C_1}{2}\cdot \left(\left(\cos\left(\ddfrac{2\pi}{12}\cdot (x_i-\phi_1)\right)+1\right)^2 - \ddfrac{3}{2} \right)  + {} \nonumber\\
      & {} ~ + C_2 \cdot \sin\left(\ddfrac{2\pi}{12}\cdot (x_i-\phi_2)\right) \quad, \label{eq:murh}
\end{align}
\noindent
in which a secondary oscillation term has been incorporated,  characterized by amplitude $C_2$ and phase $\phi_2$. Eq.~\ref{eq:murh} was used to fit the \textit{RH} medians of the seasonal cycle in Fig.~\ref{fig:monthRH}. The values that maximize the likelihood using Eq.~\ref{eq:murh} are shown in Table~\ref{tab:lresultsH}. We find the two seasonal oscillation maxima around April (month 4) and October (month 10), a mean \textit{RH} of 24\% as of 01/01/2003, and an increase of \textit{RH} of about 4\% every 10~years. The relatively unexpected magnitude of this increase has encouraged us to carry out a series of robustness tests on the data. 

\begin{figure}
\centering
\includegraphics[width=0.99\linewidth, clip, trim={0cm 0 0cm 0}]{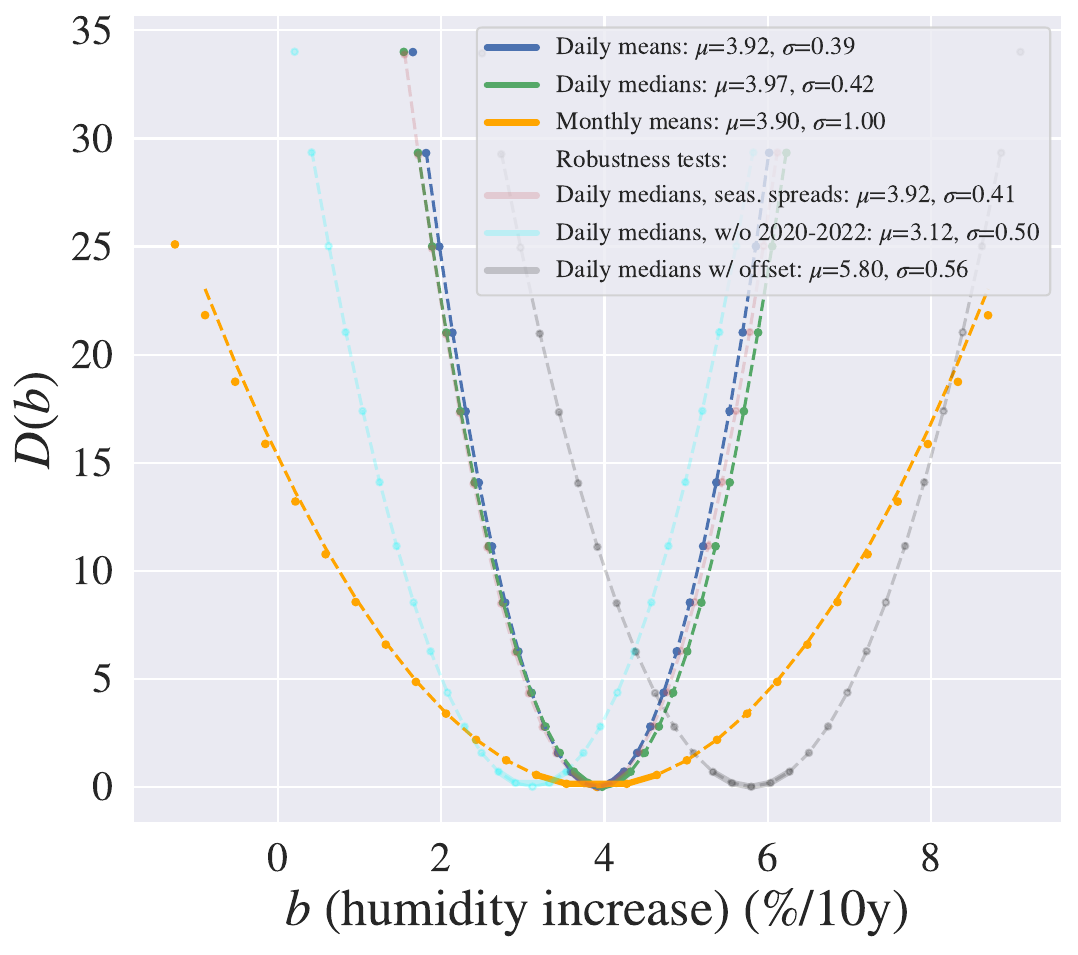}
\caption{
  \label{fig:humL}
  Profile likelihood test statistic $D(b)$ (see Eq.~\protect\ref{eq.lr1})  as a function of the \textit{RH} increase parameter $b$ for periods excluding precipitation (\textit{RH}$<$90\%). Shown are the profile likelihoods obtained from the 
  %daily mean relative humidity (green) using Eq.~\protect\ref{eq:mui}, 
  daily mean \textit{RH} (blue), daily median (green)  and monthly mean \textit{RH} (yellow), all fitted to the seasonal oscillation function Eq.~\protect\ref{eq:murh}.
   For the monthly means, a monthly coverage cut of 80\% had been applied. Daily means or medians required a daily coverage of at least 85\%. 
   Furthermore, $D(b)$ from three robustness tests are shown: with a fit using a seasonal oscillation cycle for the spreads (pink), and a fit excluding the time period from 2020 to 2022 (cyan). Finally, 
   a possible systematic sensor offset between the WS models MWS~5MV and MWS~55V was added to the list of nuisance parameters (gray). 
   All profile likelihoods have been fitted to a parabolic function $(b-\mu)^2/\sigma^2$ around the minimum, the obtained fit parameters are shown in the legend. }
\end{figure}

\textit{RH} sensors are not as accurate as pressure and temperature sensors, particularly in relation to drifts over time. Systematic drifts of $<$1.5\% per year can be expected according to the WS provider~\citep{Reinhardt:priv} and the data sheet of the HC1000 sensor used. Since we exchanged the weather station quite often with a newly calibrated system, it is \textit{a priori} not clear how a series of sensor drifts, limited in time, affect the overall measurement of an \textit{RH} increase or decrease over 20~years. For this reason, we performed toy simulations of random sensor drifts for each of the time periods shown in Table~\ref{tab:WSreplacements}. The simulations use a different sensor drift for each time period between weather station exchanges, distributed with a Gaussian width of 1.5\%, i.e.\ the maximum specified by the provider. The resulting time series were fitted to a linear function. The distributions of slopes obtained from a million such simulations are shown in Fig.~\ref{fig:sensordrift}. One can observe that in both cases, a spread and a root mean squared error of $\sim$0.8\%/decade 
% sqrt(0.17*0.17+0.77*0.77)
% 0.78854296
were obtained, independently of an assumed inherent drift direction of all stations. This number was included in the systematic uncertainty of the \textit{RH} increase parameter~$b$. 
Furthermore, we tested a possible seasonal oscillation of the Gaussian spread $\sigma_0$, exclusion of the years 2020-2022, where the apparent increase of lowest humidity has been observed, and a possible calibration offset between the MWS-55V and the MWS-5MV sensor. All these scenarios have been included in Fig.~\ref{fig:humL}, which shows the resulting profile likelihood ratio for $b$. We observe that neither moving from daily means to medians, nor including seasonal oscillation terms on the Gaussian spreads change much the behavior of the test statistic. When data from 2020-22 are excluded, the minimum for the profiled $b$ moves to slightly lower values. The calibration offset fit yields an (unrealistic) negative calibration correction of $\sim$3\% for the MWS-5MV, together with a \textit{higher} most probable \textit{RH} increase of 5.8\%/decade.
% 2.97 +- 0.51
% 3.82 +- 0.42
% 5.17 +- 0.57
% 3.64 +- 0.49 
% --> weighted total: 3.83 +- 0.24
\ccol{The test of possible individual miscalibration or drifting periods (see Sect.~\ref{sec:discussion}) yielded compatible results within the stated systematic uncertainty.}

\begin{figure}
\centering
\includegraphics[width=0.99\linewidth, clip, trim={0cm 0 0cm 0}]{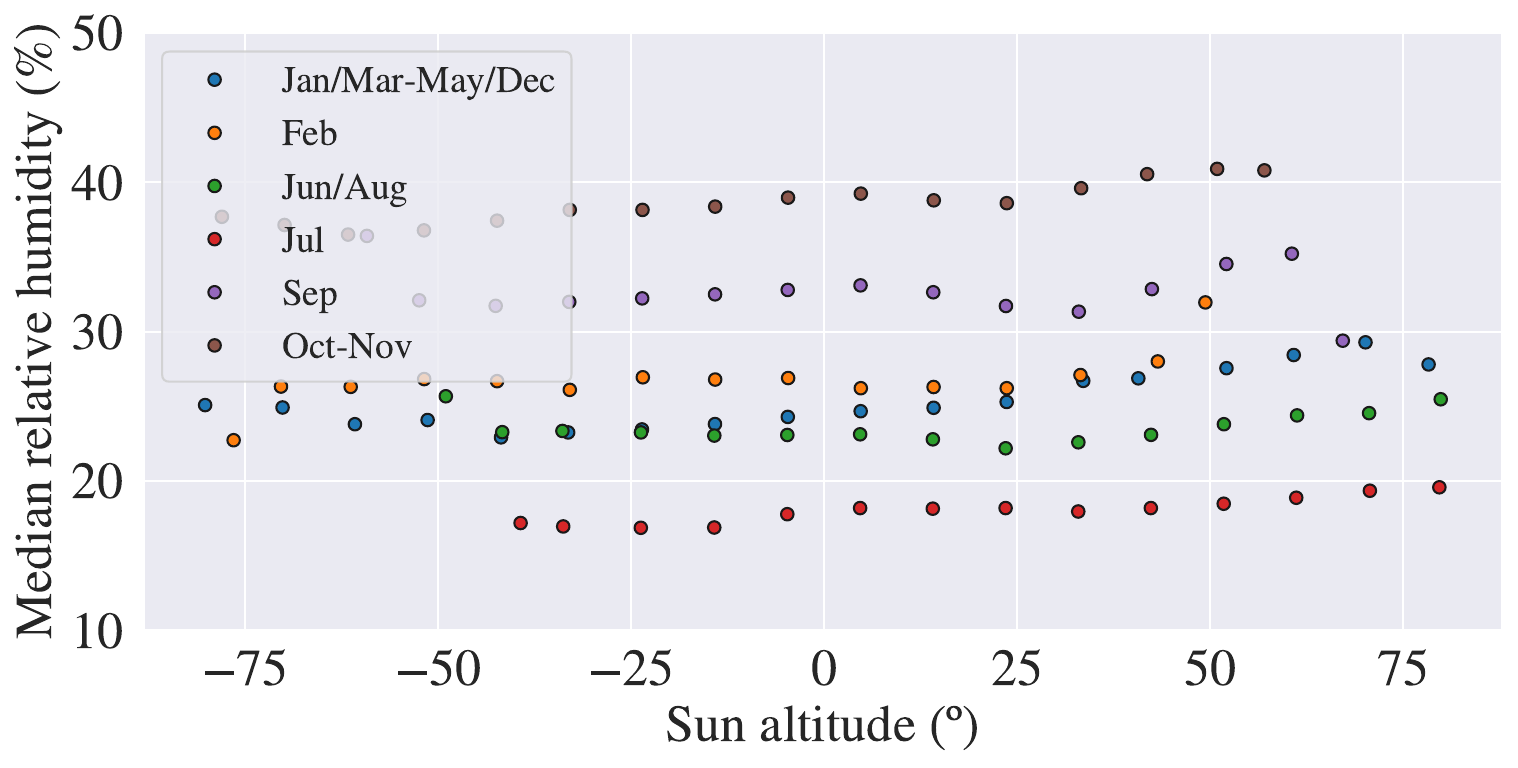}
\caption{
  \label{fig:humL:diurnal}
  Median \textit{RH} as a function of solar altitude, for different months or groups of months that have been combined for similar behavior. Only non-precipitation periods (\textit{RH}$<$90\%) have been used to produce this figure.}
\end{figure}

% sqrt(0.8*0.8+0.8*0.8) = 1.1   
We conclude that a \textit{RH} increase of about $4.0\pm 
 0.4\mathrm{(stat.)}\pm 1.1\mathrm{(syst.)}$\%/decade is observed 
 %with more than 5$\sigma$ significance 
 for periods without precipitation. Even if the time interval from 2020 to 2022 is not taken into account, the null hypothesis of no increase of $\textit{RH}$ is excluded with more than 6$\sigma$ significance. This finding may be unexpected, together with the detected increase in temperature over time; however, the trend (not the amount of change) is consistent with~\citet{Haslebacher:2022} who attribute the joint increase in temperature and humidity to possible higher evaporation rates of sea water, which in turn affects island observatories in a different way from mainland ones. This interpretation is supported by the observation that \textit{RH} during non-precipitation periods increases with solar altitude (see Fig.~\ref{fig:humL:diurnal}), contrary to the findings of \citet{lombardi2007} who claimed that \textit{RH} was higher in the first hours of the morning than in the afternoon for the CAMC site. 
Note that \citet{Hidalgo:2021} also claim that there is no difference in \textit{RH} between midnight and noon, although with a very rough binning of 10\% in \textit{RH}. 
%RH would simply change as a consequence of the temperature changes due to the
%different heights of the sun during the day. RH is higher in the first hours of
%the morning (the coldest part of the day) than in the afternoon (the warmest
%part of the day). 

\begin{table*}
\centering
\begin{tabular}{l|ccccccc|c}
\toprule
Modality  &  $\hat{a}$ & $\hat{b}$ & $\hat{C}_1$ & $\hat{\phi}_1$ & $\hat{C}_2$ & $\hat{\phi}_2$ & $\hat{\sigma}_0$ & $\chi^{2}/\mathrm{NDF}$\\ 
          &  (\%)  & (\%/10y) & (\%) & (month)  & (\%) & (month) & (\%)  & (1) \\
\midrule

Daily means    & 25.6 & 3.9 & 19.3 & 3.7 & 24.2 & 6.9 & 15.6 & 1.00 \\  
Daily medians  & 24.2 & 4.0 & 19.2 & 3.7 & 24.2 & 6.9 & 16.8 & 1.00 \\
Monthly means  & 27.3 & 3.9 & 17.9 & 3.6 & 22.9 & 6.9 & 7.4 & 1.04 \\
\bottomrule
\end{tabular}
\caption{\label{tab:lresultsH}  Parameters that maximize the likelihood, Eq.~\protect\ref{eq:ltemp} using Eq.~\ref{eq:murh} to fit the seasonal cycle for \textit{RH} excluding precipitation (\textit{RH}$<$90\%). The parameter $a$ corresponds to the average \textit{RH} as of  01/01/2003, $b$ the decadal \textit{RH} increase, $C_1$ and $C_2$ seasonal oscillation amplitudes with the respective locations of the seasonal maxima at $\phi_1$ and $\phi_2$, and $\sigma_0$ the average Gaussian spread of \textit{RH} around the fitted seasonal cycle. 
\ccol{The column $\chi^{2}/\mathrm{NDF}$ provides the sum of squared residuals of all data points with respect to the function that maximizes the likelihood, divided by the number of degrees of freedom of the fit.}}
\end{table*}

%\mg{Irregular changes in temperature and relative humidity should be due to movements of air masses and the arrival of frontal systems.}

%\citet{Haslebacher:2022} find a nonsignificant negative trend of $(-0.05 \pm 0.12)$\%/decade from NOT data taken between 2004 and 2019. 

%\mg{Might show the full correlation for a part of the data set, where we have contemporaneous data from NOT and TNG.}

\subsubsection{Long-term behavior of precipitation}

\begin{figure}
\centering
\includegraphics[width=0.99\linewidth, clip, trim={0cm 0 0cm 0}]{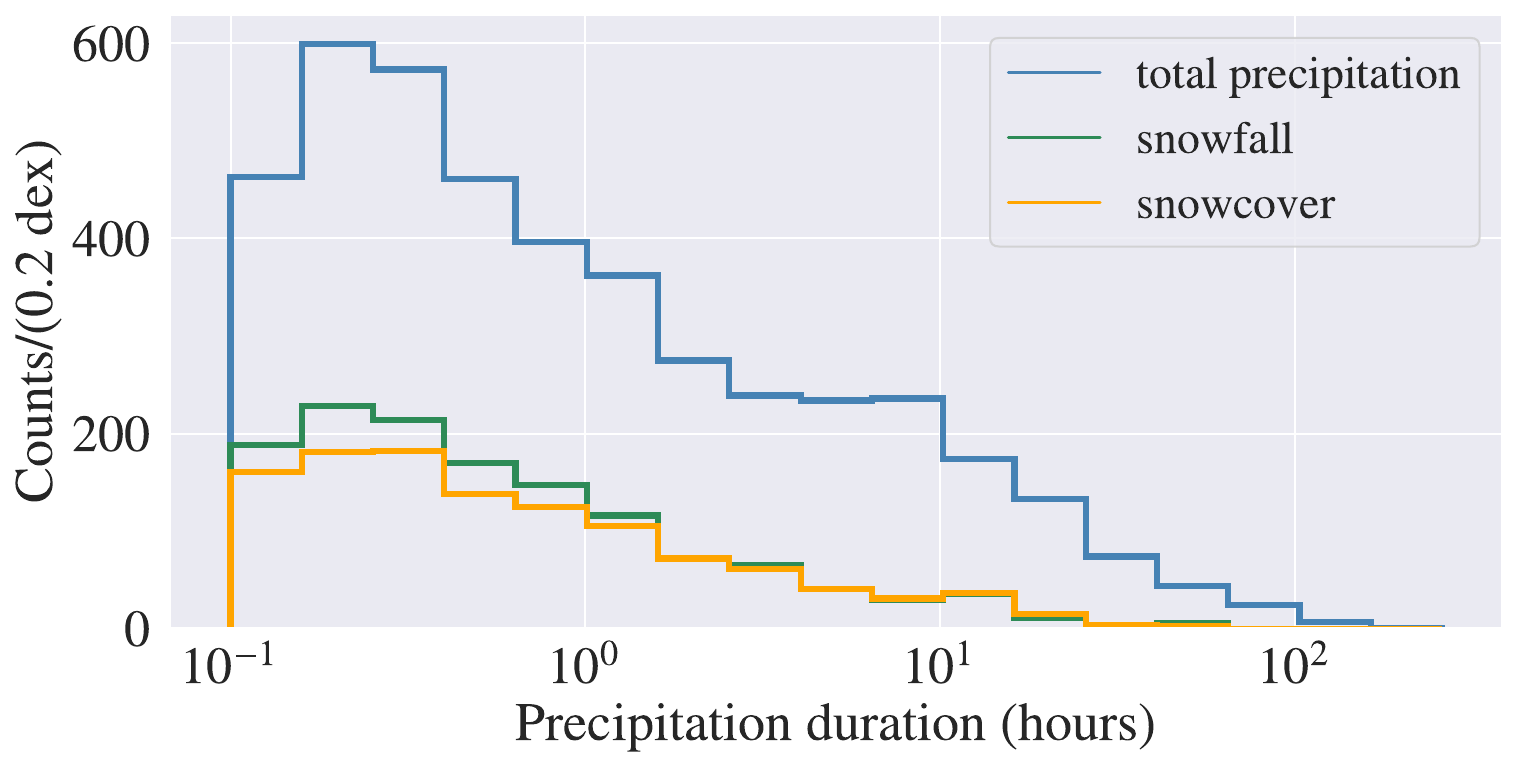}
\includegraphics[width=0.99\linewidth, clip, trim={0cm 0 0cm 0}]{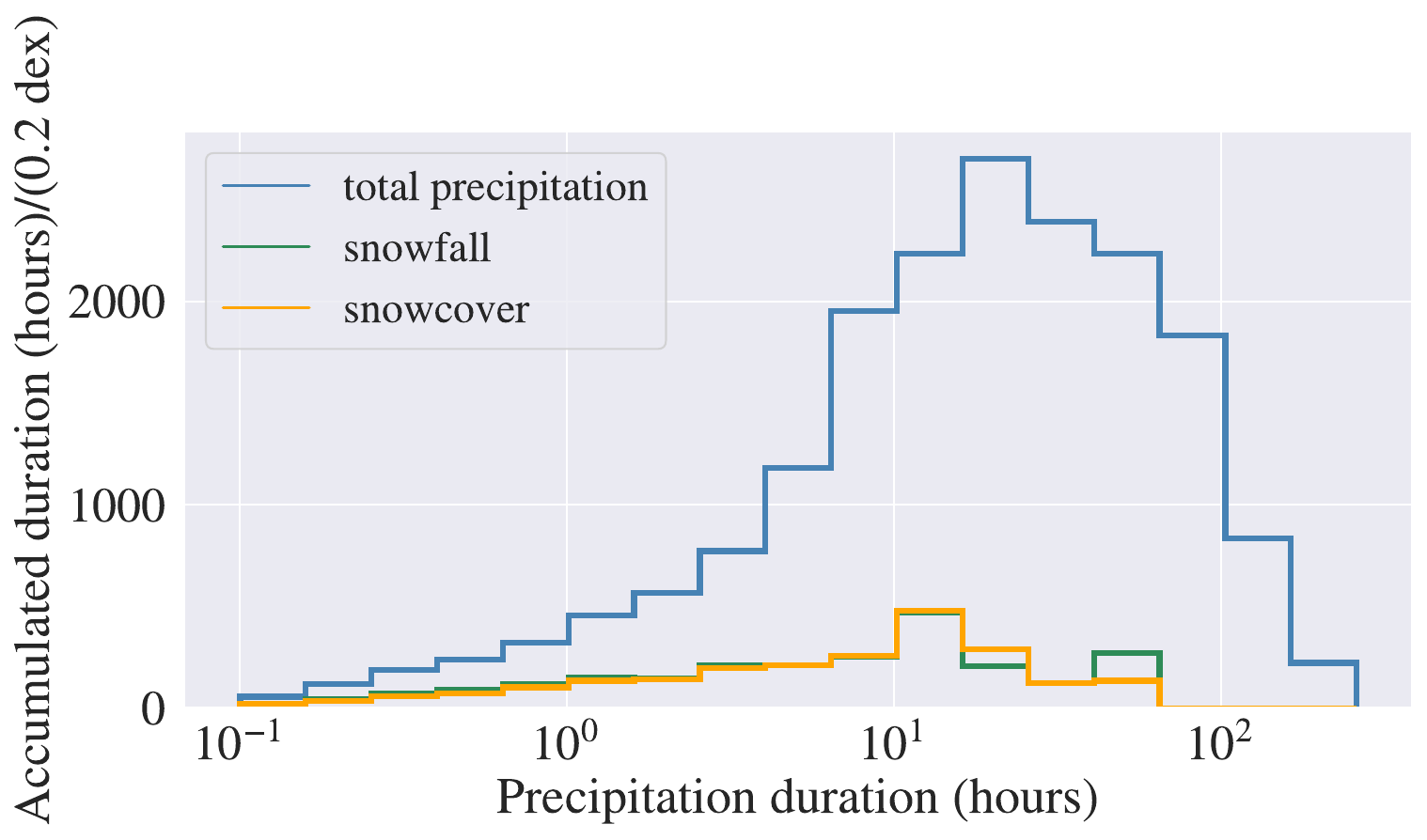}
\caption{
  \label{fig:histRain}
  Distribution of the length of precipitation periods (top) and total accumulated precipitation duration (below). Note the logarithmic horizontal time scale. The distribution may be slightly biased towards too long periods of precipitation lasting longer than 3~days (72~hours), due to some hysteresis of the humidity sensor observed under these circumstances. No attempt to correct for data gaps has been made here.}
\end{figure}

\begin{figure}
\centering
\includegraphics[width=0.99\linewidth, clip, trim={0cm 0 0cm 0}]{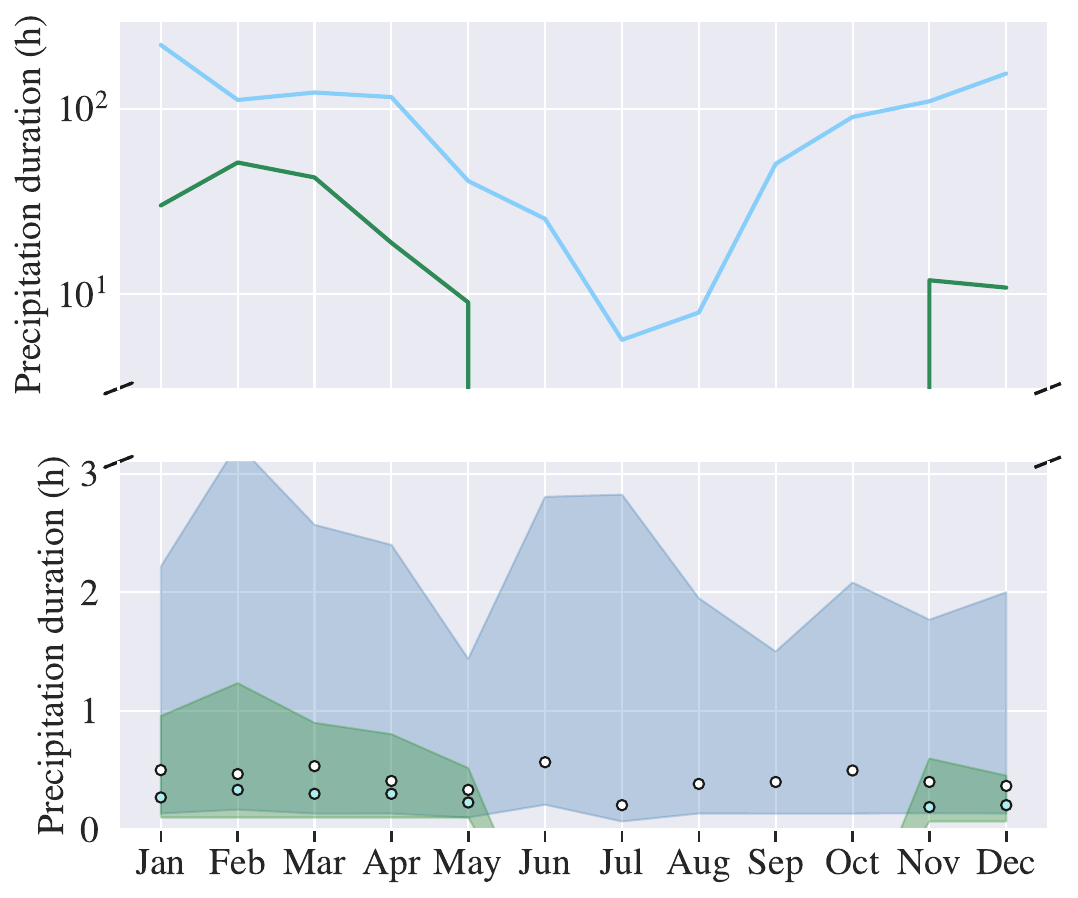}
\caption{
  \label{fig:histRainMonth}
  Seasonal cycle of precipitation duration at the MAGIC site: white circles display the month-wise medians precipitation duration, the blue shadow fills show the IQR, and the light blue lines display the month-wise maxima of precipitation duration observed. Green circles, lines and areas represent periods of snowfall. Note that from June to October no snowfall at all is observed.}
\end{figure}

\begin{figure}
\centering
\includegraphics[width=0.93\linewidth, clip, trim={0cm 0 0cm 0}]{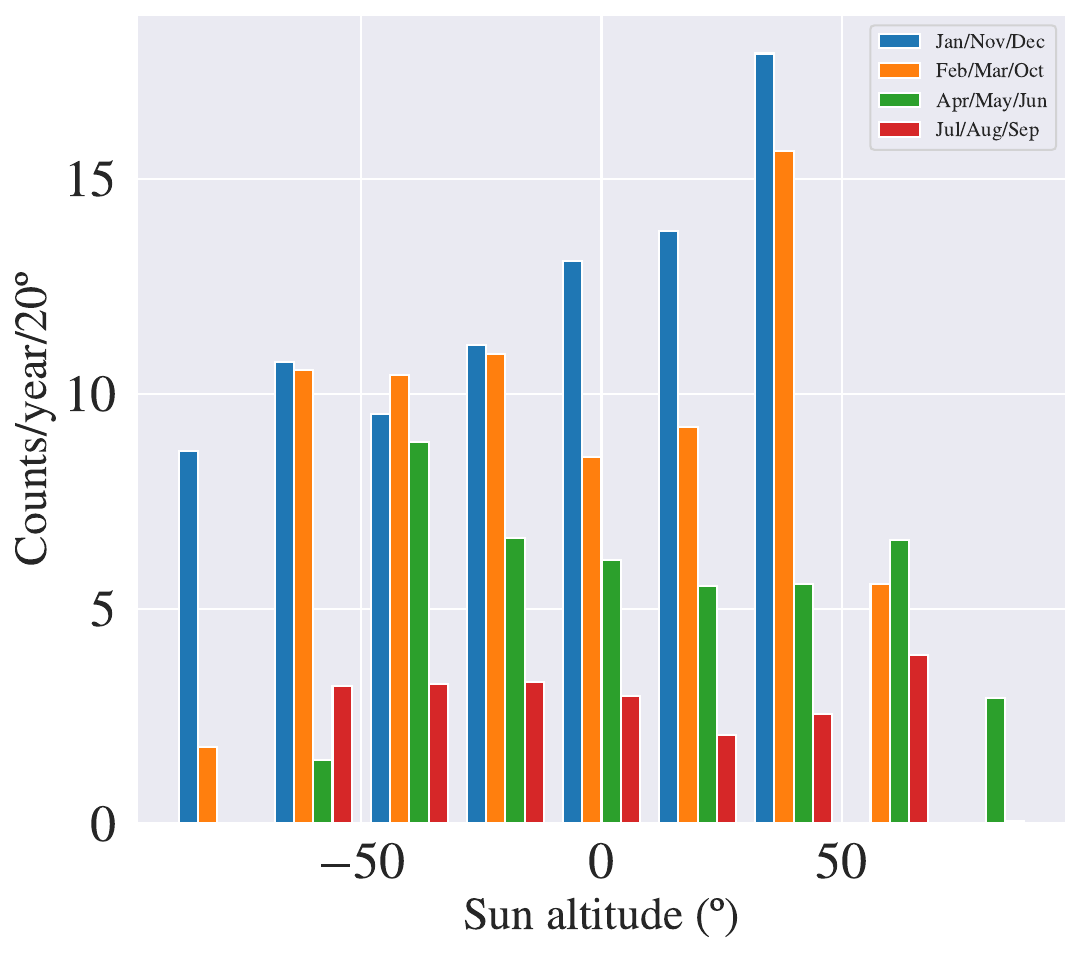}\\
\includegraphics[width=0.93\linewidth, clip, trim={0cm 0 0cm 0}]{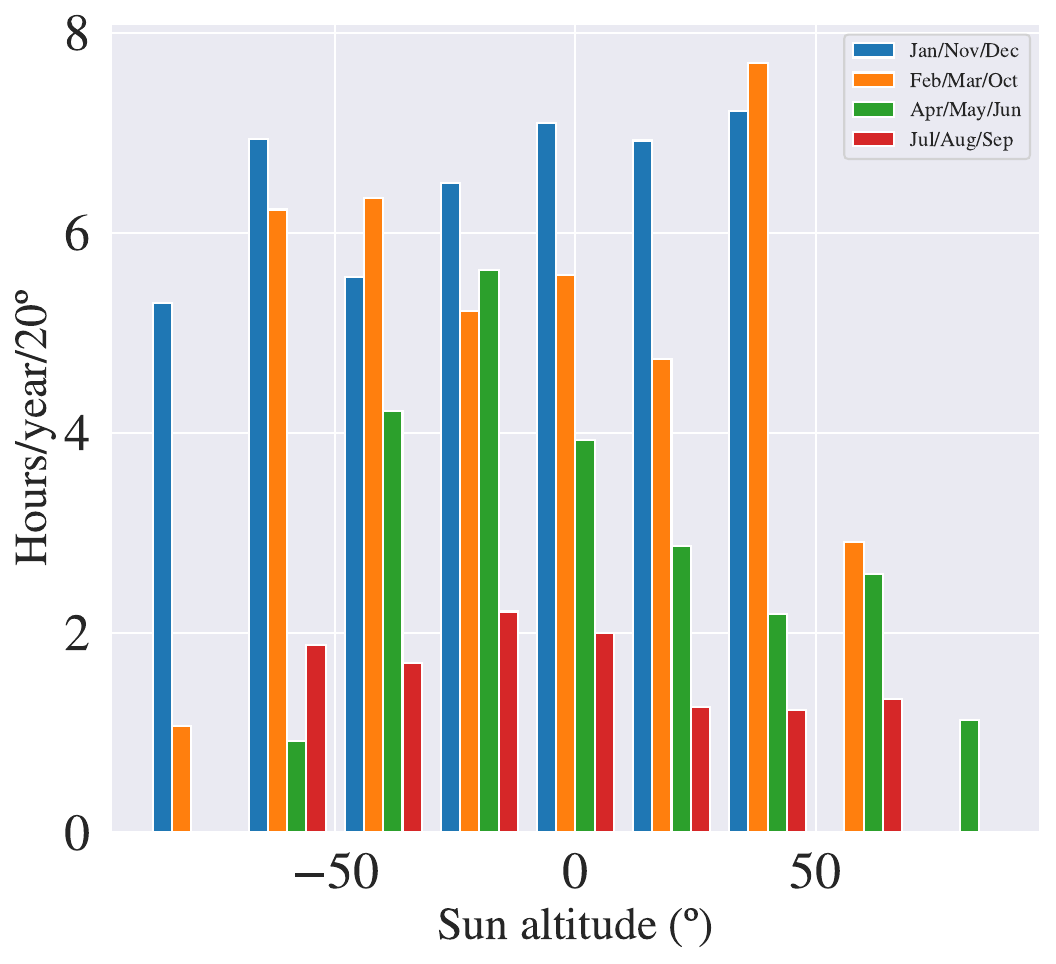}
\caption{
  \label{fig:humL:numberrains}
  Number counts (top) and total duration (below) of short ($<$3~hours) precipitation periods, as a function of solar altitude. The months have been grouped into periods of similar behavior. 
  %Note the different bin widths between the upper and the lower plot. 
  The distributions have been statistically corrected for data gaps. }
\end{figure}

We define precipitation as time intervals characterized by \textit{RH}$>90$\%. Furthermore, we calculated the wet bulb temperature and used the parameterization of~\citet{Ding:2014} to discriminate between rain and snowfall. This parameterization assigns the label \textit{snowfall} to events where the probability of snowfall is greater than 0.5, calibrated on a large set of weather data obtained across mainland China. Since it is possible that this discriminator shows some bias when used for an island in the Atlantic Ocean, we also show events characterized as snowfall in the presence of surface temperatures lower than 0\degC (henceforth labeled \textit{snowcover}). 
%In the absence of a better choice, we adhere to it baring in mind a possible over-prediction of snowfall. 

First, we look at Fig.~\ref{fig:histRain} (top), which shows the frequency distribution of the duration of the precipitation periods. 
%Although the distribution is slightly biased towards too long rains lasting longer than three days, due to observed hysteresis of the humidity sensor, 
One can observe an approximately bimodal distribution, separated by short periods of precipitation lasting less than $\sim$3~hours and longer periods lasting longer. Precipitation durations longer than 6.5~days were not found in our data sample until Jan. 2021, when an unusually long period of 9~days of continuous precipitation, including snowfall, was recorded. Long periods are responsible for the largest fraction of total precipitation duration, and therefore telescope downtime (see Fig.~\ref{fig:histRain} below). About 13\% of the total precipitation is attributed to \textit{snowfall} and 12\% to \textit{snowcover}. Hence, almost all snowfall occurred at temperatures below 0\degC. 
%; that upper limit might be even reduced to $\sim$6~days if the RH sensor hysteresis is taken into account after long rains. 

We interpret short rains as clouds moving through the site or sudden rains or ice droplets from clouds moving over the site at high altitudes, whereas long periods of precipitation are due to extended meteorological disturbances, which cover the entire observatory. 

The seasonal cycle of precipitation duration is shown in Fig.~\ref{fig:histRainMonth}. We observe that the two summer months July and August do not show long periods of precipitation at all, whereas the longest rains occur in winter. Snowfall is observed only during the winter months until May. The median duration of precipitation is shorter than or close to half an hour for all months; snowfall lasts on average about half that time.

%We have also studied the behavior of relative humidity during day and night and
%for each seasonal period. Fig.~\ref{fig:RH_DayNight} shows the comparison
%between MAGIC annual daytime (pink line), nighttime (orange line) and entire
%day (blue line) RH variations computed in wintertime and summertime.  It can be
%seen that winter period  have higher variations of RH than in summer. In both
%cases nighttime RH is higher with respect to daytime until 2007, after that
%moment, the difference between day and night is almost negligible.

%\begin{figure}
%\centering
%\includegraphics[width=0.48\linewidth]{images/hum_comparison_w.eps}
%\includegraphics[width=0.48\linewidth]{images/hum_comparison_s.eps}
%\caption[Comparison between MAGIC annual daytime, nighttime and entire day
%relative humidity variations.]{
%  \label{fig:RH_DayNight}
%  Comparison between MAGIC annual daytime (blue line), nighttime (green line)
%  and entire day (red line) RH variations. Left: Wintertime. Right: Summertime
%}
%\end{figure}

Due to their different nature, we separated, in the following, precipitation periods into short and long ones and study them separately. 

\subsubsection{Short periods of precipitation}

\begin{figure}
\centering
\includegraphics[width=0.99\linewidth, clip, trim={0cm 0 0cm 0}]{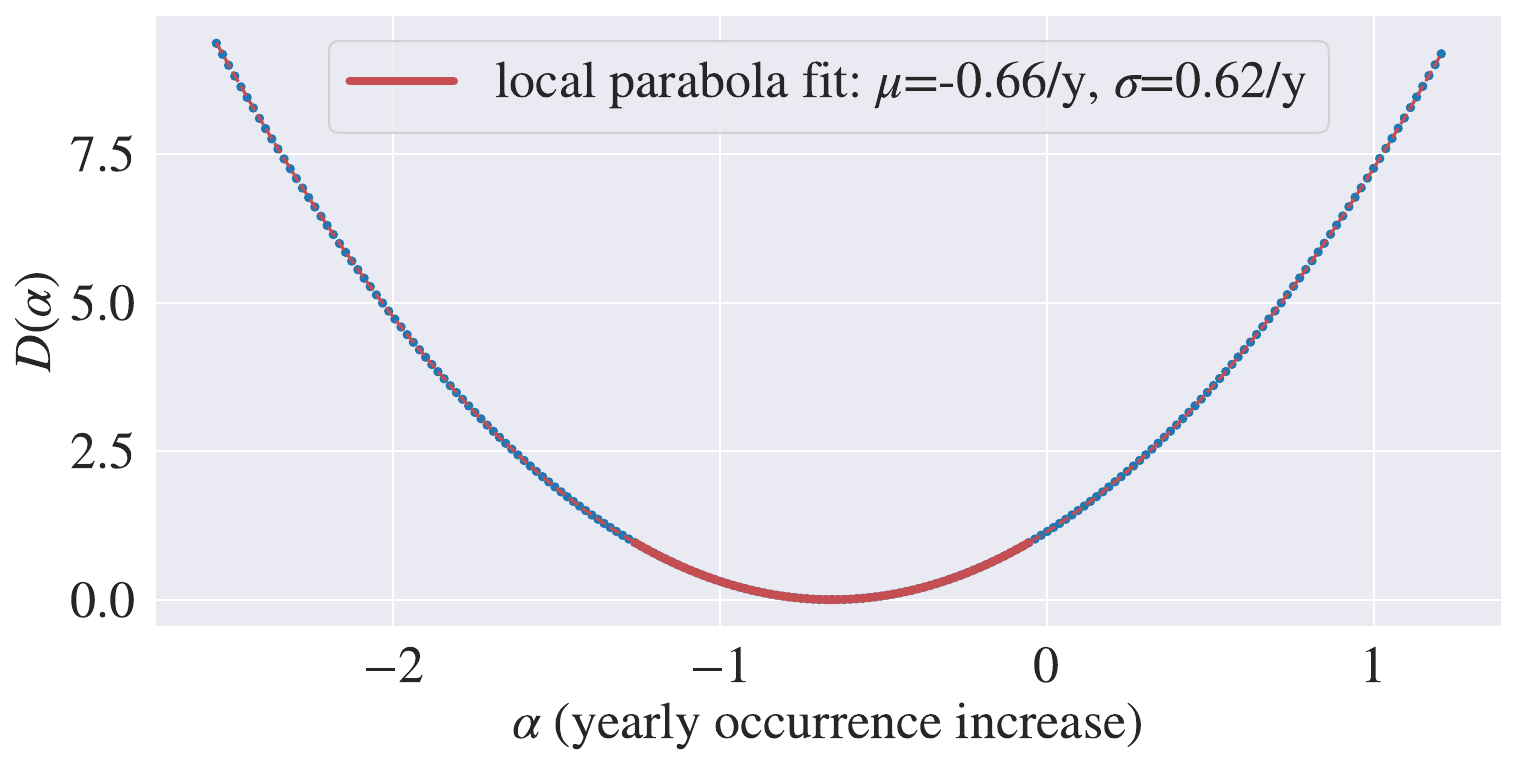}
\caption{
  \label{fig:Lshortrains}
  Profile likelihood test statistic as a function of the yearly occurrence increase of short periods of precipitation ($<$3 hours of \textit{RH}$>$90\%). 
 The profile likelihood has been fitted to a parabolic function $(b-\mu)^2/\sigma^2$ around the minimum, the obtained $\mu$ and $\sigma$ parameters are shown in the legend. }
\end{figure}

Short periods of precipitation occur during day and night, with a small preference around midday (solar altitude ranging from 30$^\circ$ to 50$^\circ$, see Fig.~\ref{fig:humL:numberrains}) during winter and fall. In contrast, the less frequent spring and summer short precipitations are approximately equally distributed throughout the diurnal cycle, with a small preference for night time.  

We tested the occurrence of short precipitation periods %with RH$>$90\%
as a function of time, using a profile likelihood
analysis based on Poissonian statistics detailed in Appendix~\ref{sec:likelihood}. 
For this purpose, we counted the number of short periods of precipitation in a given month. 
These measurements are then confronted with a Poissonian probability distribution
with a time-dependent expectation value, modeled as a linear increase or decrease over
time. Data gaps are correctly taken into account in the yearly expectation value through weights (see Appendix~\ref{sec:likelihood} for details). 
The location of the likelihood maximum yields hence
a most probable yearly occurrence of short periods of precipitation and
a yearly increase of occurrence probability, the significance of which is assessed against the null hypothesis of no increase or decrease over time. Our analysis finds a most probable yearly occurrence of 230~events and an insignificant decrease of 0.7~events/year. 
Figure~\ref{fig:Lshortrains} shows the profiled test statistic for the increase in annual occurrence. An increase of more than 0.6 / year (95\% CL) can be excluded, as well as a decrease of more than 1.9 / year (95\% CL). 

We interpret this finding as the probability of clouds passing through the observatory up- or downhill having remained constant over the past 20 years.

\begin{figure}
\centering
\includegraphics[width=0.99\linewidth, clip, trim={0cm 0 0cm 0}]{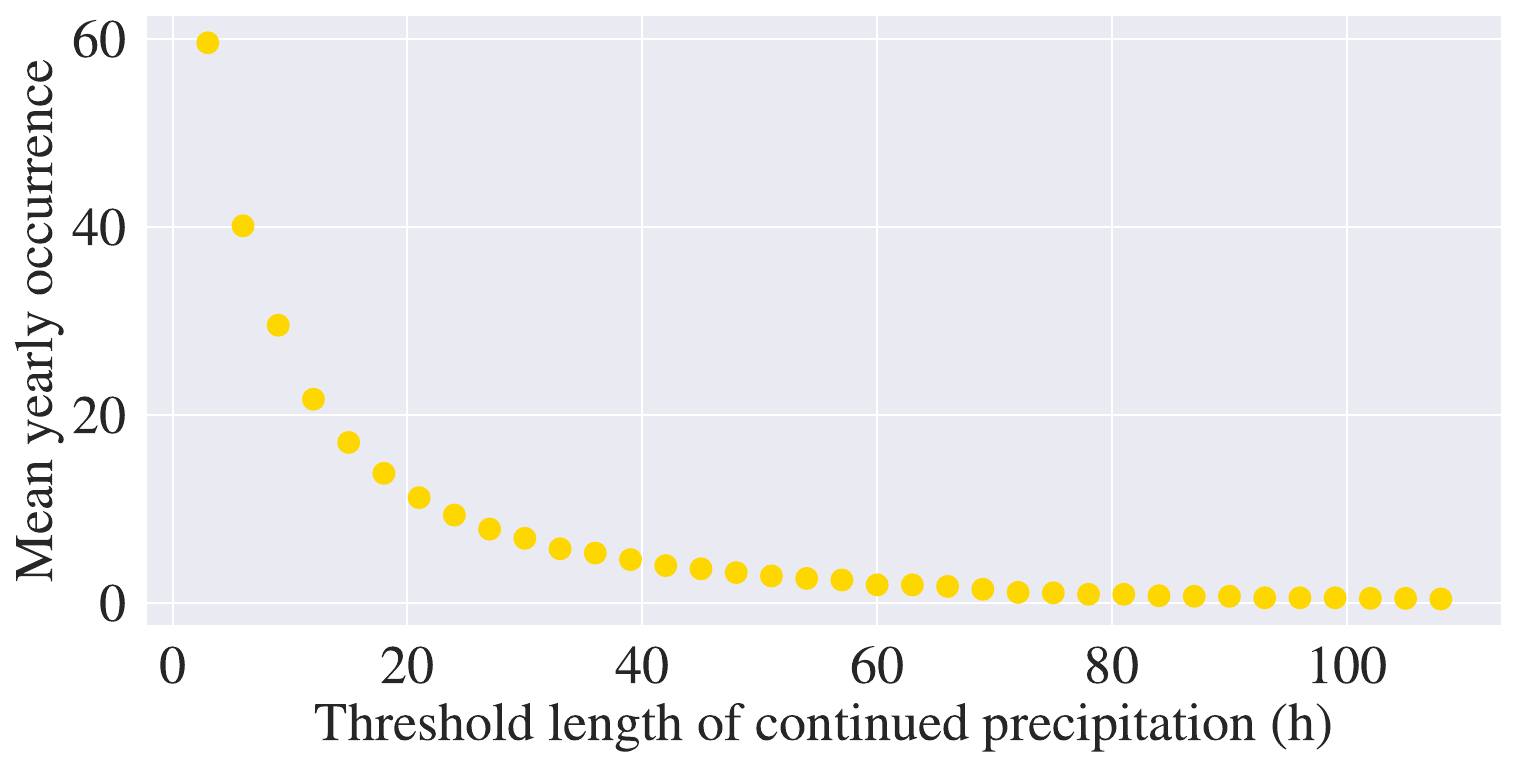}
\includegraphics[width=0.99\linewidth, clip, trim={0cm 0 0cm 0}]{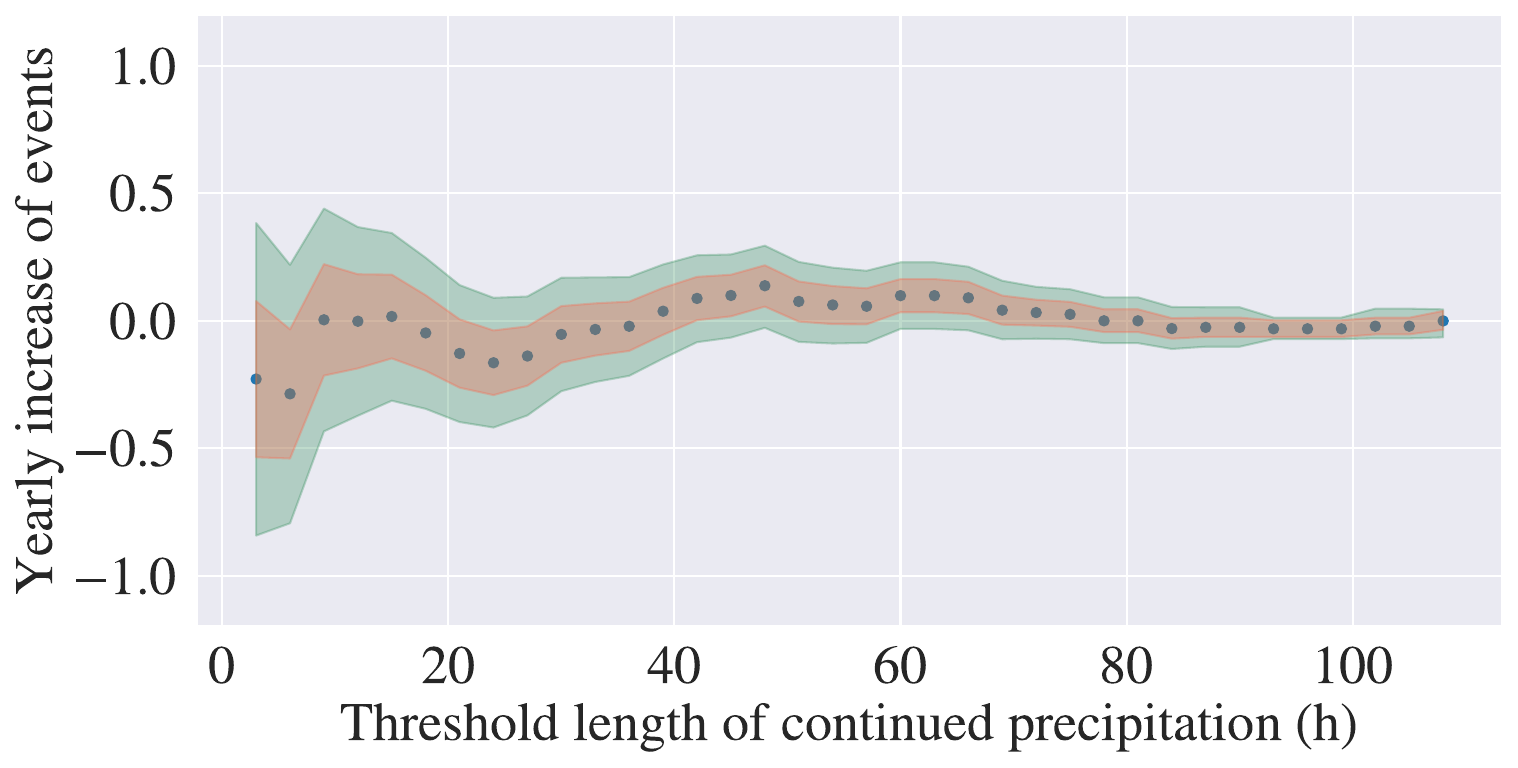}
\includegraphics[width=0.99\linewidth, clip, trim={0cm 0 0cm 0}]{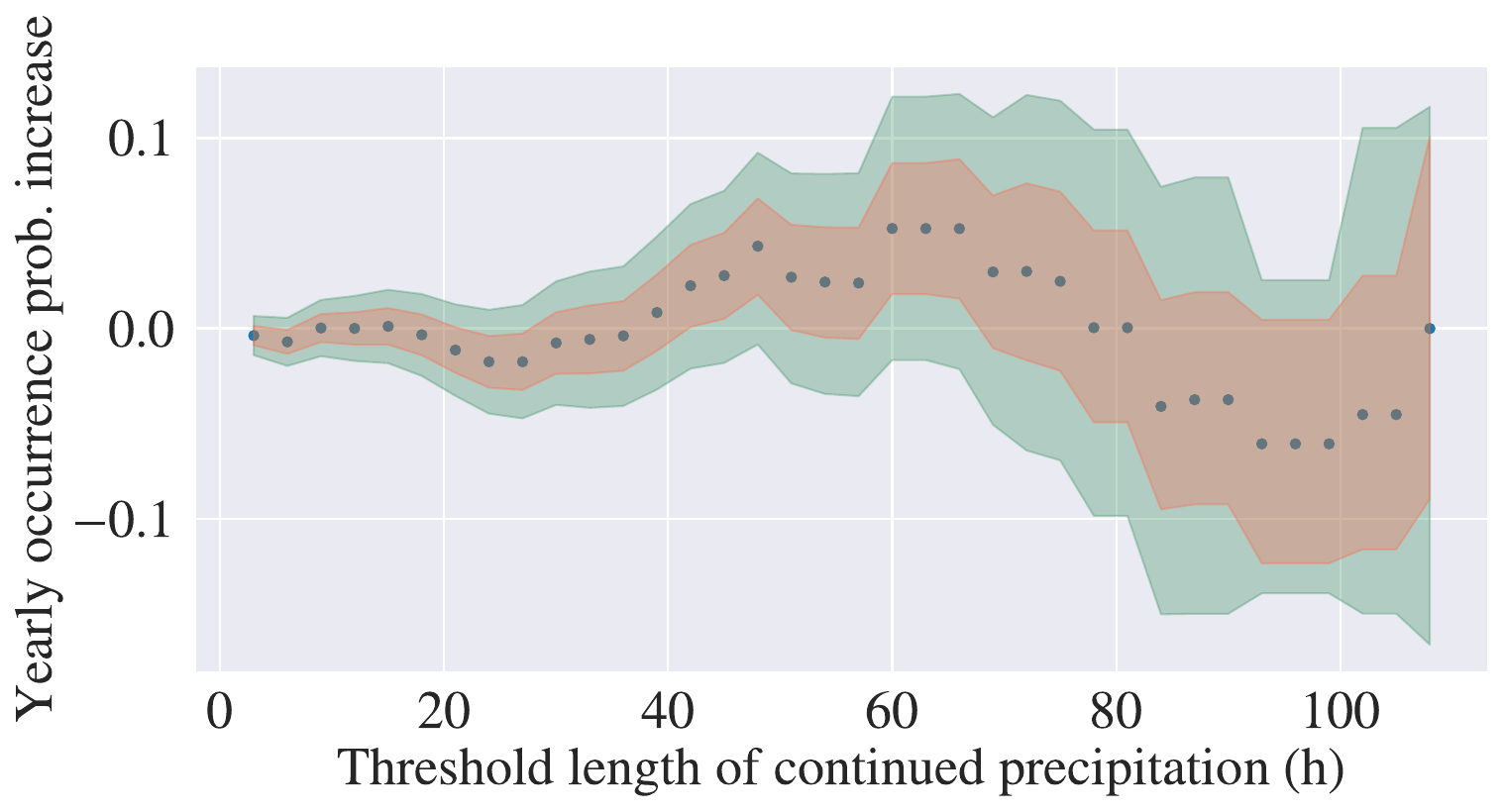}
\caption{
  \label{fig:alpha_rel_rains}
  Yearly occurrence of long periods of precipitation at the MAGIC site, derived from the likelihood analysis: On the top, the mean annual occurrence $p_0$ is shown as a function of minimum duration. In the center, the annual increase of occurrence is shown and below, the relative annual increase probability: points show the most probable value, the brown band the 68\% and the green band the 95\% confidence region of the profile likelihoods. }
  %Below, the test statistic for the  relative yearly increase or decrease of storms, tested against the null-hypothesis of no increase, is displayed. 
  %\mg{Effect of missing data still to be included!}}
\end{figure}

\subsubsection{Long periods of precipitation}

Long periods of precipitation are a major cause of telescope downtime, particularly during winter. It is therefore of major interest whether such periods change over time, either in duration or in occurrence. 

In order to test both with our data set, we calculated profile likelihoods for linear increases of the Poissonian occurrence probability (see again Appendix~\ref{sec:likelihood}), as a function of the duration of such precipitation periods. Figure~\ref{fig:alpha_rel_rains} shows the mean annual occurrences that maximize the likelihoods, the corresponding annual occurrence increases, and the occurrence probability increases. For the latter two, also the 68\% and 95\% confidence intervals have been calculated (see Appendix~\ref{sec:likelihood} for details) and are shown as brown and green bands in Fig.~\ref{fig:alpha_rel_rains}. We can observe that so far no statistically significant increase in events has occurred, independent of the duration of the precipitation, although a slight preference for an occurrence probability \textit{increase} of precipitations lasting longer than 40~hours and shorter than 75~hours is found, followed by a slight preference for an occurrence probability \textit{decrease} for even longer time intervals of precipitation. More data are needed in the future to determine whether these preferences become significant after 2023.

\subsection{Long-term Atmospheric Pressure Behavior}

\begin{figure*}
\centering
\includegraphics[width=0.99\textwidth, clip, trim={0cm 0 0cm 0}]{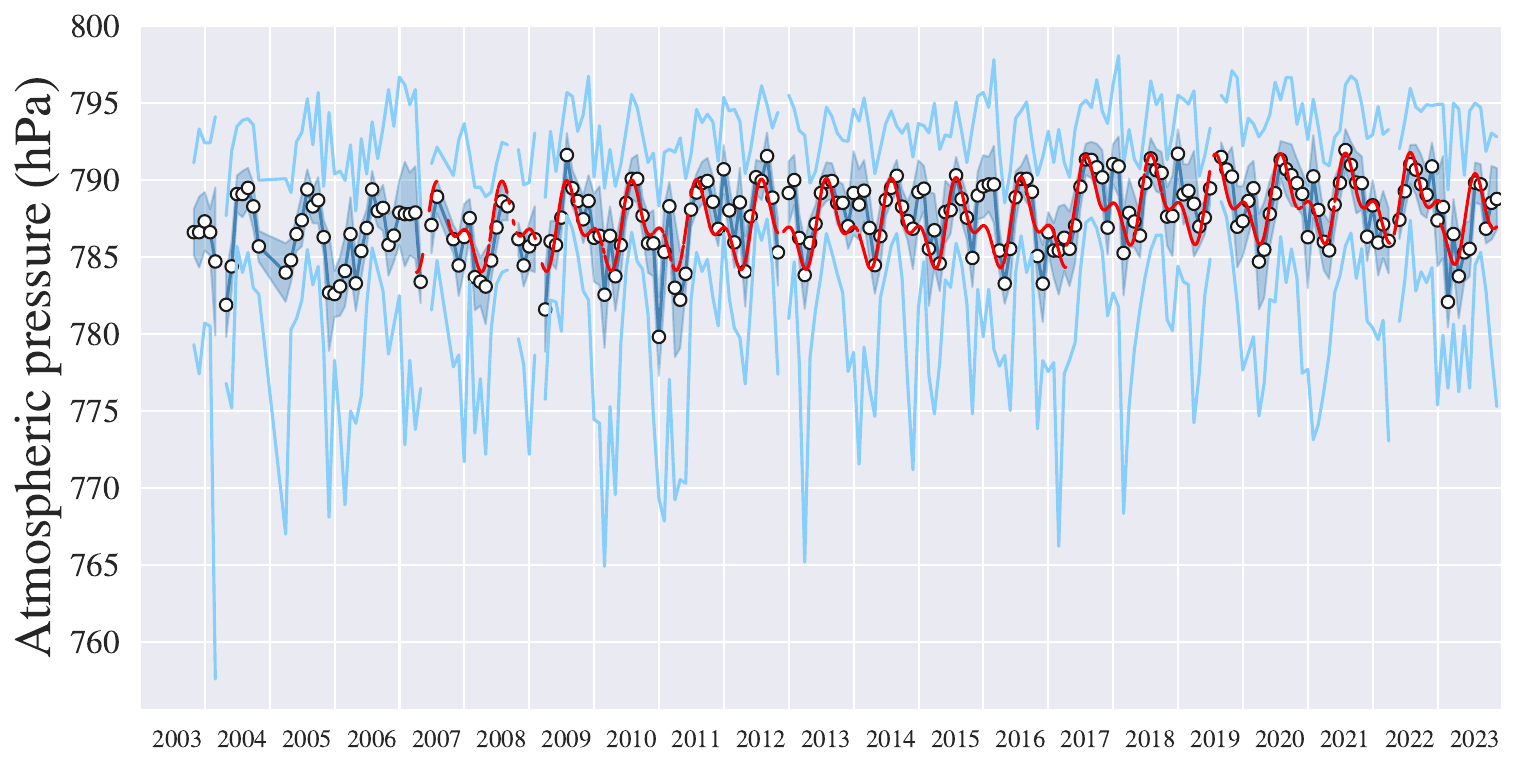}
\caption{
  \label{fig:annualP}
  Atmospheric pressure times series observed with the MAGIC weather station: the white circles show the observed monthly medians, the blue shadow fills the IQR, and the light blue lines show the monthly maxima and minima.
  The red lines show the result of the fit Eq.~\protect\ref{eq:murh} with a calibration offset for the MWS~5MV to the data. 
  Additional gaps visible in the fit are due to the (higher) daily data completeness requirement of 95\% and the exclusion of the data obtained with the early MWS5 station.}
\end{figure*}

\begin{figure}
\centering
\includegraphics[width=0.99\linewidth, clip, trim={0cm 0 0cm 0}]{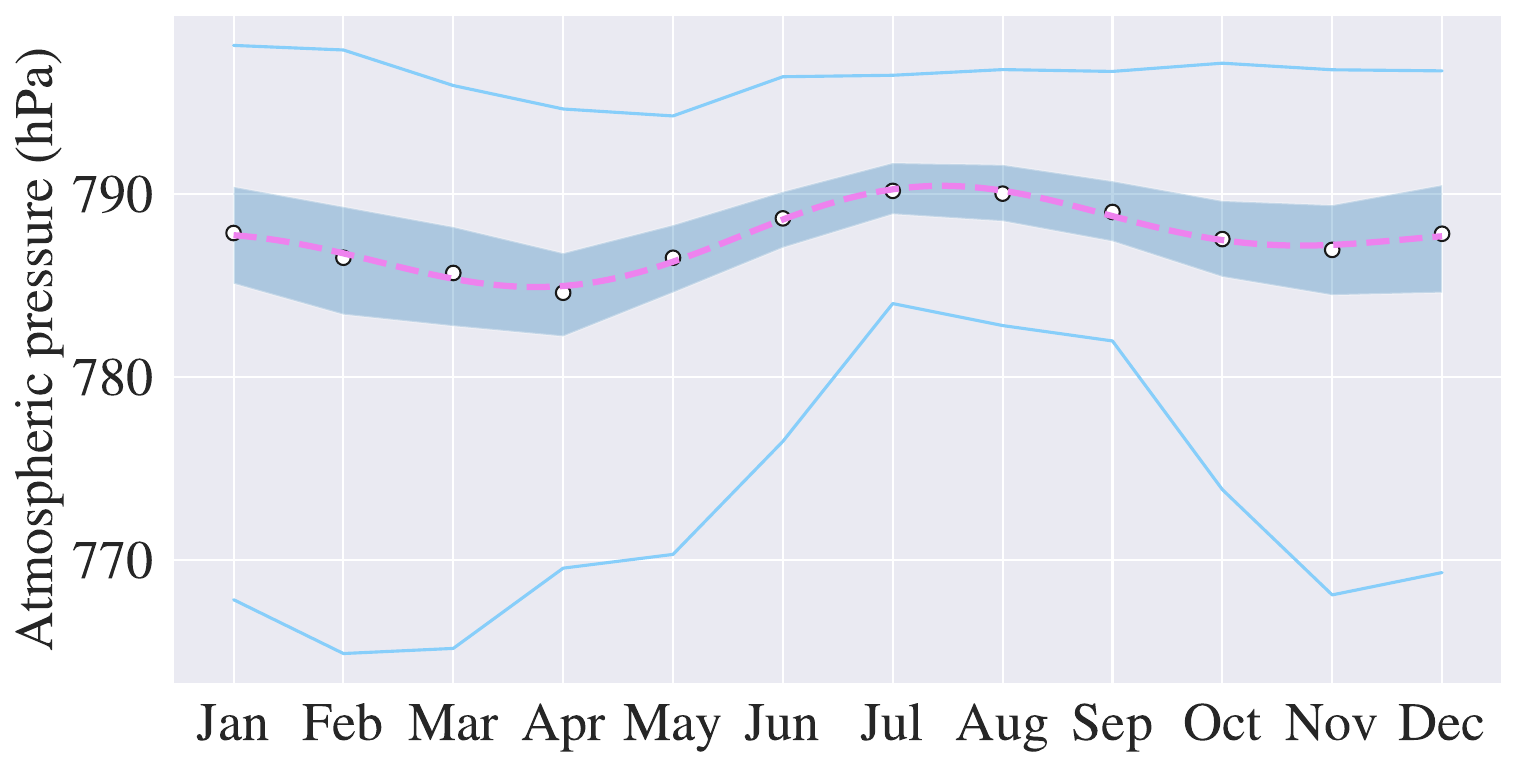}
\caption{
  \label{fig:monthP}
  Seasonal cycle of atmospheric pressure at the MAGIC site: white circles display the month-wise medians, the blue shadow fills the IQR, and the light blue lines display the month-wise maxima and minima observed.
  The magenta dashed line shows a fit to the medians obtained with Eq.~\protect\ref{eq:murh} under the assumption of no temperature increase over time ($b$=0).}
\end{figure}

\begin{figure}
\centering
\includegraphics[width=0.99\linewidth, clip, trim={0cm 0 0cm 0}]{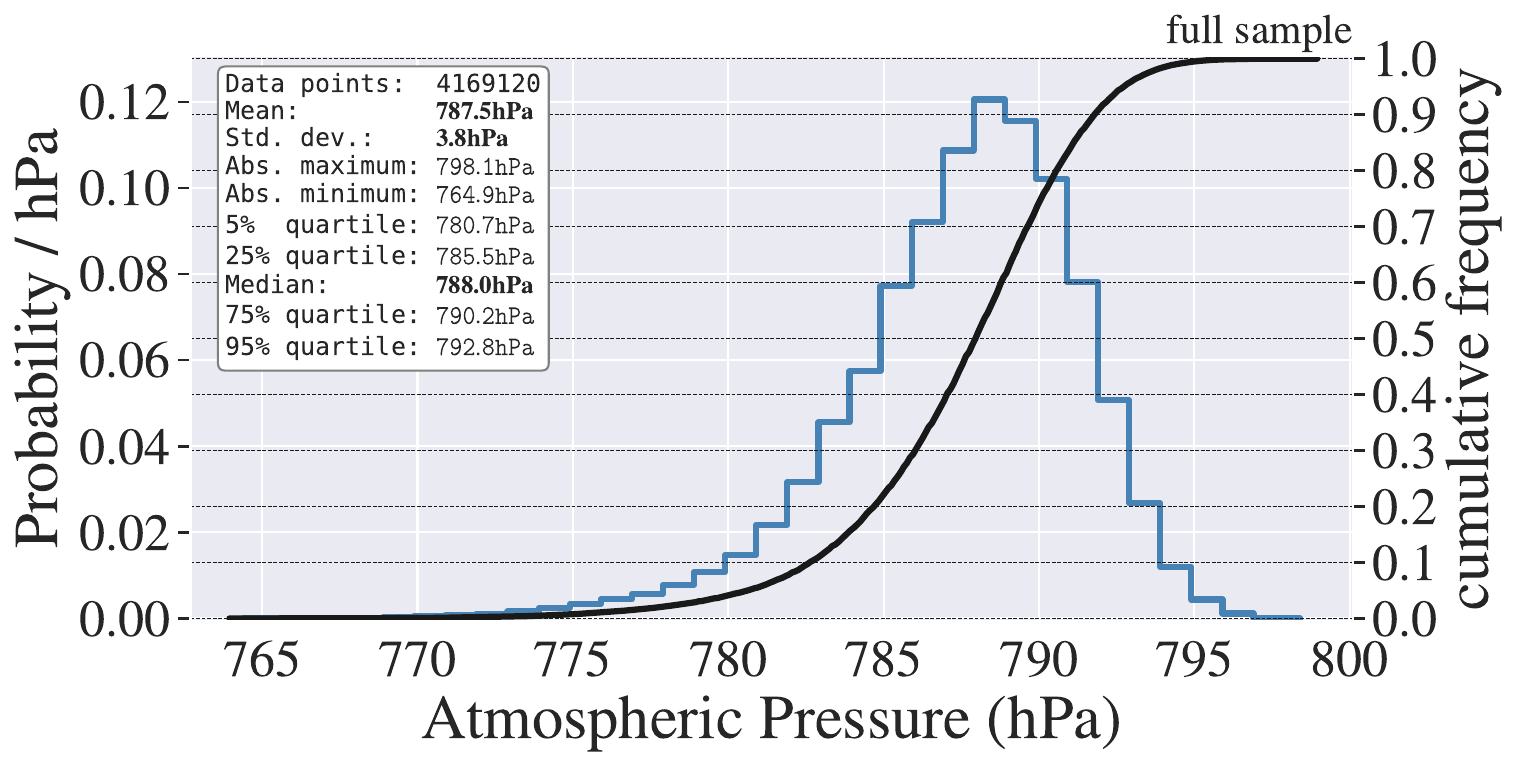}
\includegraphics[width=0.99\linewidth, clip, trim={0cm 0 0cm 0}]{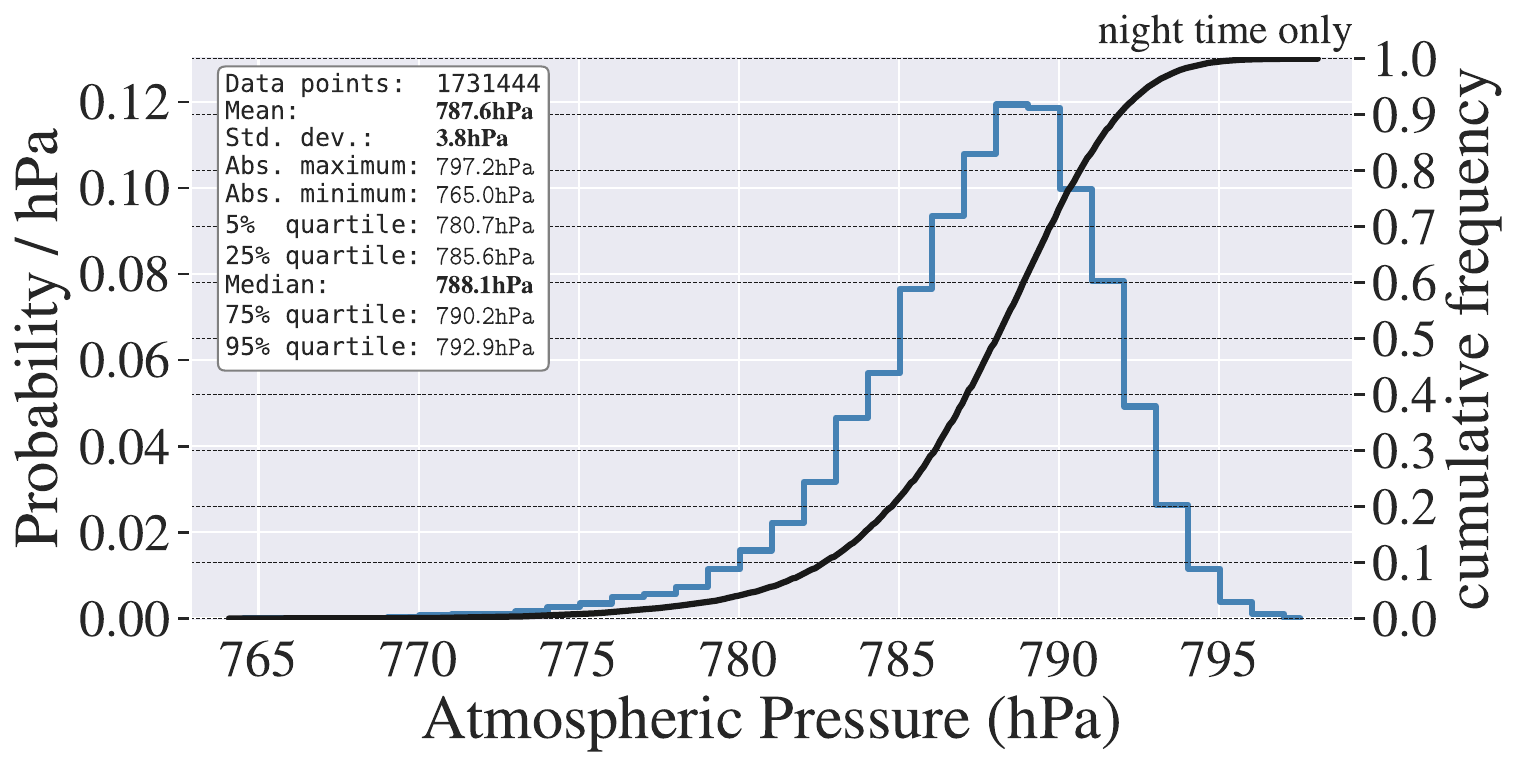}
\caption{
  \label{fig:histP}
  Distribution and statistical parameters of atmospheric pressure at the MAGIC site.}
\end{figure}

\begin{table*}
\centering
\begin{tabular}{l|cccccccc|c}
\toprule
Modality  &  $\hat{a}$ & $\hat{b}$ & $\hat{C}_1$ & $\hat{\phi}_1$ & $\hat{C}_2$ & $\hat{\phi}_2$ & $\hat{\sigma}_0$ 
 & $\widehat{\Delta P}$ & $\chi^{2}/\mathrm{NDF}$\\ 
          &  (hPa)  & (hPa/10y) & (hPa) & (month)  & (hPa)  & (month) & (hPa)  & (hPa) & (1) \\
\midrule

Daily means           & 785.9 & 1.23 & 5.6 & 6.6 & 4.2 & 9.0 & 3.1 & n.a. & 1.00 \\
Daily medians         & 785.9 & 1.30 & 5.6 & 6.6 & 4.2 & 9.0 & 3.1 & n.a. & 1.00 \\
Daily means w/offset  & 786.7 & 0.31 & 5.6 & 6.6 & 4.1 & 9.1 & 3.1 &  1.4 & 1.04 \\  
Daily medians w/offset& 786.7 & 0.13 & 5.6 & 6.6 & 4.1 & 9.0 & 3.1 &  1.4 & 1.04 \\
Monthly means         & 785.8 & 1.34 & 5.5 & 6.6 & 4.2 & 9.0 & 1.6 & n.a. & 1.04 \\
\bottomrule
\end{tabular}
\caption{\label{tab:lresultsP}  Parameters that maximize the likelihood, Eq.~\protect\ref{eq:ltemp} using Eq.~\ref{eq:murh} to fit the seasonal cycle for atmospheric pressure. The parameter $a$ corresponds to the average pressure as of 01/01/2003, $b$ the decadal pressure increase, $C_1$ and $C_2$ seasonal oscillation amplitudes with the respective locations of the seasonal maxima at $\phi_1$ and $\phi_2$, and $\sigma_0$ the average Gaussian spread of atmospheric pressure around the fitted seasonal cycle. The fitted calibration offsets $\widehat{\Delta P}$ apply to the periods in which the MWS-5MV was used (see Table~\ref{tab:WSreplacements}). 
\ccol{The column $\chi^{2}/\mathrm{NDF}$ provides the sum of squared residuals of all data points with respect to the function that maximizes the likelihood, divided by the number of degrees of freedom of the fit.}}
\end{table*}

\begin{figure}
\centering
\includegraphics[width=0.99\linewidth, clip, trim={0cm 0 0cm 0}]{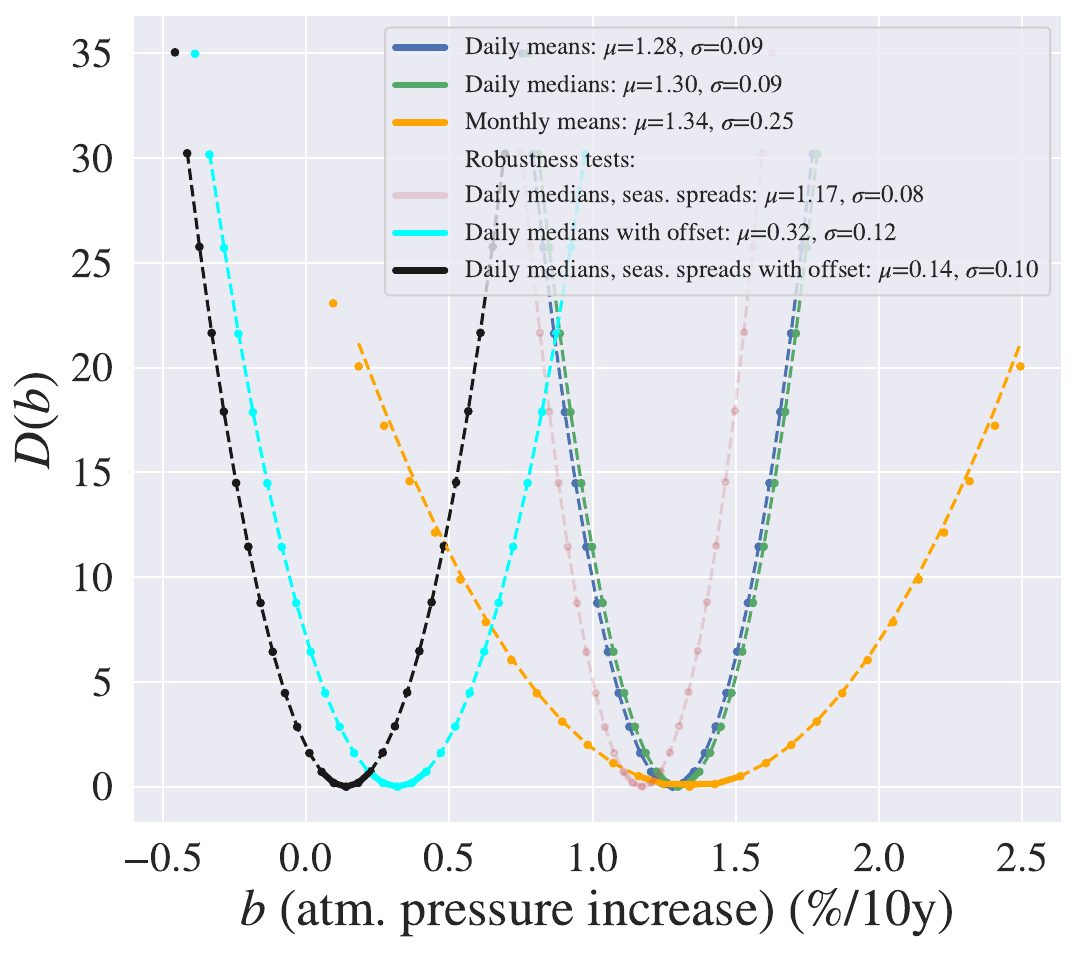}
\caption{
  \label{fig:pressL}
Profile likelihood test statistic $D(b)$ (see Eq.~\protect\ref{eq.lr1})  as a function of the atmospheric pressure increase parameter $b$. Shown are the profile likelihoods obtained from the 
  %daily mean relative humidity (green) using Eq.~\protect\ref{eq:mui}, 
  daily mean pressure (blue), daily median (green) and monthly mean pressure (yellow), all fitted to the seasonal oscillation function Eq.~\protect\ref{eq:murh}.
   For the monthly means, a monthly coverage cut of 80\% had been applied. Daily means or medians required a daily coverage of at least 85\%. 
   Furthermore, $D(b)$ from three robustness tests are shown: with a fit using a seasonal oscillation cycle for the spreads (pink) and 
   a possible systematic sensor offset between the WS models MWS~5MV and MWS~55V added to the list of nuisance parameters, with (cyan) and without seasonal oscillation cycle for the spreads (black). Note that in this case, the likelihoods, which include sensor calibration offsets yield incompatible results with the rest. 
   All profile likelihoods have been fitted to a parabolic function $(b-\mu)^2/\sigma^2$ around the minimum, the obtained fit parameters are shown in the legend.  }
\end{figure}

Figure~\ref{fig:annualP} shows the atmospheric pressure time series of the data set, in terms of month-wise statistics: median, the IQR and the month-wise pressure extremes. Seasonal variations are clearly visible, although less pronounced than in the case of temperature. On average, summer months show higher atmospheric pressure of $\sim$790\,hPa and winters lower ones from 784\,hPa to 788\,hPa on average. 
%\mg{There might be a trend to higher pressures, requires a dedicated statistical analysis.}
%\mg{Still have to re-scale the first five months, where the station was located at a slightly higher place}. 
Extreme pressure drops were observed during the winters of 2004, 2010, 2013, 2014, 2017 and 2018 and were immediately followed by winter storms~\citep{Marquez:2022}. 

The seasonal cycle of atmospheric pressure has been collapsed into month-wise statistics, shown in Fig.~\ref{fig:monthP}. The summer atmospheric pressure is higher on average and shows less variation, consistent with the stable atmospheric conditions when the Canary Islands enter the subtropical climate region. During the rest of the year, atmospheric pressure decreases on average, with a minimum reached in April. This behavior is generally observed in the mountainous regions of the Canary Islands~\citep{font-tullot}. 

The variation in pressure increases during spring and fall and reaches a maximum during winter; particularly, low-pressure extremes become more pronounced, reaching absolute minima in February and March. 
However, measurements of highest atmospheric pressure were also made in January and February.  

Finally, Fig.~\ref{fig:histP} shows the complete distribution and statistical analysis of atmospheric pressure, separately for the whole sample and only during astronomical nights (Sun elevation $<$ -12$^\circ$).  

The median atmospheric pressure is found to be about 16\,hPa higher than the one obtained on the mountain rim~\citep{Varela:2014}, consistent with the height difference.  The difference between minimum and maximum pressure observed is 70\% bigger than the one found by~\citet{Varela:2014}, but roughly consistent with the values of~\citet{lombardi2007}. We attribute this discrepancy to the larger sampling times of~\citet{lombardi2007} and our data set. 

% CAMC: 774.3 775.5 774.1 773.8 773.6 774.2 774.4 774.5 773.9 775.1 774.6 775.5 775.4 775.9 775.4 775.4 776.2 776.1 776.1 776.0
% --> Mean: 775.0
% TNG: 772.0 772.4 772.0 771.7 771.4
% --> Mean: 771.9 
% NOT: 771.3 771.1 771.4 771.6 771.8 771.8 771.5 771.3
% --> Mean: 771.5

Eq.~\ref{eq:murh} was used to fit the atmospheric pressure medians of the seasonal cycle in Fig.~\ref{fig:monthP}. 
The values that maximize the likelihood (Eq.~\ref{eq:ltemp}) using Eq.~\ref{eq:murh} are shown in Table~\ref{tab:lresultsP}. We find the two seasonal oscillation maxima around June/July (months 6 and 7) and September (month 9), a mean atmospheric pressure of 786~hPa as of 01/01/2003, and an increase of pressure of about 1.3~hPa every 10~years. 
However, here our systematic tests have revealed that a calibration offset of 1.4~hPa between the MWS-5MV and the MWS-55V is probable. The found offset is almost twice as large as the sensor accuracy given by the provider and might possibly be due to a wrongly programmed sensor height a.s.l. 
Figure~\ref{fig:pressL} shows the resulting profile likelihood ratio for the decadal increase parameter $b$. We observe that daily means and medians and the robustness test of including seasonal oscillation terms in the Gaussian spreads do not change much the behaviour of the test statistic, with a minimum found around 1.2--1.3~hPa/decade. 
The calibration offset fit yields, however, a correction of 1.4~hPa for the MWS-5MV, together with a significantly lower most probable atmospheric pressure increase of only 0.1--0.3~hPa/decade, statistically compatible with the null hypothesis. 
We conclude that in this case we cannot exclude a calibration error between the pressure sensors of the two MWS models used, possibly imitating an increase in atmospheric pressure. 

\subsection{Wind Speed and Direction}

Trade winds (locally called "Alisio") are the main driver of the climate of the Canary Islands, prevalent in more than 90\% of summer and 50\% of winter~\citep{font-tullot}. 
Whereas the predominant direction of the "lower" trade winds over the Sea is from the NE, driven by the Azores high-pressure system~\citep{Carrillo_2015}, north-westerly "upper" trade flows dominate above the temperature inversion layer in the free troposphere~\citep{Azorin-Molina:2018}, with different seasonal variations due to a seasonal North-South migration of the Azores anticyclone. During winter, cold air from polar air masses may occasionally intrude into the area through large-scale synoptic winds from the north-west~\citep{font-tullot}. Occasionally, tropical and subtropical storms or the aftermath of hurricanes hit the Canary Islands, particularly the island of La Palma~\citep{Marquez:2022}.

Several authors have analysed the wind patterns and speeds at the ORM, 
claiming prevailing (next-to-prevailing) wind directions from N(NNE) at the CAMC site~\citep{Jabiri:2000}, WNW(NW) at the GTC~\citep{Mahoney:1998}, E(NE) at NOT and TNG~\citep{lombardi2007} and NE(W) at the EST site~\citep{Hidalgo:2021}. These differences suggest that the predominant wind direction is greatly influenced by local orography. 
None of these studies separated wind directions between trade winds and tropical storms.

%with the main direction from the NW at the level of the observatories. 
%The subtropical high-pressure belt (the "Azores anticyclone"), together with the Canary Current, drives the trades in an NS direction. 
%\citet{Azorin-Molina:2018} pointed out, however, that above and below the temperature inversion layer different seasonal variations are found, with decoupled wind speed variability and trends.  

%Wind vector has been analysed considering its daytime and nighttime behavior.
%As usual, daytime data have been defined in the range 10:00--16:00, while
%nighttime data in the range 22:00--4:00. From each raw data series we have
%computed the hourly averages. From each set of wind speed hourly averages, we
%computed the monthly averages and then the annual percentage of hours in which
%the wind came from the different directions $D$. The wind rose has been divided
%into eight mean directions (N, NE, E, SE, S, SW, W, NW) corresponding to (0\deg,
%45\deg, 90\deg, 135\deg, 180\deg, 225\deg, 270\deg, 315\deg) and the percentages
%of hours are calculated into intervals defined as
%$[D-22.5^{\circ},D+22.5^{\circ}]$.

\begin{figure}
\centering
\includegraphics[width=0.99\linewidth]{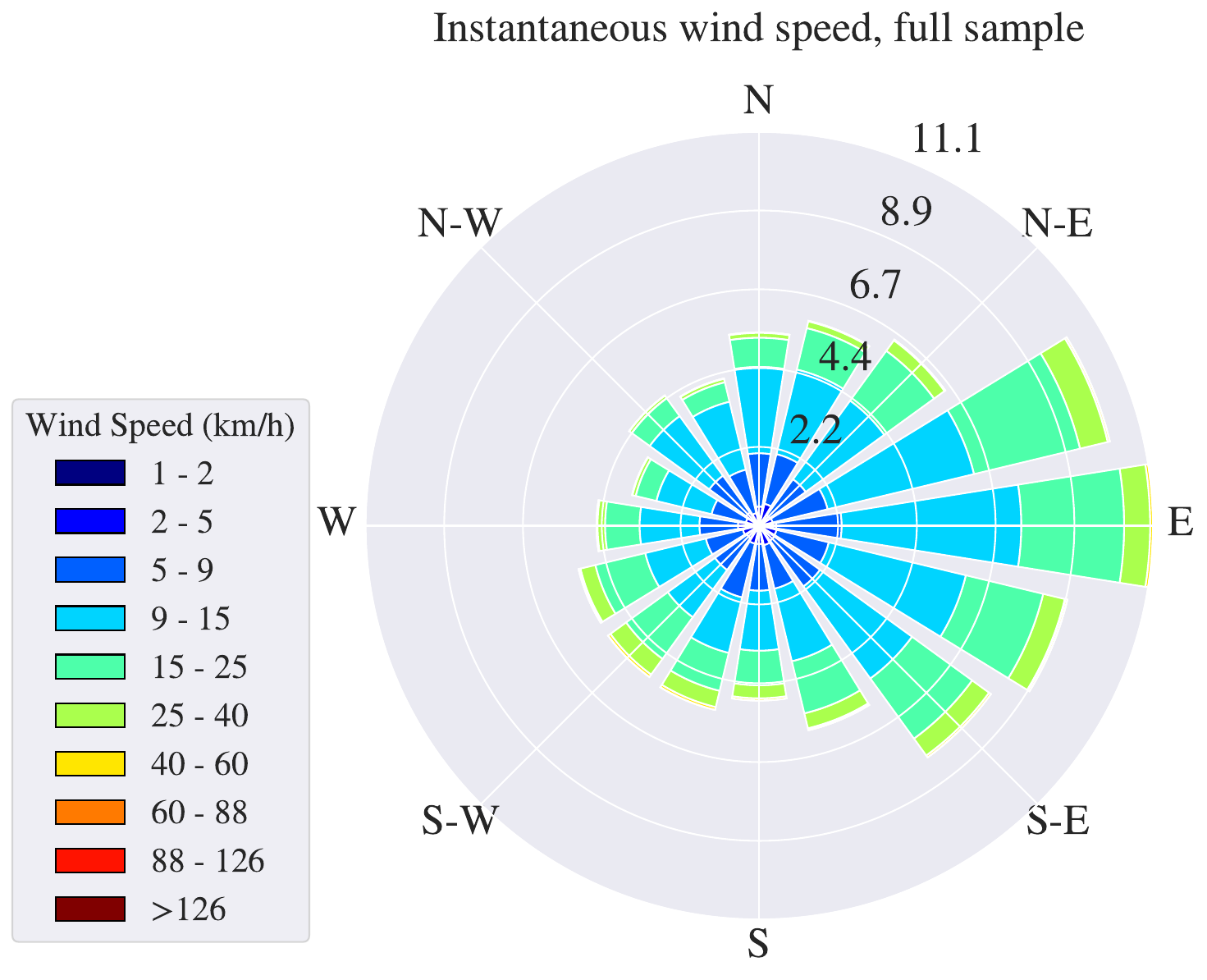}\\  \vspace{5mm}
\includegraphics[width=0.99\linewidth]{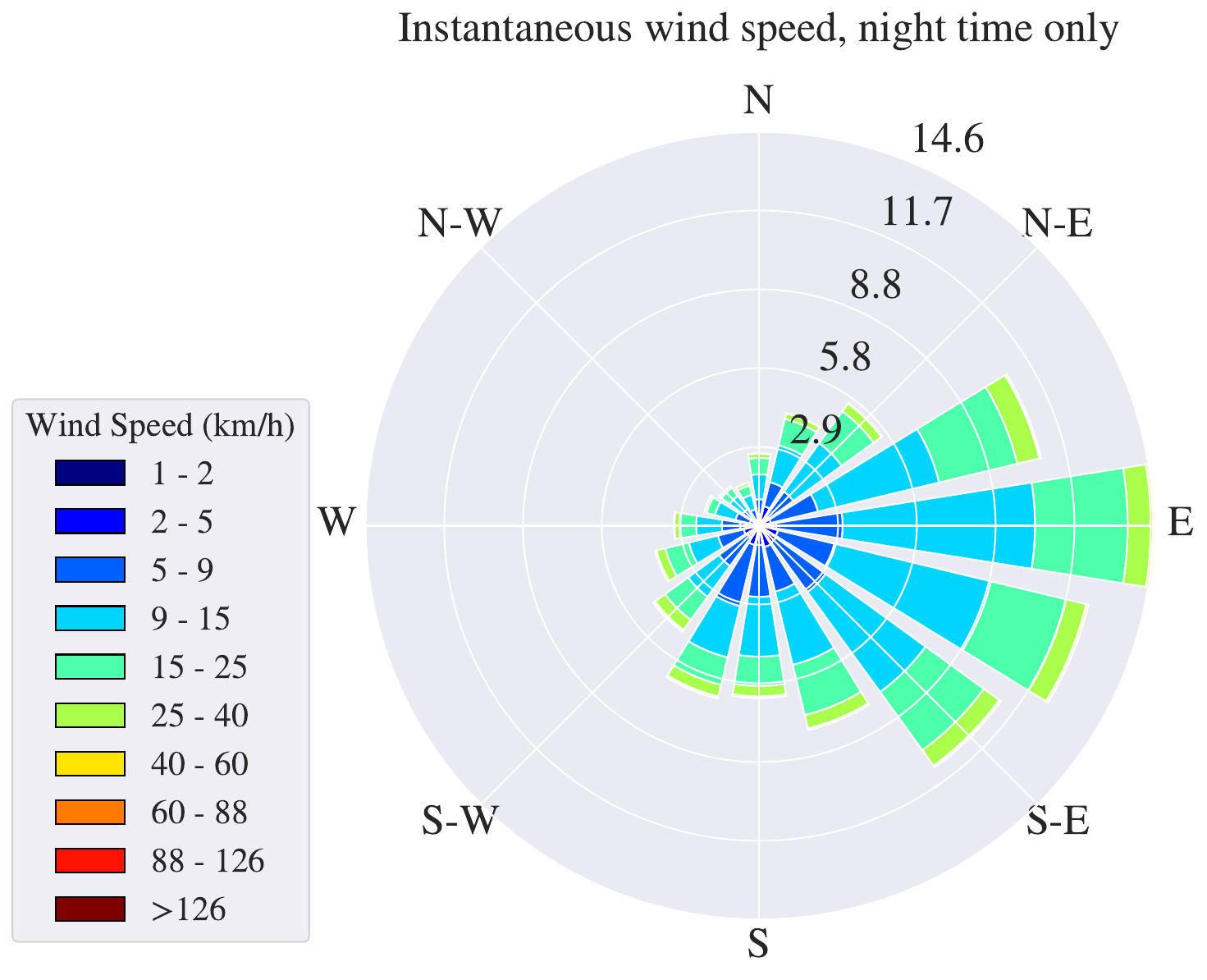}\\ \vspace{5mm}
\includegraphics[width=0.99\linewidth]{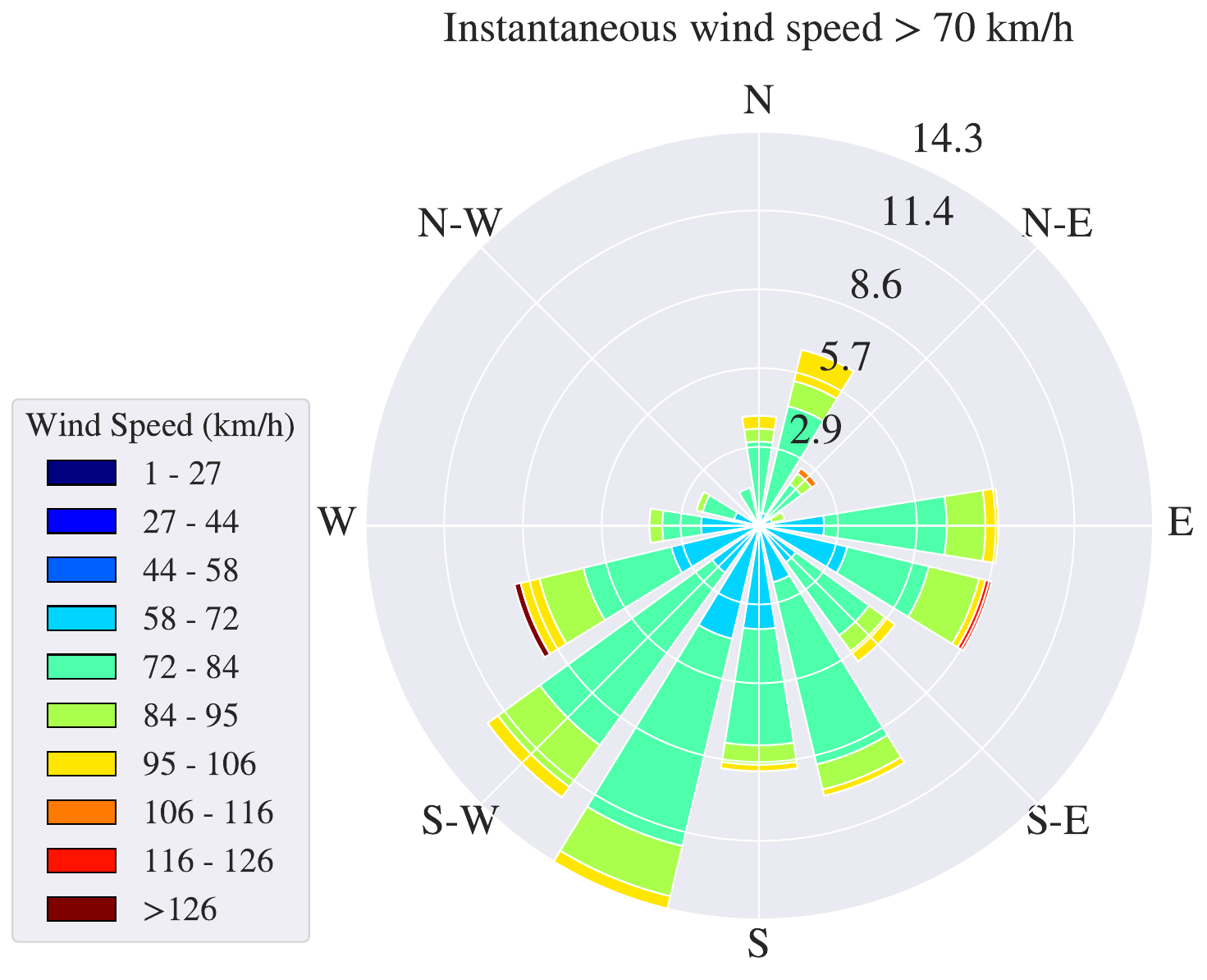}
\caption[Wind rose for MAGIC site in the period 2003--2023.]{
  \label{fig:windrose}
  Wind roses for the MAGIC site. The values on the polar axes show the probability to fall into the corresponding angle bin. Colors denote wind speed bin in km/h. Top: instantaneous wind speeds for all data, center: only astronomical night-time data, below:
  instantaneous wind speeds greater than 70\,km/h. Note the different color scales between the upper two and the lower wind rose. 
}
\end{figure}

Figure~\ref{fig:windrose} shows the wind roses for the MAGIC
telescope site, separately for the full data sample and only for astronomical nights. Below, the storms are shown with an additional lower wind speed cut applied. We observe that the MAGIC site shows a predominant wind direction from the east, followed by ENE during the day and ESE during astronomical night. Whereas westerly nocturnal winds from SW to NE are almost absent, about less than half of the daytime winds blow from an almost arbitrary westerly direction comprised within
NNE and S. 

The picture changes considerably when only strong storms are considered: here the predominant direction is SSW, followed by SW and further southern directions. However, it is interesting to observe that the absolutely strongest events, with wind speeds exceeding 110\,km/h, did not follow this pattern and blew from the WSW, ESE or NE, suggesting that such extreme winds can blow from any direction, except the fourth quadrant.
%in nighttime and a less prevailing
%wind direction from W to E in daytime. 
%The mean wind speed is 12.8\,km/h, lower
%than the ones registered at TNG (16.6\,km/h) and NOT (25.9\,km/h) but higher
%than the CAMC mean wind speed (7.9\,km/h)~\cite{lombardi2007}. The maximum wind
%speed measured at MAGIC is 136.1\,km/h in November 2005.

\begin{figure*}
\centering
\includegraphics[width=0.99\linewidth, clip, trim={0cm 0 0cm 0}]{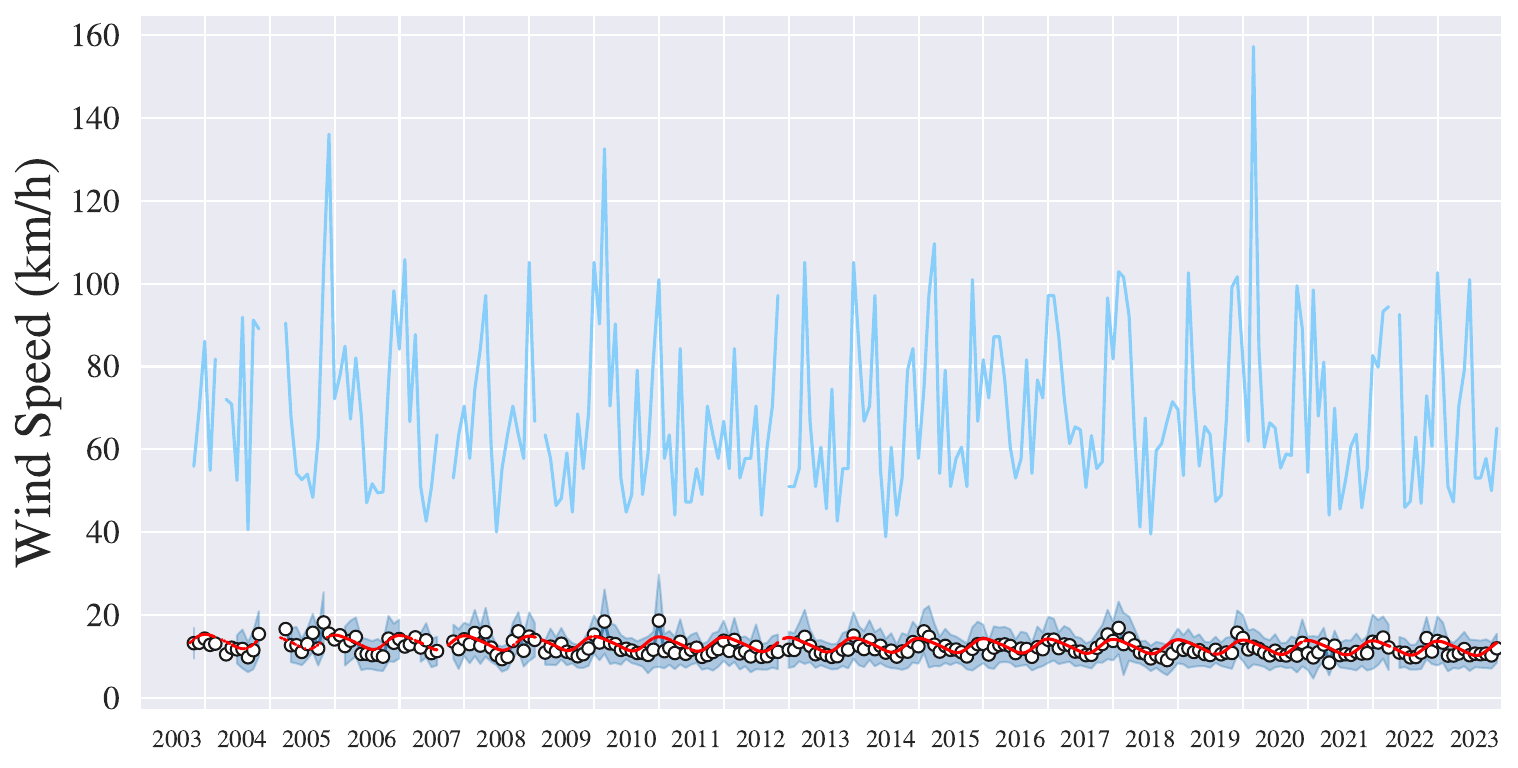}
\caption{
  \label{fig:annualW}
 Wind speed times series observed with the MAGIC weather station: white circles show the monthly medians of 10-minute wind speed averages, the blue shadow fills the IQR, and the light blue lines show the observed monthly wind gust maxima. The red line displays the result of the fit of daily wind speed medians to Eq.~\ref{eq:murh}. Additional gaps visible in the fit are due to the (higher) daily data completeness requirement of 95\%. }
\end{figure*}

\begin{figure}
\centering
\includegraphics[width=0.99\linewidth, clip, trim={0cm 0 0cm 0}]{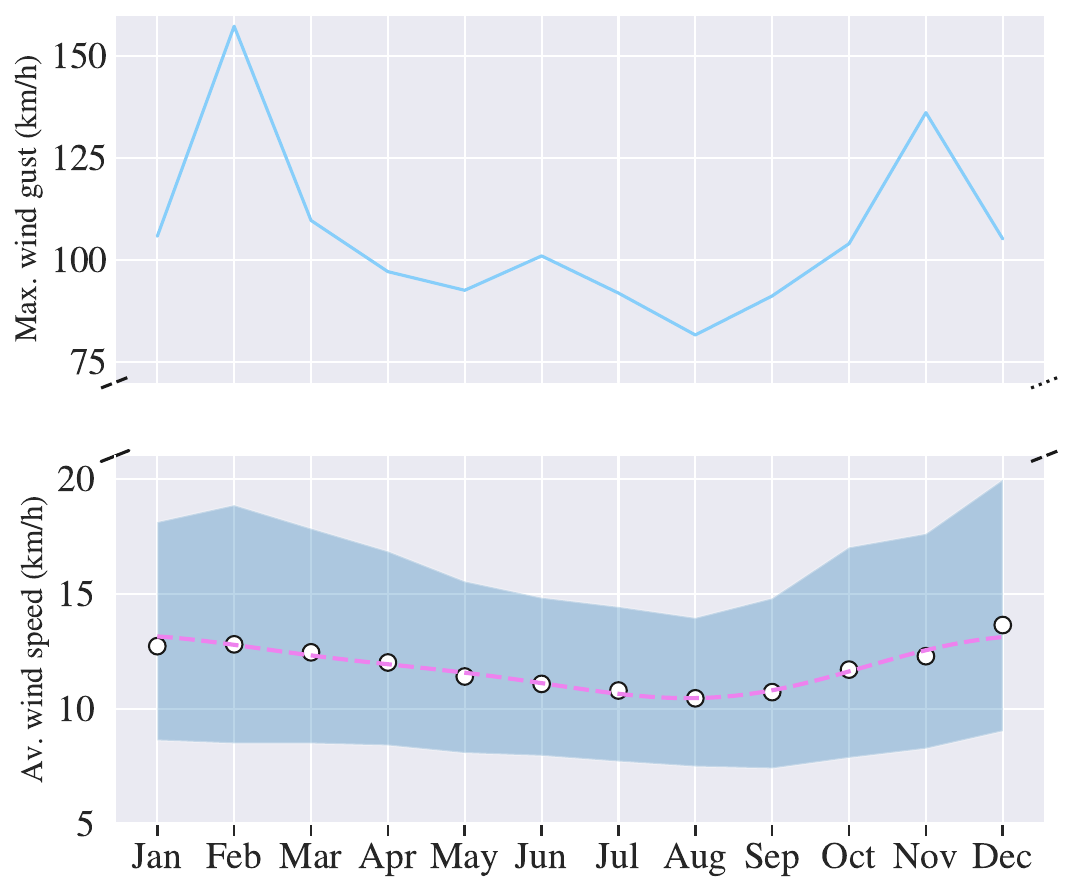}
\caption{
  \label{fig:monthW}
 Seasonal cycle of wind speeds at MAGIC: white circles display the month-wise medians of the 10-minute wind speed averages, the blue shadow fills the IQR, and the light blue lines display the month-wise wind gust maxima observed. The magenta dashed line shows a fit to the medians obtained with Eq.~\protect\ref{eq:murh} under the assumption of no temperature increase over time ($b$=0).}
\end{figure}

\begin{figure}
\centering
\includegraphics[width=0.99\linewidth, clip, trim={0cm 0 0cm 0}]{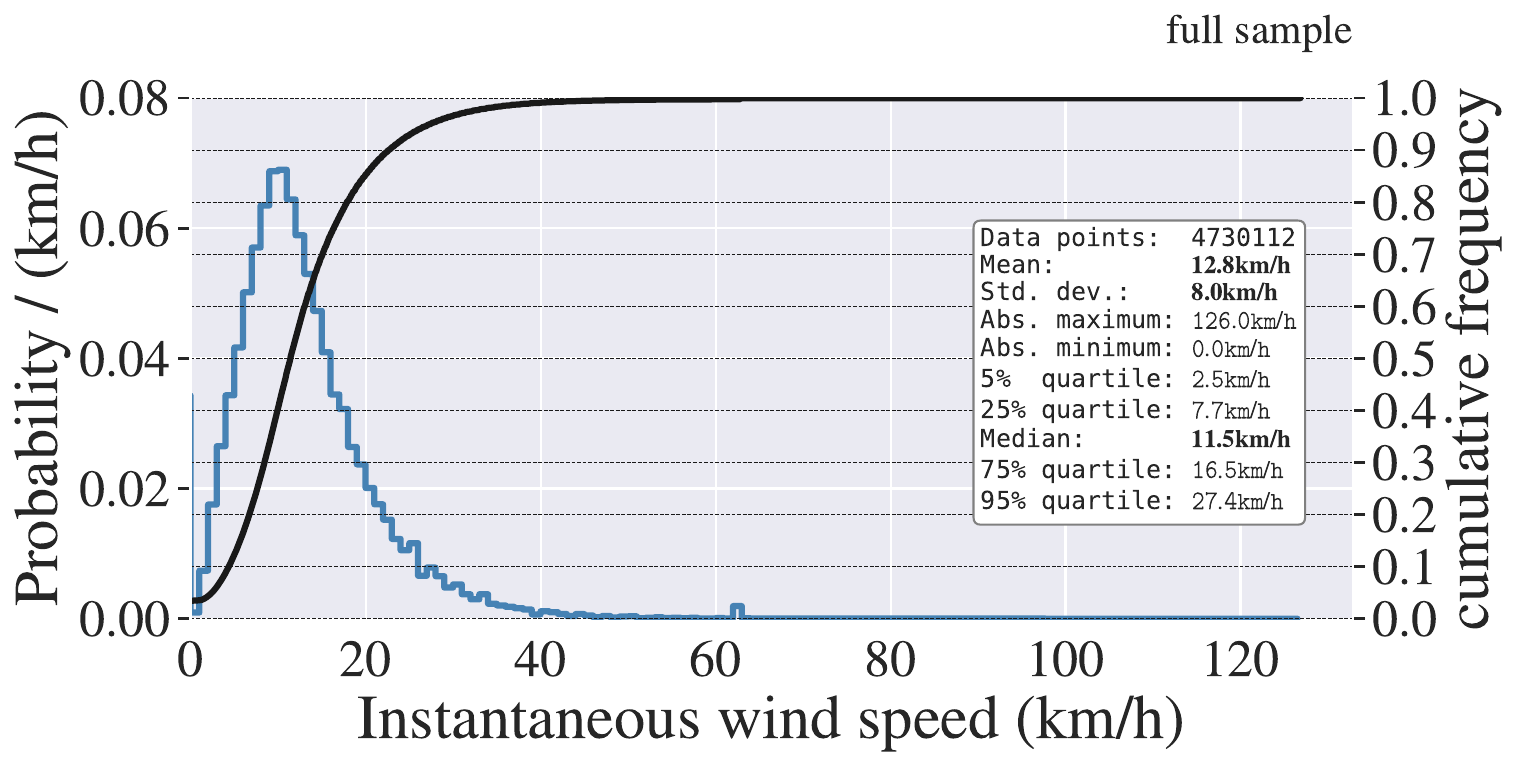}
\includegraphics[width=0.99\linewidth, clip, trim={0cm 0 0cm 0}]{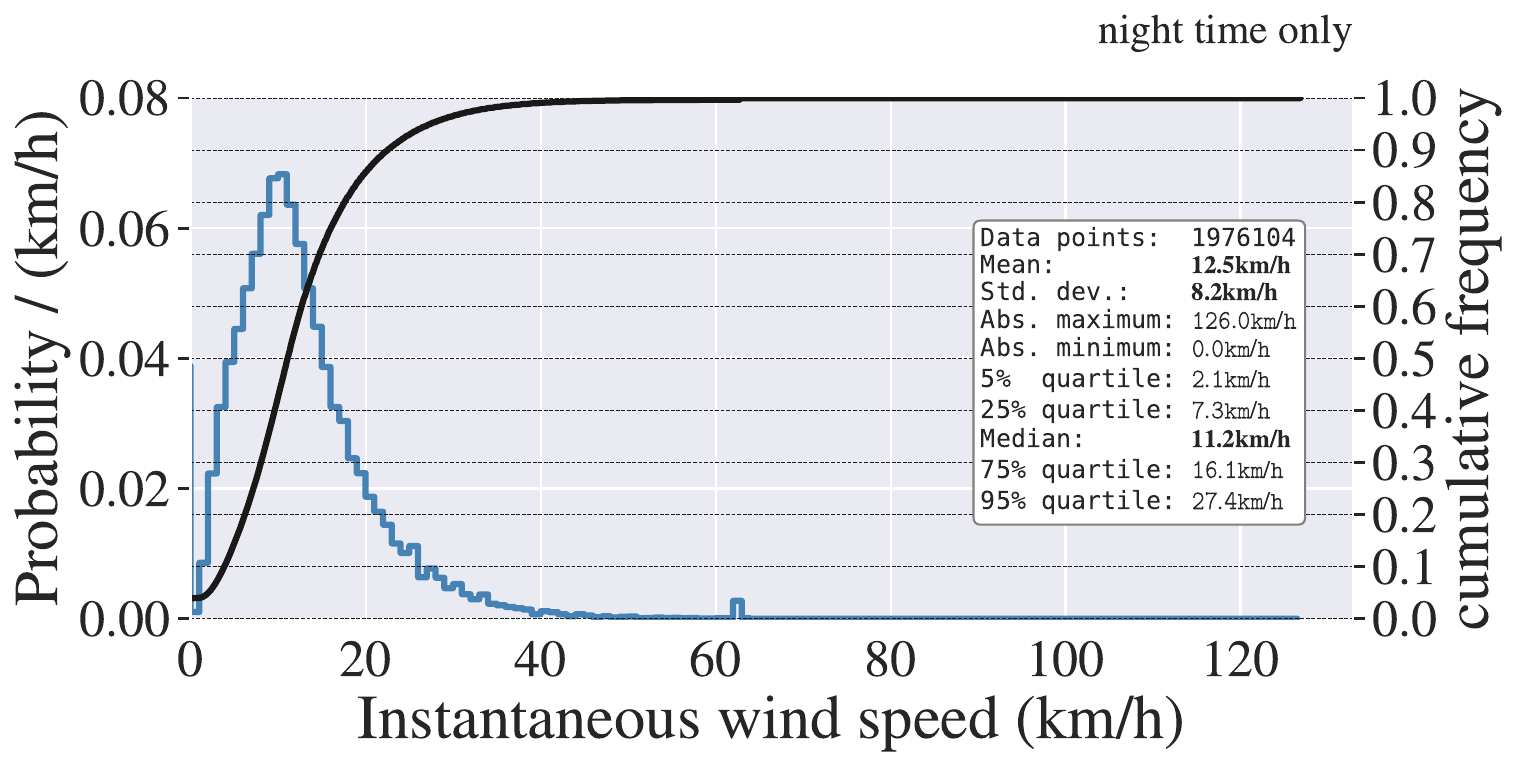}
\caption{
  \label{fig:histW}
  Distribution and statistical parameters of instantaneous wind speed at the MAGIC site.}
\end{figure}

\begin{figure}
\centering
\includegraphics[width=0.99\linewidth, clip, trim={0cm 0 0cm 0}]{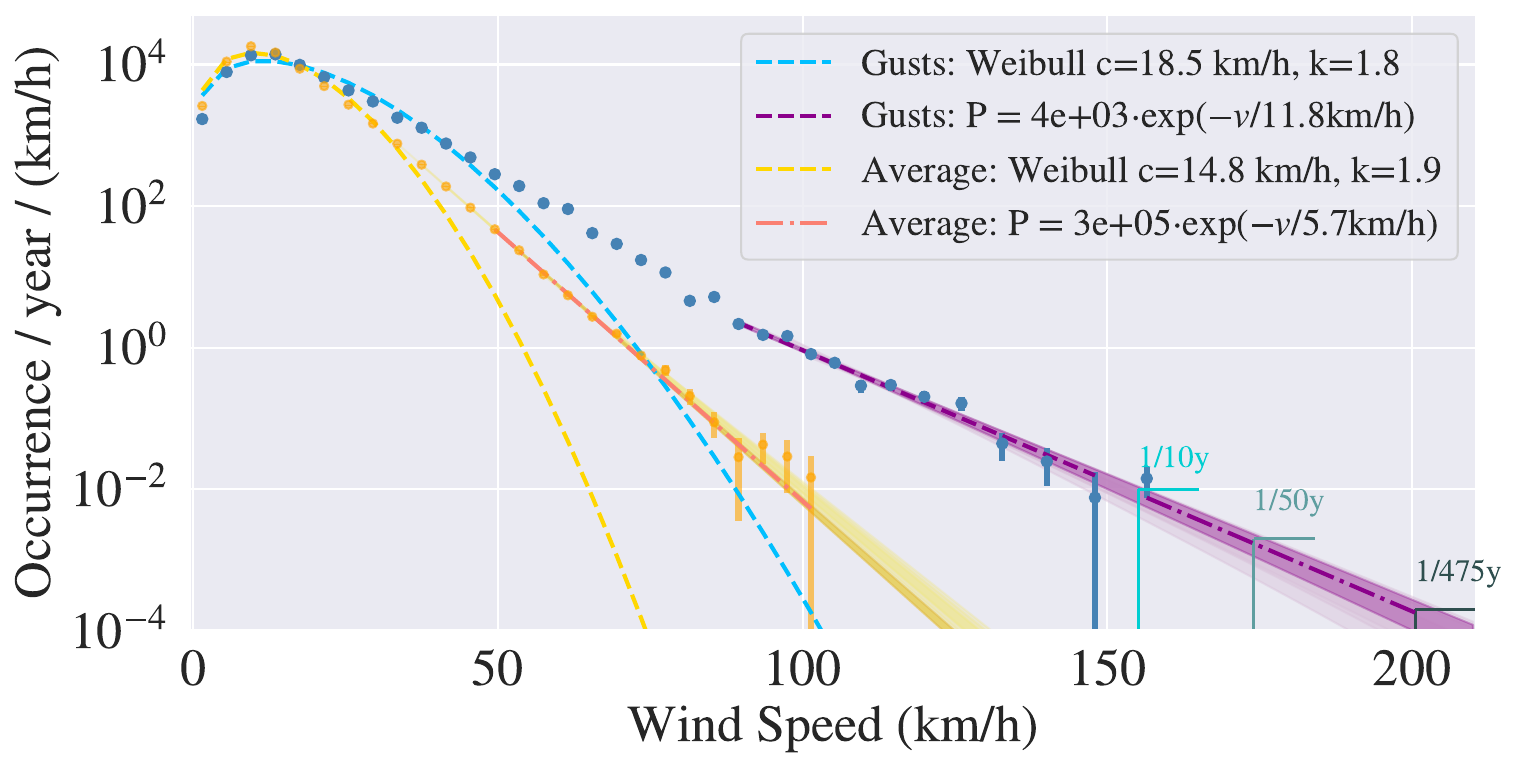}
\caption{
  \label{fig:histWG}
  Annual occurrence of wind speeds at the MAGIC site: In orange, 10-minute wind averages are shown and in blue, wind gusts. Note the logarithmic scale of the vertical axis. 
  Data gaps have been taken into account by building month-wise distributions and joining these individual (normalized) distributions according to the number of days of each month in a year. Error bars show the propagated Poissonian uncertainties. The dashed light blue and yellow lines show the result of a fit to a Weibull distribution~\protect\citep{Stevens:1979,Seguro:2000,Petkovic:2014}, the violet and orange dashed dotted lines show a fit to an exponential starting from where the $\chi^{2}/\mathrm{NDF}<1.5$. The violet area denotes the 1$\sigma$ fit uncertainty, which was obtained by taking into account the complete covariance matrix of the fit parameters, whereas the underlying pale violet area shows the range of fits obtained from varying the fit ranges (all within $\chi^{2}/\mathrm{NDF}<1.5$). The small rectangles denote return periods of 10, 50 and 475~years, obtained from an extrapolation of the most probable exponential fit to the wind gusts. }
\end{figure}

\begin{table*}
\centering
\begin{tabular}{l|ccccccc|c}
\toprule
Modality  &  $\hat{a}$ & $\hat{b}$ & $\hat{C}_1$ & $\hat{\phi}_1$ & $\hat{C}_2$ & $\hat{\phi}_2$ & $\hat{\sigma}_0$ & $\chi^{2}/\mathrm{NDF}$\\ 
          &  (km/h)  & ((km/h)/10y) & (km/h) & (month)  & (km/h) & (month) & (km/h)  & (1) \\
\midrule
Daily means    & 13.9 & -0.80 & 0.97 & 10.7 & 1.4 & 10.8 & 5.2 & 1.00 \\  
Daily medians  & 13.8 & -0.90 & 1.12 & 10.7 & 1.3 & 11.1 & 5.3 & 1.00 \\
Monthly means  & 13.6 & -0.60 & 1.19 & 11.1 & 1.3 & 11.2 & 1.5 & 1.04 \\
\bottomrule
\end{tabular}
\caption{\label{tab:lresultsW}  Parameters that maximize the likelihood, Eq.~\protect\ref{eq:ltemp} using Eq.~\ref{eq:murh} to fit the seasonal cycle for winds excluding storms (average wind speed $<$50\,km/h). The parameter $a$ corresponds to the average wind speeds as of 01/01/2003, $b$ the decadal increase in wind speed, $C_1$ and $C_2$ seasonal oscillation amplitudes with the respective locations of the seasonal maxima at $\phi_1$ and $\phi_2$, and $\sigma_0$ the average Gaussian spread of wind speeds around the fitted seasonal cycle. \ccol{The column $\chi^{2}/\mathrm{NDF}$ provides the sum of squared residuals of all data points with respect to the function that maximizes the likelihood, divided by the number of degrees of freedom of the fit.}}
\end{table*}

\begin{figure}
\centering
\includegraphics[width=0.99\linewidth, clip, trim={0cm 0 0cm 0}]{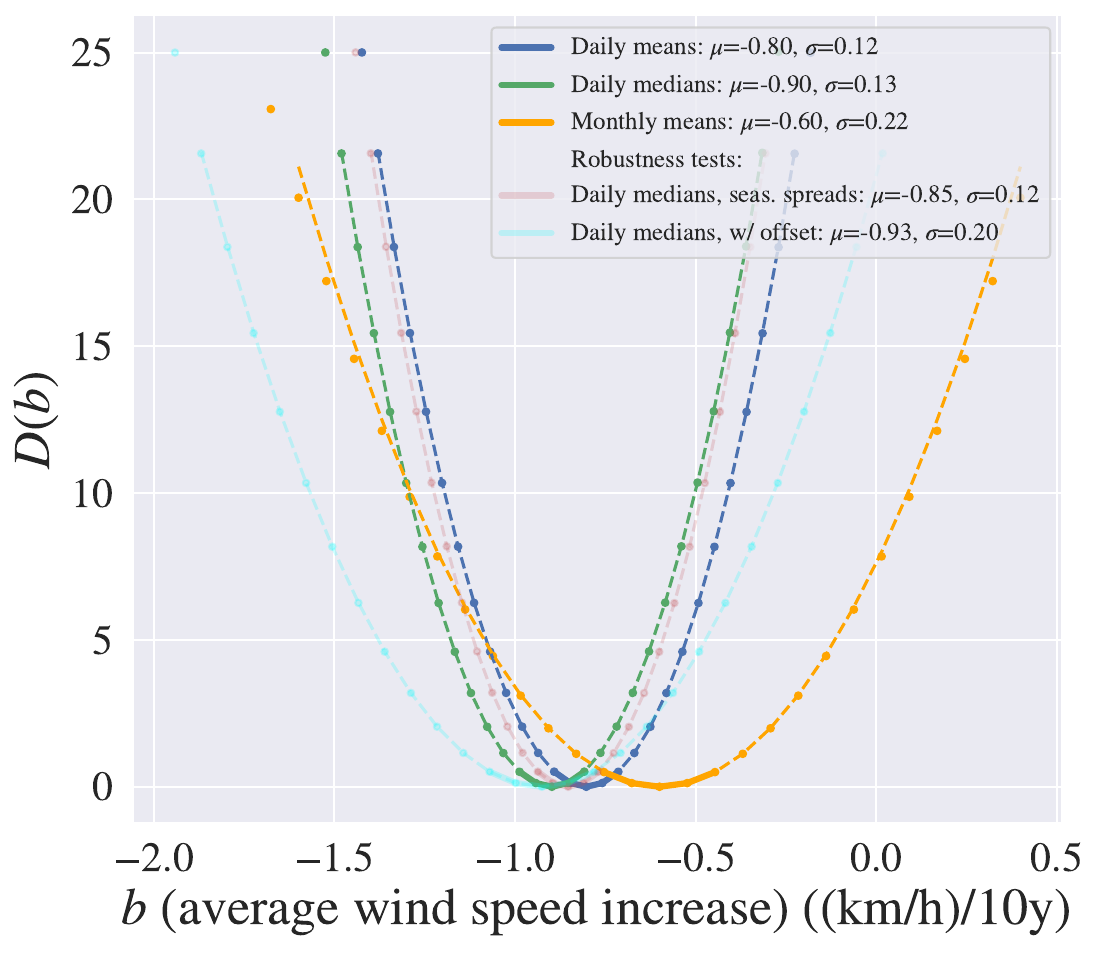}
\caption{
  \label{fig:windL}
  Profile likelihood test statistic $D(b)$ as a function of the average wind-speed-increase parameter $b$ for periods excluding storms (average wind speed $<$50\,km/h). Shown are the profile likelihoods obtained from the 
  %daily mean relative humidity (green) using Eq.~\protect\ref{eq:mui}, 
  daily mean (light blue) and the median (dark blue) average wind speed  fitted to Eq.~\protect\ref{eq:murh}, and a fit using varying seasonal spreads including (red). For reference, the profile likelihood of monthly means (yellow) are also shown. For the latter, a monthly coverage cut of 80\% had been applied. Daily means or medians required a daily coverage of at least 85\%.  
  All profile likelihoods have been fitted to a parabolic function around the minimum, the obtained $\mu$ and $\sigma$ parameters are shown in the legend. }
\end{figure}

Figure~\ref{fig:annualW} shows the time series
of 10-minute wind averages, in terms of month-wise statistics: median, IQR, and extremes. Note that the WS also constantly calculates the maxima of a 3-second sliding average and keeps the value in memory until reading, or until a new maximum is reached.  The month-wise extremes of the such defined \textit{wind gusts} are shown as light blue lines. 
We observe that wind speeds exceeding 100\,km/h are common in winter. 
Particularly strong tropical storms exceeded even 130\,km/h and were observed during the winters of 2005/06 (the \textit{Delta storm}), 2009/10 (the \textit{Xynthia}), and a strong south-easterly \textit{calima} sandstorm in the winter of 2019/20~\citep[see also][]{Marquez:2022}. 

The seasonal cycle of both parameters has been collapsed into month-wise statistics, shown in Fig.~\ref{fig:monthW}. Summer months show slightly weaker winds on average and much lower wind gust extremes. The highest average wind speeds are found in December, but the strongest storms occur in February, followed by November, consistent with the literature~\citep{font-tullot,Varela:2009,Hidalgo:2021}.

Finally, Fig.~\ref{fig:histW} shows the complete distribution and statistical analysis of instantaneous wind speeds, separately for the entire sample and only during the astronomical night.  
The latter, with a median wind speed of 11.2\,km/h (mean 12.5\,km/h), are found to be slightly softer than during the day (median: 11.5\,km/h, mean: 12.8\,km/h for the full sample). 
%Note that also completely wind-still periods account for about 4\% of the night-time. 
The observed median and mean wind speeds are lower by a factor of about two than the values obtained at the ELT and EST test sites and the NOT~\citep{Varela:2014,lombardi2007,Hidalgo:2021}, but only 30\% lower than the mean wind speed observed at TNG~\citep{lombardi2007} and 20\%-30\% higher than those obtained at CAMC~\citep{Jabiri:2000,lombardi2007}.
It seems that telescopes located directly on the mountain rim (the TNG and the NOT) experience much stronger winds on average than those located further downhill, of which the MAGIC site has the lowest altitude.

To understand the importance of less frequent, but potentially harmful, storms, we show the annual occurrence of average wind speeds and wind gusts in Fig.~\ref{fig:histWG}. Both distributions have been fitted to a Weibull \ccol{probability distribution} function, typically used for wind analysis~\citep{Petkovic:2014}:
\begin{equation}
\ccol{P(v) = \ddfrac{k}{c} \cdot 
\left(\ddfrac{v}{c}\right)^{(k-1)} \cdot \exp\left(-(v/c)^k\right)\quad,}
\end{equation}
\ccol{where $v$ denotes the wind speed, $c$ is a scale parameter and $k$ determines the shape or asymmetry of the distribution.}
One can immediately see that the Weibull fit describes only the trade wind part of the distribution, not the storms. To obtain a quantitative estimate, we used an exponential fit to the occurrence of wind gusts, starting from the first bin at 90\,km/h until the last but one wind gust entry, which corresponds to the maximum wind speed that the sensor can measure and may show signs of signal pile-up. For that reason, it has been excluded from the fit. Fits that yield an acceptable $\chi^2/\mathrm{NDF}\lesssim 1.3$ are shown as a violet dashed line in Fig.~\ref{fig:histWG}\footnote{\ccol{Here, the $\chi^2$ is defined as the sum of squared reduced residuals (i.e., measured in terms of standard deviations) of all data points with respect to the fit function, divided by the number of degrees of freedom of the fit.}}. The fit was extrapolated to return periods similar to or greater than the time coverage of the MAGIC weather station. These estimates are of particular importance for the CTAO and are listed in Table~\ref{tab:returnperiods}, which shows the most probable values obtained for different return periods of interest, together with the fit uncertainty itself and the ranges of values obtained from exponential fits to shorter, but still acceptable ($\chi^2/\mathrm{NDF} < 1.5, \mathrm{NDF}>6$) wind gust ranges starting at wind gust speeds $>$90\,km/h. When we include data points with even lower wind gust speeds in the fit, we obtain unacceptable fit qualities ($\chi^2/\mathrm{NDF}>4$), suggesting that the exponential model is not sufficiently accurate in the transition region between trade winds and storms. One can see that the statistical uncertainty intervals are asymmetric and considerably larger for negative values, what concerns the wind gusts. The varying fit ranges also yield asymmetric ranges of the same order, shifted to lower values.

\begin{table}
\centering
\begin{tabular}{c|ccc}
\toprule
Return  & Maximum    & Fit  & Uncertainty from \\ 
period & wind speed & uncertainty & varying fit ranges \\ 
 (years)    &  (km/h)   & (km/h) & (km/h)\\
\midrule
\multicolumn{4}{l}{Wind gusts} \\
\midrule
10   & 155   &  -3, +3 & -4, +1 \\  
50   & 174   &  -5, +4 & -5, +1\\
475  & 201   &  -7, +4 & -7, +2 \\
\midrule
\multicolumn{4}{l}{Average wind speeds} \\
\midrule
10   & 94.6    &  $\pm$0.6 & -1, +2 \\  
50   & 103.8   &  $\pm$0.7 & -1, +3\\
475  & 116.6   &  $\pm$0.8 & -1, +4 \\
\bottomrule
\end{tabular}
\caption{\label{tab:returnperiods} Expected maximum wind speed expected to occur during different return periods, obtained from the exponential fits to the occurrence of wind speeds (see Fig.~\protect\ref{fig:histWG}). The return period of 475~years corresponds to a 10\% occurrence 
 probability during 50~years. 
The second column shows the result of the most probable fit, with its statistical uncertainties shown in the third column. The fourth column displays the range of fit results to different wind speed ranges with  acceptable $\chi^2$'s. See Fig.~\protect\ref{fig:histWG} and text for more details. }
\end{table}
 
In the following, we divide our wind speed sample into those considered trade winds (average wind speed $\lesssim$50\,km/h), and storms. 
We will study possible long-term changes of both for the 20~years of our data sample.

\begin{figure*}
    \centering
   \includegraphics[width=0.8\textwidth]{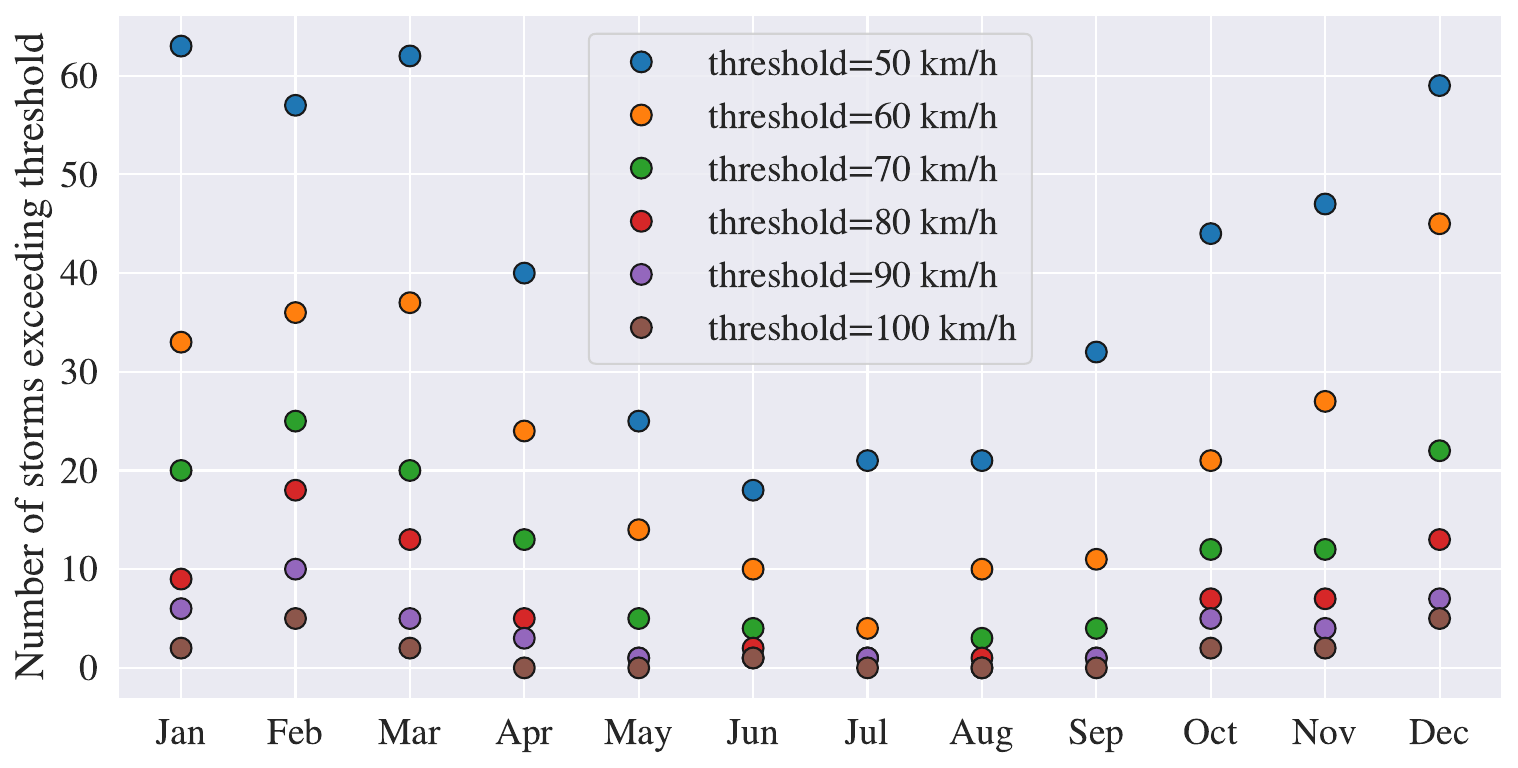}
    \caption{Seasonal cycle of total storm counts for different wind gust thresholds. Data gaps have not been taken into account.}
    \label{fig:storm_count}
\end{figure*}

%TENEMOS LA WS ORIENTADA MAL?? AL SUR HAY LA TORRE DEL LIDAR Y LA MONTAÑA.
%PORQUE ESTE COMPORTAMIENTO DISTINTO ENTRE DIA Y NOCHE?

%\begin{figure}
%\centering
%\includegraphics[width=0.99\linewidth, clip, trim={0cm 0 0cm 0}]{Limgs/Huracans_tall_50.pdf}
%\includegraphics[width=0.99\linewidth, clip, trim={0cm 0 0cm 0}]{Limgs/Huracans_tall_100.pdf}
%\caption{
%  \label{fig:huracans}
%  Two examples of the yearly occurrence of storms at the MAGIC site: On top, for wind speeds larger than 50\,km/h, below for very strong storms with wind speeds larger than 100\,km/h. The red dashed lines show the results of the most probable expectation values following a linear model. }
%\end{figure}

\begin{figure}
\centering
\includegraphics[width=0.99\linewidth, clip, trim={0cm 0 0cm 0}]{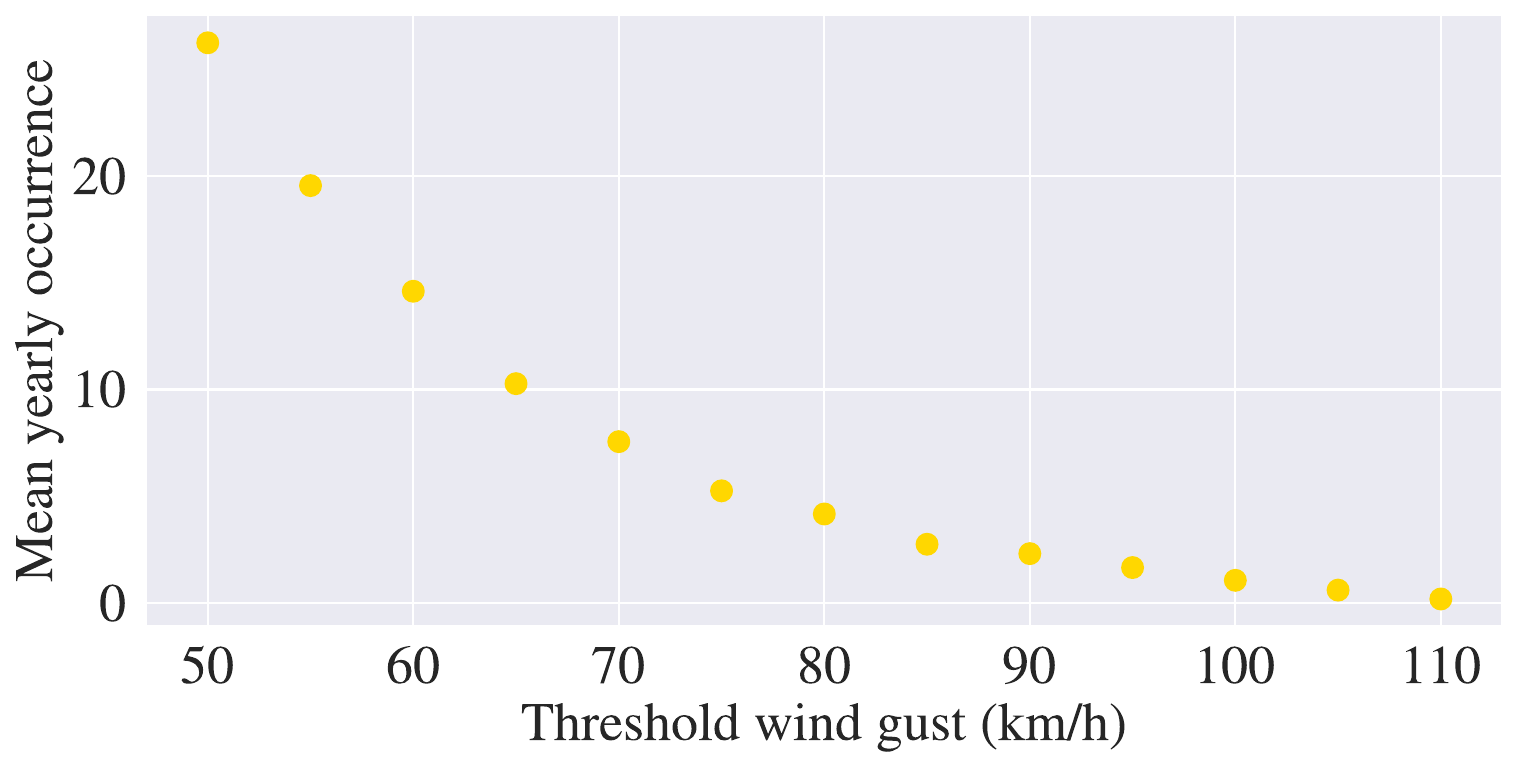}
\includegraphics[width=0.99\linewidth, clip, trim={0cm 0 0cm 0}]{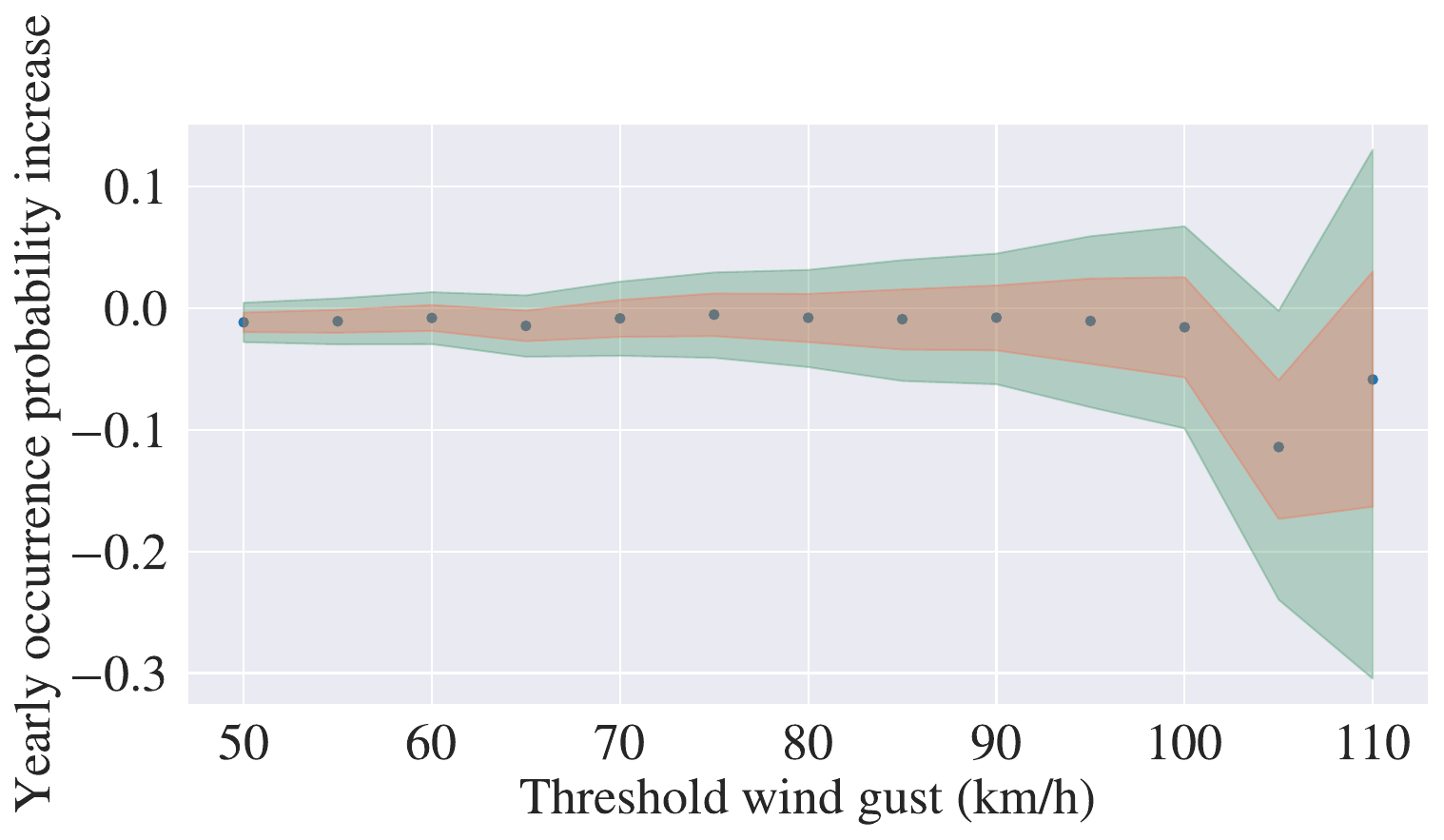}
\includegraphics[width=0.99\linewidth, clip, trim={0cm 0 0cm 0}]{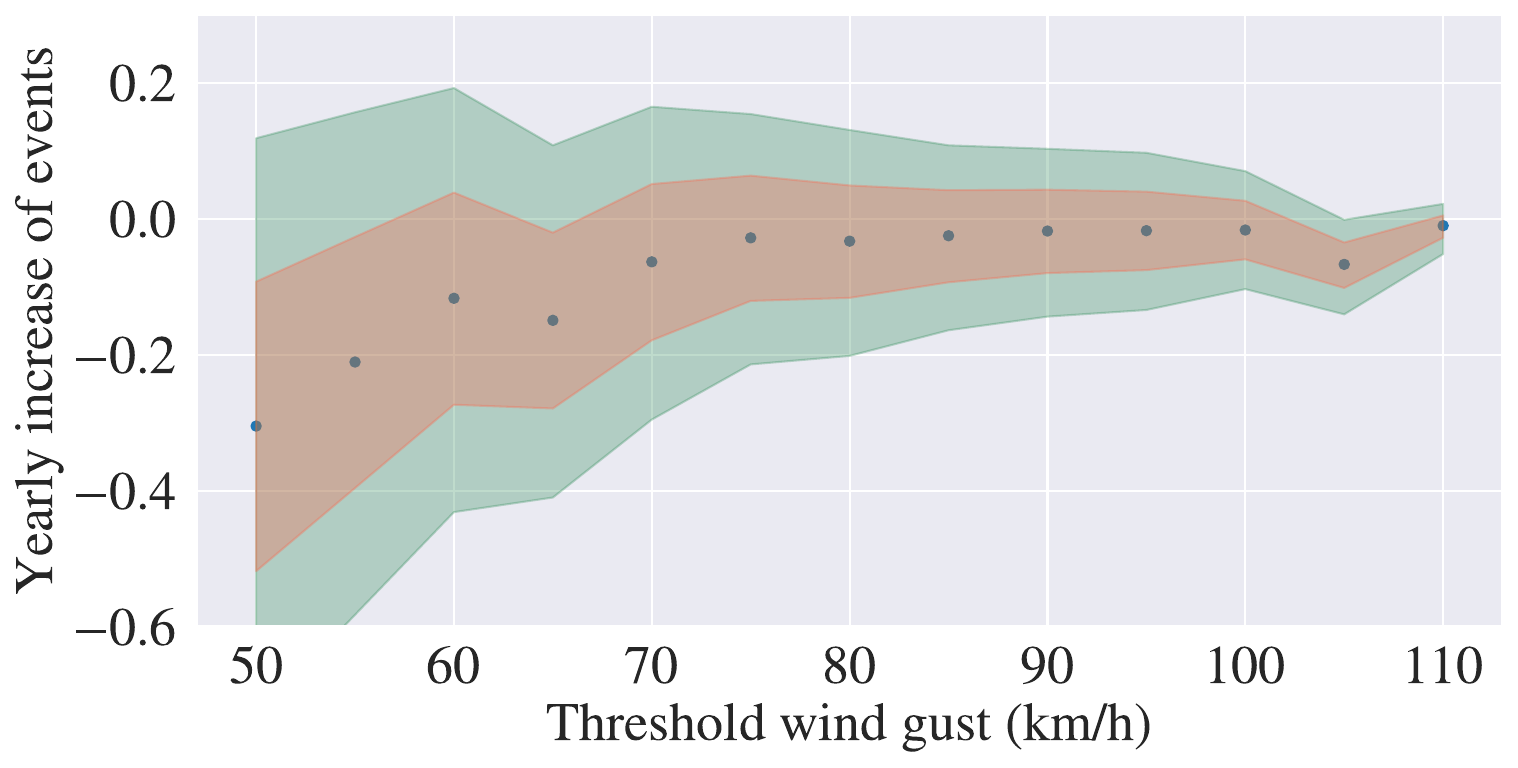}
\caption{
  \label{fig:alpha_rel}
  Yearly occurrence of storms at the MAGIC site, derived from the likelihood analysis: On top, the mean yearly occurrence $p_0$ is shown as a function of minimum wind gust speed applied. In the center, the relative yearly increase of storms: points show the most probable value, the brown band the 68\% confidence and the green band the 95\% confidence region of the profile likelihood. Below, the test statistic for the  relative yearly increase or decrease of storms, tested against the null-hypothesis of no increase, is displayed. }
\end{figure}

\subsubsection{Long-term evolution of trade winds}

We used Eq.~\ref{eq:murh} for the likelihood analysis (Eq.~\ref{eq:ltemp}) of average wind speeds $<$50~km/h to study possible a long-term evolution of trade wind speeds (expressed through the parameter $b$, increase in wind speed per decade). Table~\ref{tab:lresultsW} shows the combination of parameters that maximize the likelihood, and
Fig.~\ref{fig:windL} shows the obtained profile likelihood test statistic for $b$. We observe that neither moving from daily means to medians, 
%nor including a secondary oscillation term to the median seasonal cycle (Eq.~\ref{eq:murh}, also shown in Fig.~\ref{fig:annualW}) 
nor including seasonal oscillation terms on the spreads of the seasonal cycle nor allowing for a possible calibration offset between the MWS~5MV and the MWS~55V change much the behavior of the profiled test statistic: an average wind speed \textit{decrease} of  0.85$\pm$0.12(stat.)$\pm$0.07(syst.) (km/h)/decade is observed. 
%with more than 5$\sigma$ significance. 
If only monthly means are used, the detection significance drops to 2.7$\sigma$, with a best fit for $b$ of 0.60$\pm$0.22(stat.) (km/h)/decade.  This finding is in contradiction with \citet{Azorin-Molina:2018}, who reported a significant \textit{decrease} in \textit{annual} wind speeds below the inversion layer, but an \textit{increase} of about 2.2\,(km/h)/decade for Iza\~na (2373\,m a.s.l., Tenerife) using data from 1989 to 2014. The reason for the discrepancy might lie in the different influence of local orography, the different time period investigated, or the possibility that the MAGIC site sometimes lies below the second temperature inversion layer~\citep{Carrillo_2015}, whereas Iza\~na lies always above it. 
In this context, it is interesting to note that \citet{Hellemeier:2019} find a very significant \textit{decrease} of upper troposphere wind speeds of 2.4~(km/h)/decade at 200~hPa for La~Palma.

\subsubsection{Long-term evolution of storms and hurricanes}

We tested the occurrence of storms for different wind gust thresholds as a function time, using the profile-likelihood analysis based on Poissonian statistics (Appendix~\ref{sec:likelihood}). To do so, we counted the number of storms, defined as the number of nonconsecutive days in which at least one measured wind gust exceeded a given threshold speed; see Fig.~\ref{fig:huracans} for two examples and Fig.~\ref{fig:storm_count} for a rough seasonal cycle of storm counts, here still without taking into account data gaps. As expected, the strong prevalence of storms in winter is reproduced correctly. The measurements are then confronted with a Poissonian probability distribution with a time-dependent expectation value modelled as a linear increase or decrease over time. The expectation value does correctly take into account all data gaps through appropriate weights (see Appendix~\ref{sec:likelihood} for details). The location of the likelihood maximum yields hence a mean yearly occurrence of storms (see Fig.~\ref{fig:alpha_rel} top), a yearly increase of occurrence probability (Fig.~\ref{fig:alpha_rel} center), and a yearly increase of the total number of storm events (Fig.~\ref{fig:alpha_rel} below). 
%the significance of which is assessed against the null hypothesis of new increase or decrease with time  
We find only a marginal significance of about $1.5\sigma$ for a yearly decrease in storms below a maximum wind gust speed of 50\,km/h, compatible with our previous finding on trade winds. For higher wind speeds, no significant increase or decrease has been found.  We can exclude a change in storm occurrence with wind gusts stronger than 100~km/h by more or less than 1~event per decade at 95\%~CL. 

\subsubsection{Influence of nearby obstacles}

\ccol{To assess a possible influence of the nearby obstacles: telescopes and the LIDAR tower on wind measurements, we calculated a turbulence intensity, normalized to its expectation value for different wind speeds above 20~km/h, and study that \textit{excess turbulence intensity} as a function of wind direction (see Appendix~\ref{sec:TI}). No correlation has been found with any of the obstacles.}

\section{Impact of weather on the duty cycles of MAGIC and CTAO}
\label{sec:dutycycle}
%\ob{Responsible is Oscar Blanch}

\begin{figure}
\centering
\includegraphics[width=0.99\linewidth, clip, trim={0cm 0 0cm 0}]{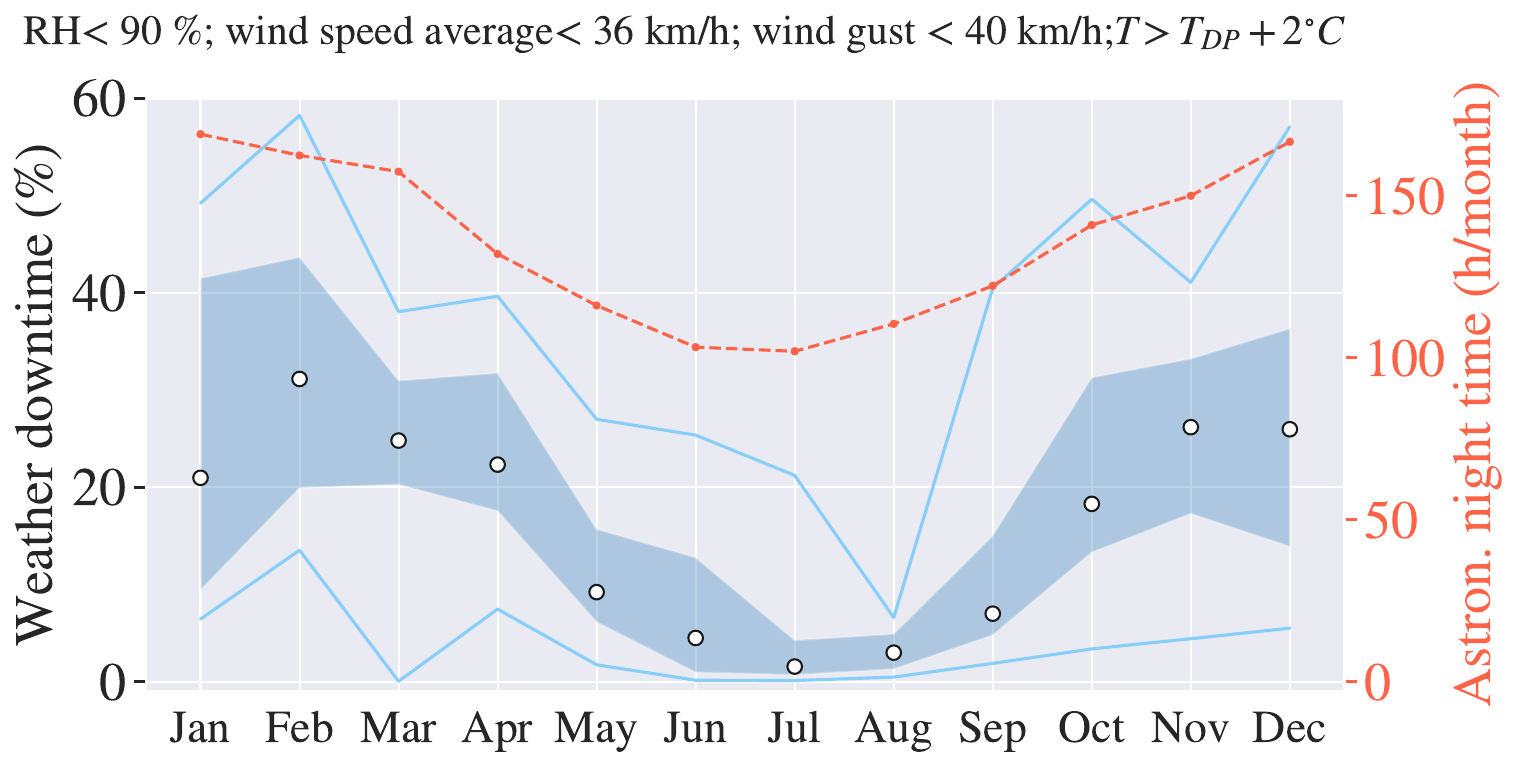}
\caption{
  \label{fig:monthDown}
  Seasonal cycle of hypothetical weather-related nightly downtime at the MAGIC site, if the observing criteria in the title are applied. Note that the condition on temperature lying 2\degC above the dew point is a CTAO requirement for mirrors not allowed to mist. For each wind gust, a waiting time of maximally 20 minutes has been applied. White circles display the month-wise medians, the blue shadow fills month-wise medians plus/minus median absolute deviation, and the light blue lines display the month-wise maxima and minima obtained. The dashed red line shows the total amount of dark night hours available for each month.}
\end{figure}

\begin{figure}
\centering
\includegraphics[width=0.99\linewidth, clip, trim={0cm 0 0cm 0}]{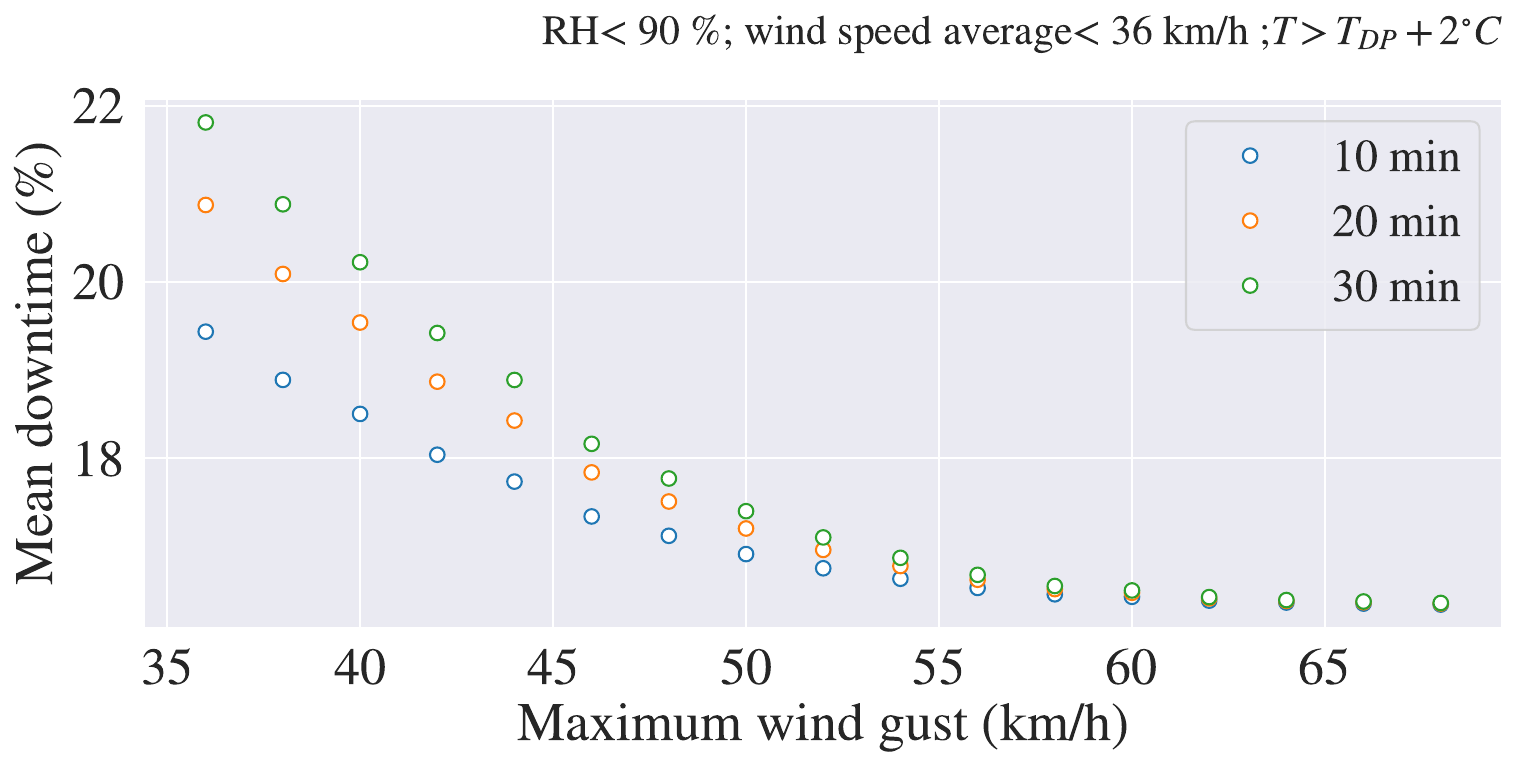}
\caption{
  \label{fig:gustDown}Mean annual downtime as a function of wind gust limits and waiting times, after all other weather-related operation limits have been applied. See text for details on the waiting time.   }
\end{figure}

\subsection{Safety Limits at the MAGIC telescope}

Correct and safe operation of the telescopes requires ensuring not only that
each element of the telescope can work in a safe way, but also that weather
conditions are appropriate for operation. Within the safety limits, the mechanical
and electronic components of MAGIC work correctly and are not at risk of
being damaged. 
%These components include the camera protection
%lids, the camera low voltage (LV) supply, and the camera photomultiplier
%tubes (PMT) high voltage (HV).
Additionally, the telescopes must
be parked and secured in case of severe weather events. The following
limits guarantee that the telescope hardware is not damaged: 

\begin{enumerate}[label=(\roman*)]
  \item Average and peak wind speed $<$40\,km/h,
  \item Relative humidity in air $<$90\%,
 % \item Relative humidity inside the camera $<$ 60\%,
 % \item Relative humidity inside the LV supply box $<$ 60\%,
 % \item Average PMT current $<$ 20\,$\mu$A,
 % \item Individual PMT current $<$ 47\,$\mu$A,
%  \item Zenith angle $>$ 1.5\deg.
\end{enumerate}

In case of bad weather, the camera lids get closed automatically if any of the
limits (i)--(ii) are violated, and the high voltage of the camera is switched
off in case limit (ii) is surpassed. 
%High humidity can result in short
%circuits in the electronics of the camera or the Low Voltage (LV) supply box, %which is
%attached below the camera frame. Both the camera and the LV box are designed to be waterproof, and only the humidity inside the electronics is important. Still, with outside humidity %approaching saturation, the high voltage is turned off as
%a safety measure. 
%The LV can be operated even under rain.

The part of the
telescope most affected by wind are the camera lids. They extend
on either side of the camera when opened, and the motor that controls them can be
easily damaged by wind pushing against the lids (the design was improved with a shutter for the LSTs, see~\citet{Inome:2014}). 
Therefore, the lids are closed even for short wind gusts that exceed limit (i). 
The 17\,m diameter mirror has a
similar effect on the structure as the sail of a ship. High wind speeds put
great stress on the elements of the telescope, which can lead to
positioning and pointing problems, or even bending or structural breaks. To
avoid this, the telescope incorporates a security system consisting of a number of heavy bolts that anchor the entire mobile structure to the base of the telescope, as
well as a mechanical break for the elevation drive. The telescopes are parked
and secured this way at the end of the nightly observations and if the average
wind speed exceeds limit (i).

Finally, human (operator) safety can only be guaranteed under low wind, i.e., sustained wind speeds $<$40~km/h and frequent wind gusts $<$55~km/h\footnote{see, e.g., \url{https://www.weather.gov/mlb/seasonal_wind_threat}.}. 
Safety limits need to be lower than that in order to allow operators to safely park the telescope and move from the counting house to the ORM residence. 

\subsection{Estimation of weather downtime}

Calculating average weather downtime is a surprisingly difficult task (see~\citet{garciagil2010} for a comprehensive discussion on topic). Apart from automatically (or semi-automatically) applied operation limits, a decision needs to be taken on when or whether to resume operation once the weather has improved. That decision depends on the operator's judgement of how the weather will most probably evolve during the rest of the night and whether the risk of worsening weather conditions may be taken or not. Often, this decision is influenced by the remaining amount of time scheduled for observations, after the weather has improved.  
In addition to that, catastrophic events can lead to considerable downtime after the event, required for inspection, tests, or repairs~\citep{Marquez:2022}. Operator logs are often not precise enough to disentangle with sufficient accuracy which part of the observation time losses were due to purely technical time, weather-related or weather-induced technical issues. 

We will present, in the following, a purely statistical analysis of weather downtime, applicable to an imaginary telescope system considered perfectly recovered and ready for observations right after each improvement of weather conditions, and confront this with a less precise, but probably more accurate, analysis of MAGIC observation logs. For the much more automatized Cherenkov Telescope Array Observatory (CTAO), the available observation duty cycle will certainly lie somewhere between both results.

Figure~\ref{fig:monthDown} shows the seasonal cycle of weather downtime, obtained from the complete data set with monthly coverage greater than 80\% and a set of typical operation limits foreseen for the CTAO. Data gaps have been correctly taken into account through the use of weights in this analysis. For comparison, the amount of monthly astronomical dark time is also shown. This figure, when compared to Fig.~9 of~\citet{garciagil2010}, shows the same trends, but a generally lower median downtime, particularly during the winter months. The difference may be due to the issues in assessing downtime mentioned in the previous paragraph.

The MAGIC Telescopes are operated in such a way that a wind gust exceeding 40\,km/h halts operation for at least 20 minutes. Only if during that time no other wind gust of the same magnitude or greater has occurred, observations can be resumed. We have taken into account that limit in the calculation of Fig.~\ref{fig:monthDown} (obviously, the limit does not apply if the weather conditions do not allow observation for another reason). In order to study the impact of that particular operation scheme (probably applied with different thresholds and waiting times by the CTAO),  we show, in Fig.~\ref{fig:gustDown}, the overall weighted downtime as a function of limiting wind gust speeds for three possible waiting times. The overall hypothetically achievable minimum weather downtime for the conditions applied at MAGIC (\textit{RH}$<$90\%, wind gust $<$40\,km/h, waiting time of 20\,min after wind gusts) comes out to be $\sim$19.5\%. Relieving the observation limits to wind gusts of $<$60\,km/h allows one to gain about 3\% on observation time on average. 
In that sense, 16.5\% weather downtime may be considered the absolutely lowest limit achievable for the CTAO at its northern site.

The reasons for the loss of science data taking time have been 
%why no physics data is taken is being 
logged and evaluated by the MAGIC Collaboration only since the beginning of 2010. Yet, there is still no fully automatic procedure for collecting these data. 
%\aha{Maybe: "THowever, there is no fully automated procedure for evaluating the various sources of information collectively."} 
The evaluation is made every day based on the operator logbooks, exchange of information within the MAGIC Collaboration, the weather information on site, and a fast look at the data recorded. On the basis of that, the dark time is classified as: 

\begin{enumerate}
\item Observed Dark Time (\textit{ODT}): Time during which science data are recorded.
\item Technical Time (\textit{TT}): Time during which non-scientific data are taken.
\item Overhead Operation Time (\textit{OOT}): Additional time needed for telescope operation: pointing, calibration, etc.
\item Bad Weather Time (\textit{BWT}): Time during which no data could be taken due
  to bad weather conditions. 
\item Technical Problems (\textit{TP}): Time during which no data could be taken due to technical issues. 
\end{enumerate}

The amount of data in each category is summarized for every data collection period, which is defined as the days that span two
consecutive full moons. The estimated precision on the number of hours collected 
in each category is about 5\%. In
general, time is tagged as \textit{BWT}, whenever the weather conditions are such that they do not allow data taking. 
However, this is not the case
when serious technical losses occur, which basically means that data
cannot be taken for a technical reason during more than a week. 
For this reason, a major source of uncertainty stems from the unrecorded contribution of bad weather to \textit{TT} and \textit{TP}. We have therefore evaluated two different scenarios: one that assumes that technical time is taken only under good weather conditions and technical problems occur independently on weather
(hence the bad weather fraction is obtained as \textit{BWT}/(\textit{ODT+OOT+BWT+TT})), and a second scenario where technical tests and problems are assumed to be scheduled or to occur independently of weather conditions, with a bad weather fraction of \textit{BWT}/(\textit{ODT+OOT+BWT}). 
This scenario applies to technical tests carried out with closed camera or inside the counting house, where the readout and trigger electronics, and the DAQ are located~\citep{magicperformance1}.
%but the technical problems occur that bad weather contributes by the same proportion to technical time and problems as to observed dark time and included 
%the variations caused by this assumption into the systematic uncerainty.

%\ob{I think we do not need the
%    table: The hours assigned to each category for each period from
%January 2010 to December 2014 can be seen in Table 1}. 

%The fraction of time available for observations depends on the epoch of the year. 

\begin{comment}
\mg{remove the next two sentences?}
Two periods have been selected for this analysis to avoid periods during which the telescopes were upgraded and to have the same statistics for all
epochs of the year. The selected periods  
are from 2~January 2010 to 17~January 2011 and from 1~November 2012 to 8~November 2014. 
\end{comment}

The two calculation methods yield then an average annual weather-related downtime of
% syst. error dominated by 5% on BWT: 0.05*18 and 0.05*20
18.0$\pm$0.5(stat.)$\pm$0.9(syst.)\% and 20.0$\pm$0.6(stat.)$\pm$1.0(syst.)\%, compatible with the 19.5\% obtained from the weather data analysis alone. 

These results are found to be compatible with the weather downtime obtained at CMT, but 6\% lower than those at NOT, WHT and even 10\% lower than the Liverpool Telescope (LT) and the TNG downtimes~\citep{garciagil2010}. The differences may be partly explained by the more conservative \textit{RH} limits for the latter two telescopes and the stronger winds blowing on the mountain rim. Furthermore, telescopes protected by a dome tned to be more conservative in their decisions to resume observations, implying reopening of the dome) than the free-standing IACTs like MAGIC and CTAO.

\section{Fractal time series analysis of meteorological parameters}
\label{sec:fractal}

Fractal analysis can be a meaningful tool for characterising meteorological time series. An example is found in~\citet{jiang2016multi}, where long records of air and ground temperature time series show changes in their multifractal characteristics, which decrease with increasing latitude in South China and are reported to be stronger for data acquired along the coast.

%\mg{I would like to move this part, until END into the introductory section.}
Several methodologies for time-series analysis exist, depending on the features to be extracted from the data. If we need to establish how much a given oscillatory mode contributes to the variability of the signal, the Fourier spectrum can be used. However, if missing data are present, the Lomb Scargle periodogram should instead be used~\citep{Lomb76}. If data are both nonlinear and nonstationary, adaptive algorithms can be used to obtain the oscillatory modes embedded in the time series, a technique that has been found useful for noise mitigation in gravitational wave interferometers~\citep{Longo_2020sc,Longo:2020wpg,bianchi2021gwadaptive_scattering,Longo:2021avq,longo2023scattered}. Time-series analyses also have applications in wind speed forecasting~\citep{li2010comparing,de2021adaptive}. To investigate the presence of self-similarity and persistence in the data and to characterise fluctuations at different scales, algorithms for fractal analysis such as Detrended Fluctuation Analysis (DFA) are widely used, for example, in physiology~\citep{Peng_1995} and seismometer array data analysis~\citep{Longo_2019,longo2021local}. Investigation of time-series variability across temporal scales is also of relevance in climate physics~\citep{franzke2020structure}. A fractal index, known as the Hurst exponent $H$, estimated by DFA, quantifies the persistence of a time series. The persistence of a time series is closely related to its long-range correlation, a property relevant, for example, in space weather data analysis~\citep{sharma2011extreme}. 
%\mg{END}

Generally, time series can be monofractal or multifractal. %\citep{Ihlen_2012}
A monofractal time series is characterised by a single value of $H$. If instead data persistence changes over time, and both fluctuations of large and small magnitude are present in the data, the time series is said to be multifractal. This behaviour can be highlighted using multifractal DFA (MFDFA)~\citep{kant,Ihlen_2012}, in which DFA is extended to higher orders $q$. A generalised Hurst exponent $h(q)$ is calculated, from which the multifractal spectrum can be obtained. The width of the multifractal spectrum provides information on the degree of multifractality in the data. The time series of meteorological parameters from the MAGIC weather station  were analysed using fractal analysis 
%to characterize fluctuations in the data when computed at different scales and to evaluate the presence of multifractality. The results were obtained using a fast fractal analysis
with a software called \emph{fathon}~\citep{Bianchi_2020_fathon}. 
%The adopted methodology is described in detail in section \ref{Hurst}, while the results are presented in section \ref{results}.  

\subsection{Methodology}
\label{Hurst}

%\begin{comment}
%Fractal time series analysis allows to investigate the presence of self-similarity and %to characterise fluctuations in the data. The Hurst exponent $H$ is often used in this %regard, a fractal index which is obtained using the Detrended Fluctuation Analysis %(DFA) algorithm~\citep{Peng_1995}.
%\end{comment}

%\mg{I would move this methodology section in the appendix.}

The Hurst exponent $H$ is related to the spectral index $\beta$ of a power-law process and is hence a measure of long-range correlation in the data. A process is said to be a power law if its frequency spectrum is related to the frequency by $P(f)\sim f^{-\beta}$, $\beta$ being the spectral index. $H$ is closely related to $\beta$ by the following equation~\citep{buld}:
\begin{equation}
H=\frac{\beta+1}{2}\quad.
\label{Eq:spectral}
\end{equation}
For a flat spectrum, i.e. white noise, ($\beta=0$) $H=0.5$  while for pink noise ($\beta=1$) and red noise ($\beta=2$) $H=1$ and $H=1.5$, respectively. For $H>1$ the time series is nonstationary \citep{Ihlen_2012}. 
$H$ defines the scale-invariant structure of a time series:
\begin{equation}
x(ct) \overset{d}{=} c^{H}x(t)\quad,
\label{hurst_dist}
\end{equation}
where $\overset{d}{=}$ stands for distribution equality. For that reason,
$H$ quantifies the roughness of a time series, which will appear smooth or very noisy if $H$ is high or low, respectively. If a time series can be described using one value of $H$, it is said to be monofractal, and $H$ can be estimated by DFA. If the time series is characterised by both large and small fluctuations, and its persistence varies on different time scales, it is said to be multifractal and is better described by a set of Hurst exponents. To quantify multifractality in the data, MFDFA can be used, which computes the generalised Hurst exponent $h(q)$ and the multifractal spectrum \citep{Ihlen_2012}. Finally, Detrended Cross Correlation (DCCA) analysis can be used to evaluate correlation between two time series performing detrending and using windows of different scales. The two time series can also be non-stationary. To test the significance of the DCCA results, DCCA has also been performed on a thousand couples of different random noises, and the 0.27 and 99.73 percentiles of the obtained correlations have been chosen as the thresholds for a significant negative and positive correlation, respectively. The distribution of these values has also been fitted with a binomial distribution in order to evaluate the probability that a random noise could give the same correlation value as the significant one. The DFA, MFDFA and DCCA algorithms are described in more detail in Appendices~\ref{sec:DFA} and~\ref{sec:H}. 

\begin{figure}%[t!]
\centering
\centerline{
\includegraphics[scale=0.25]{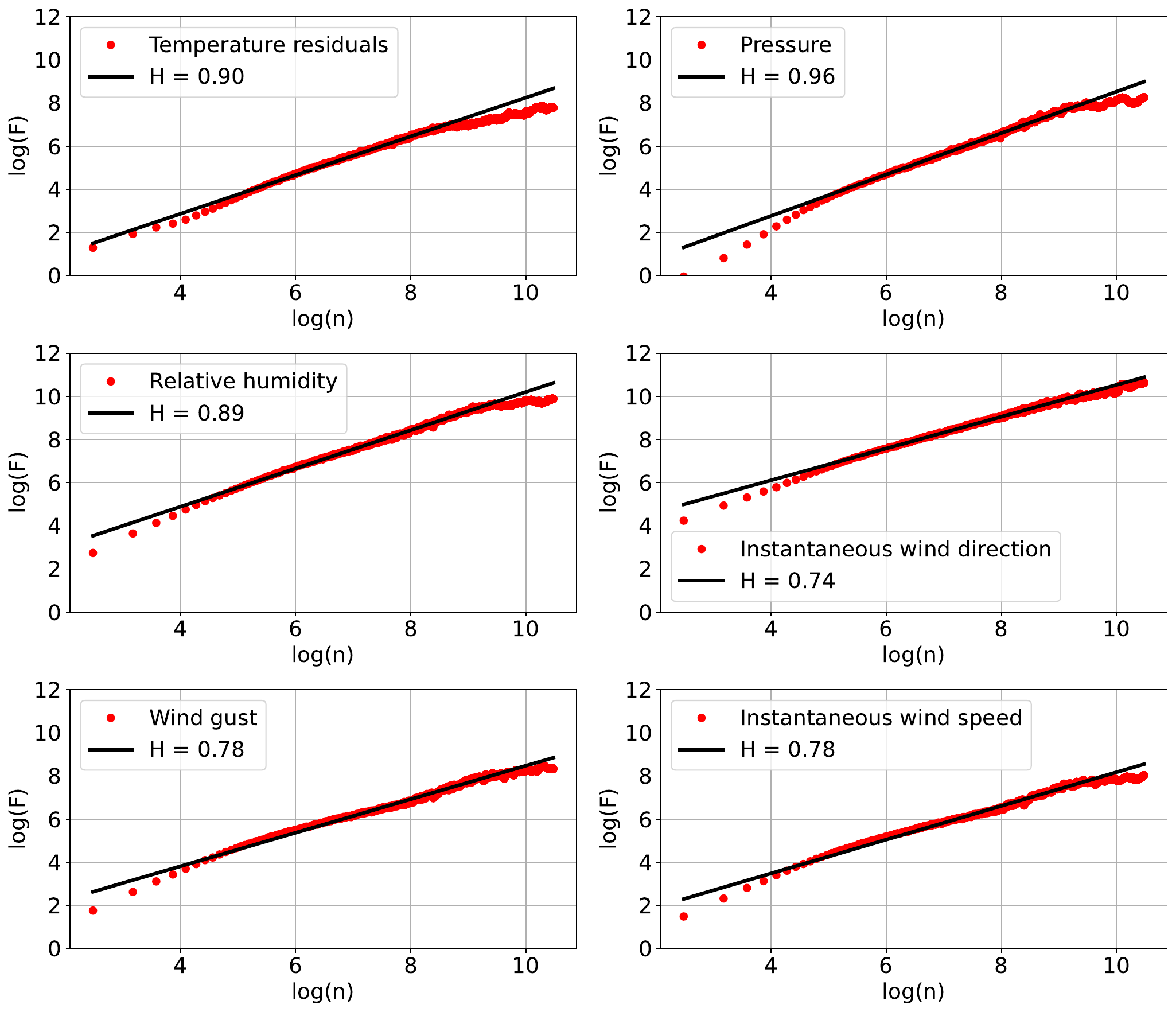}}
\caption{Double logarithmic plot of the fluctuation function $F(n)$ at different scales $n$, for the different meteorological parameters. Scaling is present over a wide range of scales, with the fluctuations following a power law relationship of the type $F(n)\sim n^{H}$. $H$ is the Hurst exponent. \label{fig:DFA}}
\end{figure}

\begin{figure}%[t!]
\includegraphics[scale=0.42]{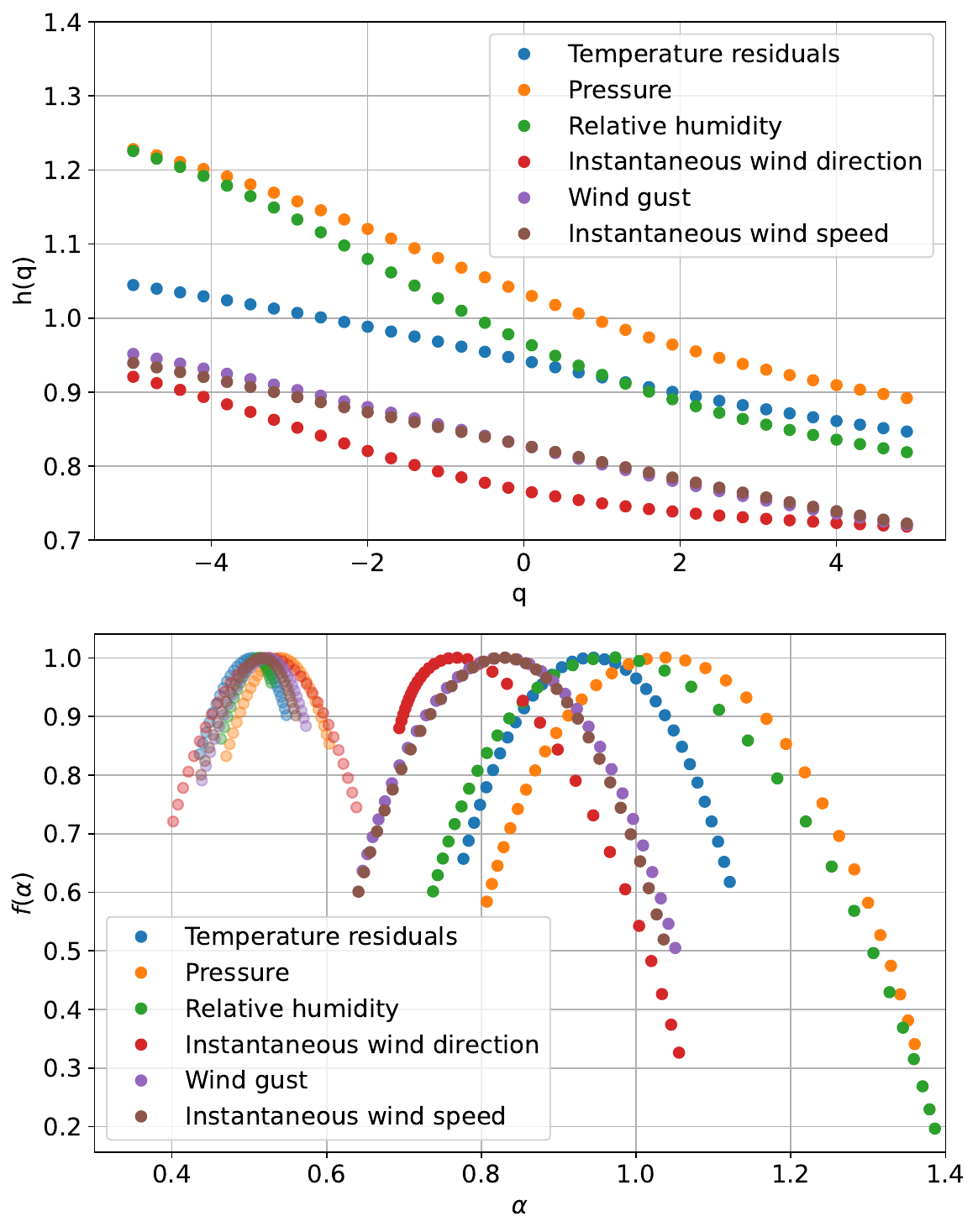}%
\caption{Top: $q$-order Hurst exponents for the meteorological parameters. Values of $q$ span from -5 to 5. General multifractality is observed for all the meteorological parameters. Bottom: Multifractal spectra of the actual meteorological time series (right), and of the shuffled meteorological time series (left). The wide spectra on the right show different degrees of multifractality, while the narrow spectra on the left confirm that multifractality depends on the temporal structure of the data. \label{hq_mf}}
\end{figure}

\begin{figure}%[t!]
\centering
\centerline{
\includegraphics[scale=0.45]{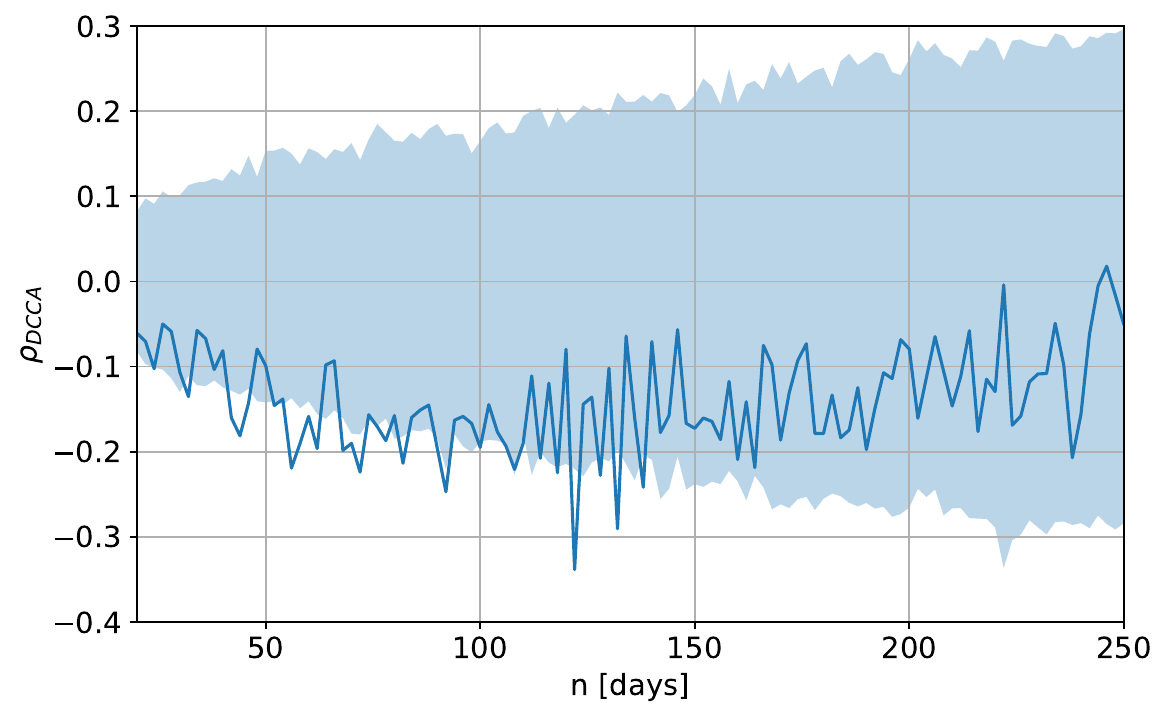}}
\caption{$\rho_\mathrm{DCCA}$ computed between the NAOI and temperature anomalies recorded at the MAGIC telescope location. Temporal scales in the range $n = [20, 250]$ days were used. The shaded area represents the $\pm 3 \sigma$ confidence intervals. A significant correlation is found only at small temporal scales, up to $\sim 130$ days. Regarding the value of correlation obtained using $n\sim 130$, the probability to obtain the same value correlating two random white noises is $8.75\mathrm{E}^{-6}$. It was estimated fitting a binomial distribution on data obtained correlating 1000 different couples of random white noises. \label{fig:DCCA}}
\end{figure}

\subsection{Results of the fractal time series analysis}
\label{results}

Fractal analysis was performed using hourly means computed from the original dataset. Since fractal analysis requires continuous data, due to long periods of missing data, the period considered started from 2007/10/01. First, the Hurst exponent $H$ of each time series was computed. For temperature only, the pronounced yearly oscillation was removed before the analysis was carried out. A linear trend plus the sum of two sinusoidal terms was used to better match the phase of the yearly oscillation.
%\mg{In fact, all parameters have yearly oscillation patterns, and the residuals have been computed. --> for temperature only we removed the strong yearly oscillation which if not removed affects DFA computation.}
Figure~\ref{fig:DFA} shows that all meteorological parameters exhibit long-range correlations, i.e., are persistent, with a Hurst exponent greater than 0.5 (the value corresponding to white noise). This means that on average, if a value in the time series lies above the mean, the following value has also high probability to lie above the mean. Temperature, pressure and relative humidity exhibit higher values of $H$ ($\sim 0.9$), while the three parameters related to the wind show similar Hurst exponents ($H < 0.8$). It is worth mentioning that for the temperature data, the obtained value of $H=0.9$ is consistent with what is reported in Fig.~10 of~\citet{franzke2020structure} and Fig.~2 of~\citet{fraedrich2003scaling} for locations near the ocean, such as one of the MAGIC telescopes (in the two references, anomaly temperature data were analysed with DFA using a second-order polynomial fit to perform detrending). 

Since DFA cannot distinguish between monofractal and multifractal time series, MFDFA can be used. If the time series is multifractal, large and small fluctuations behave differently on different time scales. This behaviour can be highlighted by computing the generalised Hurst exponent $h(q)$ at different orders $q$, using Eq.~\ref{Eq:qFn}.
The values obtained for $h(q)$ are shown in Fig.~\ref{hq_mf}, upper panel, where the generalised Hurst exponent $h(q)$ was calculated for different values of $q$ ranging from -5 to 5. It can be seen that all the parameters analysed exhibit multifractal behaviour, that is, $h(q)$ has a slope different from zero. In case of absence of multifractality (monofractal behaviour), the value of $h(q)$ would have been almost the same for each $q$. Pressure and relative humidity exhibit stronger multifractality, compared to the other parameters, which exhibit a wider range of $h(q)$ values. The bottom panel of Fig.~\ref{hq_mf} shows the multifractal spectrum, which quantifies the degree of multifractality from the width of the spectrum and identifies which scales contribute the most to the multifractal behaviour. The group of spectra on the right refers to the actual time series analysed. All of these spectra are generally wide, which confirms the presence of multifractality. This finding is consistent with \citet{franzke2020structure} and references therein, where multifractal behaviour was found in temperature, wind speed and relative humidity time series. The largest width is obtained for the pressure and relative humidity time series, which means that different structures are present on a wide range of scales. The widths of each spectrum, calculated as $\alpha_\mathrm{max} - \alpha_\mathrm{min}$, are listed in Table~\ref{amax_amin}.

\begin{table}
\centering
\begin{tabular}{l|c}
\toprule
Parameter & $\alpha_\mathrm{max}$ - $\alpha_\mathrm{min}$\\
\midrule
Temperature residuals & 0.34 \\
Pressure & 0.55 \\
Relative humidity & 0.65 \\
Instantaneous wind direction & 0.36 \\
Wind gust & 0.40 \\
Instantaneous wind speed & 0.39 \\
\bottomrule
\end{tabular}
\caption{Width of the multifractal spectrum for each analysed meteorological parameter. \label{amax_amin}}
\end{table}

Pressure, relative humidity, wind gust, and wind direction have longer tails on the right side of the spectrum, which means that multifractality is prominent on small scales. 
%\mg{How did you treat the periodic behaviour of the wind direction parameter? --> We didn't, only for temperature which has a strong and evident seasonality we removed it.}
Notably, the wind direction spectrum has a very short tail on the left side, and this is a sign of a more regular structure of the time series periods with high variations. This can also be seen in the $h(q)$ curve for wind direction, which tends to flatten after $q = 0$, meaning the almost absence of multifractality. This is due to the definition of $f(\alpha)$ (Eq.~\ref{mf_spectrum}), proportional to $-h(q)$. The other parameters' spectra are symmetric, instead. Finally, wind gust exhibits a structure that is very similar to the one of wind speed. Multifractality in a time series can be due to a broad probability density function or long-range correlations for small and large fluctuations \citep{baranowski2015multifractal}. 

To test that these results depend on the temporal evolution of the meteorological parameters, the time series have been shuffled, removing temporal correlations, and the multifractal spectra computed again for each shuffling trial. The group of spectra on the left side of Fig.~\ref{hq_mf} refers to the shuffled data. It can be seen that each of these spectra does not exhibit multifractal properties; instead, they are very narrow and centered around $\alpha = 0.5$. This confirms that the results obtained in Figs.~\ref{fig:DFA} and~\ref{hq_mf} depend on the particular temporal structure of the time series, since multifractality is lost when the  temporal order of the data is changed. 

\subsection{Influence of the NAO}

The North Atlantic Oscillation (NAO) is the dominant mode of atmospheric circulation in the North Atlantic region~\citep{Barnston:1987}. It is defined as the difference in surface sea level pressure between the subtropical (Azores) High and the subpolar (Islandic) Low, with one center located over Greenland and the other center of opposite sign spanning the central latitudes of the North Atlantic between 35\deg and 40\deg.  Strong positive phases of the NAO tend to be associated with above-normal temperatures across northern Europe and below-normal temperatures often across southern Europe and the Middle East. The NAO exhibits considerable interseasonal and interannual variability, and prolonged periods of both
positive and negative phases of the pattern are common.

The US National Oceanic and Atmospheric Administration (NOAA) provides a daily NAO index\footnote{\url{https://www.cpc.ncep.noaa.gov/products/precip/CWlink/pna/nao.shtml}} (NAOI), which has been used to investigate possible correlations of weather-related parameters with the NAO. 

\citet{lombardi2006} had claimed a correlation between the annual averages of the NAO index and the annual mean temperatures of CAMC, with a significance of 86\%. 

\begin{comment}
We performed three tests: Correlate the daily average temperature fit residuals from Eq.~\ref{eq:mui} with the daily NAOI, 

correlate monthly averages of the fit residuals with monthly averages of the NAOI, and finally yearly averages, as done by \citet{lombardi2006}. In any of the three tests, we could find a significant correlation ($r$ = -0.095, 0.014, and -0.009). 

Similarly, we tested 

possible correlations of the DTR with the NAOI and finally the wind speeds. 

and failed again ($r$=0.004, 0.022 and \mg{TBD}).
\end{comment}

According to \citet{Azorin-Molina:2018}, wind speed at Iza\~na, Tenerife, is negatively and significantly correlated with the NAO, mainly in winter and spring. This fact may have explained their findings of a higher frequency of extratropical storms affecting subtropical latitudes, and thus in the increasing wind speed trend observed at Iza\~na, correlating with   the tendency of the NAOI toward a more negative phase (mainly in winter and spring) during 1989-2014.

To investigate whether changes in the North Atlantic Oscillation Index (NAOI) affect the temperature recorded at the MAGIC telescope location, the temperature fit residuals with respect to Eq.~\ref{eq:mui} with the parameters that had maximized the likelihood (see Sect.~\ref{sec:temperature}) were correlated with the NAOI. Detrended Cross Correlation Analysis (DCCA) was used to this end. The stationarity of both temperature anomalies and NAOI time series was evaluated by means of the Augmented Dickey-Fuller test. Both time series are stationary according to the test.
The temperature anomalies were obtained after subtracting the main seasonality (Eq.~\ref{eq:mui}) from the data. 
%\mg{Suggest to move all explanations of NAO to either where they appear first, or the discussion}
 As described in~\citet{tatli2019detrended}, during the positive phase of NAO, in the south of Europe, a cold and dry period prevails. In the negative phase of NAO, warm and rainy conditions prevail over the Mediterranean basin. 
 %\mg{until HERE}
 A negative correlation is expected between the NAOI and the MAGIC temperature residuals. The DCCA method \citep{vassoler2012dcca} is described in more detail in Appendix~\ref{sec:DCCA}.

Figure~\ref{fig:DCCA} shows the correlation coefficient $\rho_\mathrm{DCCA}$ from DCCA between temperature residuals from the fit of Eq.~\ref{eq:mui} and the NAOI. A negative correlation between these two time series can be expected. This is confirmed by the values of $\rho_\mathrm{DCCA}$ found in Fig.~\ref{fig:DCCA}, significant on temporal scales up to 130~days. $\rho_\mathrm{DCCA}$ has also been computed between NAOI and the residuals of the other meteorological parameters DTR, relative humidity, and wind speed, but no significant correlations have been found.

\section{FAIRification of the MAGIC Weather Station Data}
\label{sec:fairification}

%As we will prove in this contribution that the measurements from the weather station of the MAGIC telescope can have impact to many different scientific domains like the understanding of local climates with difficult orography,site characterization for new generation Cherenkov Telescopes,assessment of climate change  etc, 
In this section, we focus on the delivery of the data for further re-usage in different domains. 
The implementation of the \textit{Findable, Accessible, Interoperable, and Reusable} (FAIR) principles~\citep{Wilkinson:2016} as well as a high quality of the data 
 is fundamental for the reuse and long-term preservation of all scientific data.
The publication of the data and their metadata in a FAIR data repository makes 
them findable and accessible. Metadata will contain a description of 
the data in a machine actionable format based on 
linked open data (LOD) recommendations ~\citep{bizer2008linked}
in order to enable interoperability and 
additional information (like a standardized license agreement) to enable controlled reusage of the dataset. 
The FAIR digital object (FDO) approach~\citep{de2020fair}, 
which provides the dataset and its metadata as one object,  
enables the reusage of the data in any scientific domain and new use cases even beyond the silo of the original scientific domain.

We first describe the quality assurance efforts on the dataset and show then how we publish the data in accordance with the FAIR principles.

\subsection{Quality checks of the WS data}
\label{sec:qualitychecks}

Before the publication of the WS data set, detailed studies were carried out
in order to provide high-quality data following the FAIR principles. 
%The MAGIC weather station (WS) data were taken in principle with one
%dedicated weather station system every 2 seconds, resulting in a time series of. 
%different data fields. 
%
%  HK: moved this paragraph as it is about quality only. 
%
Technical problems of the WS and the data acquisition systems have led to missing
individual measurements. 
Additionally, different firmware versions resulted in different measured data 
fields over time. The first date of occurrence of a data field
can be seen in Table~\ref{tab:parameters}. In the published data set, not measured parameters or those delivered with a bogus value (like -9999) have been replaced by a 'NaN' value. 

\begin{table*}
\centering
\begin{tabular}{l|cccccc}
  \toprule
  Parameter 	& unit & data type & measure type & first measurement & \#measurements & \#qualityMeasurements \\
  \midrule
  	TimeStamp (index)			& 				& float 	& M & 2003-01-30 15:01:20 & 4770374 	& \\
 %       mjd 						& MJD-55000		& float 	& D & 2003-01-30 15:01:20 & 4513228 & \\
 % 	sun\_alt					& \textdegree	& float 	& D & 2003-01-30 15:01:20 & 4513228 & \\
 % 	sun\_az						& \textdegree	& float 	& D & 2003-01-30 15:01:20 & 4513228 & \\
  	temperature  				& \degC 	    & float 	& M & 2003-01-30 15:01:20 & 4770374 & 4770015 \\
  	temperature\_reliable 	    &               & boolean 	& D & 2003-01-30 15:01:20 & 4770374 & \\
  	pressure     				& hPA 			& float 	& M & 2003-01-30 15:01:20 & 4770374  & 4684421 \\  
        pressure\_reliable     		&    			& boolean 	& D & 2003-01-30 15:01:20 & 4770374 &  \\  	
        humidity  					& rel.\% 		& float 	& M & 2003-01-30 15:01:20 & 4770374  & 4723988 \\
  	humidity\_reliable    		&               & boolean 	& D & 2003-01-30 15:01:20 & 4770374 & \\
%	dewpoint 					&  \degC 		& float 	& D & 2003-01-30 15:01:20 & 4460598 & \\
 %	dewpoint\_reliable           &               & boolean 	& D & 2003-01-30 15:01:20 & 4513228 & \\
  	windSpeed  					& km/h 		    & float 	& M & 2003-01-30 15:01:20 & 4759617 & 4657666\\
  	windDirection				& $^\circ$	 & float 	& M & 2003-01-30 15:01:20 & 4759617 & 4657666\\
  	windGust			 	    & km/h 		    & float 	& M & 2003-01-30 15:01:20 &  4759617  & 4657666\\
  	windSpeedAverage			& km/h	        & float 	& M/D & 2003-01-30 15:01:20 & 4759617 & 4657666\\
  	windDirectionAverage		& km/h 		    & float 	& M/D & 2010-12-19 01:21:42 & 4759617 & 4657666\\
        wind\_reliable              &               & boolean 	& D & 2003-01-30 15:01:20 & 4759617 & \\
	rain 						&   			& int   	& M & 2018-06-25 16:12:01 &  1025244 & 1025244 \\
 \bottomrule 
\end{tabular}
\caption{The columns of the WS data and their properties including the units of the measurements, 
the data type, the first valid measurements, the number of all measured values and the number of 
quality measurements after applying the corresponding quality flag. The column \textit{measure type} denotes 
whether a direct measurement has been made ('M'), or a derive quantity calculated ('D'). Entries labeled
'M/D' contain both direct measurements and derived quantities: WindSpeedAverage was calculated from 2007-03-28 16:27:16 through 2010-12-19 00:49:33 and measured before and after; WindDirectionAverage was calculated before 2010-12-19 01:21:42 and measured afterwards. \label{tab:parameters} } 
\end{table*}

Apart from software issues detected and documented, which led to removal of the corresponding data from the set, the following abnormalities have been observed and marked as unreliable: 

\begin{enumerate}
\item period (2003-01-30 15:01:20 2004-03-01 00:00:00): \\
    During the first year of operation, the weather station was located on the 
    lightning rod, at a different altitude than its final location. Atmospheric pressure 
    measurements during this period have been corrected for the different height, 
    but marked as unreliable for pressure.\\ 
\item period (2009-03-25 00:00:00 2009-03-25 05:20:00): \\
    During a rainy night with relatively high wind speeds, two sudden increases in temperature are detected, followed by an exponential decline. One possible explanation could be a lightning strike. The period has been marked as unreliable for temperature. \\
\item period (2010-10-11 09:30:00 2010-10-11 13:00:00): \\
    A sudden rise in temperature is detected, and further studies showed that during the installation of MAGIC-I telescope mirrors, sunlight was reflected on the WS, leading to a sudden increase in temperature. 
    The temperature values are not correct, and a general malfunction of the WS is probable. \\
 \item period (2010-12-22 03:50:00 2010-12-22 04:30:00): \\
    During a rainy night with high wind speeds, a sudden increase in temperature is detected, followed by an exponential decline. A possible explanation could be a lightning strike. The period has been marked as unreliable for temperature. \\
% \item period (2014-02-04 14:00:00 2014-02-04 15:40:00): \\
%    A sudden rise in temperature is detected from the direction of the MAGIC-II telescope. 
%    It is probable that sunlight was focused on the weather station, forcing a sudden increase in temperature. 
%    The temperature values are not correct, and a general malfunction of the WS is probable. 
\item period (2014-11-20 12:25:00 2014-11-20 13:10:00): \\
    A sudden rise in temperature is detected during 100\% relative humidity and relatively strong winds. A possible explanation could be lightning strike. 
    The short period has been marked as unreliable for temperature. \\
\item period (2014-11-24 00:00:00 2015-01-27 15:29:00): \\
    The humidity sensor started to show random jumps in the measurements. Shifters noted frequent disagreements with humidity measured by other weather stations at the ORM. The station was then replaced by a spare. 
    The short period has been marked as unreliable for humidity. \\
%\item period (2020-04-17 10:40:00 2020-04-17 11:40:10:00): \\
%    A sudden rise in temperature is detected from the direction of the MAGIC-I %telescope. 
%    It is probable that sunlight was focused on the weather station, forcing a %sudden increase in temperature. 
%   The temperature values are not correct, and a general malfunction of the WS is probable. 
%    \item period (2020-03-24 08:30:00 2020-03-24 13:00:00): \\
%    A sudden rise in temperature is detected from the direction of the MAGIC-I telescope. 
%    It is probable that sunlight was focused on the weather station, forcing a sudden increase in temperature. 
%    The temperature values are not correct and a general malfunction of the WS is probable. 
%  \end{enumerate} 
%  \item the range of the measured humidity values should be similar over all measurements
%  \begin{enumerate}
%\item period (2020-01-01 00:00:00 2023-01-16 14:00:00): \\
%    With the beginning of year 2020, an increasing cutoff of low humidity values is observed.  This small effect was discovered at the beginning of 2023. The effect is a hint of a drift of the humidity measurement in this period.  The period has been marked as unreliable for humidity.
\end{enumerate}

In order to provide as many data as possible and only data with approved quality, 
we introduced two new boolean columns in the data set: 
\begin{description}
  \item[temperature\_reliable:\xspace] 
  	This value is set to \textit{False} for all measurements of the periods mentioned above with anomalies in temperature and pressure % (see enumeration \ref{enum_temp})
   and \textit{True} for all other measurements. 
  \item[humidity\_reliable:\xspace] 
  	this value is set to \textit{False} for all measurements of the above mentioned periods 
  	with anomalies in humidity %(see enumeration \ref{enum_temp}) 
  	and \textit{True} for all other measurements. 
   \item[wind\_reliable:\xspace] 
   If one wind parameter (e.g., windGust) shows a bogus value, but the rest of wind-related parameters lie in the expected range of values, then this flag was set to \textit{False}. These situations happened particularly with the first WS model replaced in 2007.
\end{description} 

The WS data are a time series of the measurements of different parameters, aka data fields.
All data are contained in one single file stored in the Hierarchical File System. 
Table~\ref{tab:parameters} lists the columns of the WS data set and their properties.

% ----------------------------------------------------
% ----------------------------------------------------
\subsection{Publication of FAIR and machine actionable data}

\begin{table*}
\centering
\begin{tabular}{|l||lcll|}
  \toprule
  parameter & Class (type of)  &  hasDataType & [OM2]:hasUnit & [OM2]:hAF   \\
  \midrule
  TimeStamp (index)
    & [CF]:time
    &  				
    & 							 	
    &	\\
% --------------------------------------------
  temperature  				
    & [CF]:air\_temperature 
    & Float 	
    & [OM2]:CelsiusTemperatureUnit	
    &  	\\
% --------------------------------------------
  pressure
    & [CF]:air\_pressure 
    & Float
    & [OM2]:millibar
    &  	\\
% --------------------------------------------  	
  humitidy  					
    & [CF]:relative\_humidity
    & Float
    & [OM2]:PercentageUnit			
    &  	\\
% --------------------------------------------
  windSpeed  					
    & [CF]:wind\_speed 
    &  	Float
    & [OM2]:kilometrePerHour 		
    &  	\\
% --------------------------------------------
  windDirection
    & [CF]:wind\_from\_direction     
    & Float
    & [OM2]:degree
    &   \\
% --------------------------------------------
  windSpeedGust
    & [CF]:wind\_speed\_of\_gust
    & Float
    & [OM2]:kilometrePerHour       
    &  	\\
% --------------------------------------------
  windSpeedAverage
  & [CF]:wind\_speed 	
  &  Float
  & [OM2]:kilometrePerHour  		
  & [OM2]:average \\
% --------------------------------------------
  windDirectionAverage		
  & [CF]:wind\_from\_direction 
  & Float
  & [OM2]:degree 					
  & [OM2]:average  	\\
% --------------------------------------------
  rain
  &  [CF]:rainfall\_amount
  &  Integer
  & 
  & 	\\ 
  \bottomrule 
\end{tabular}
\caption{\label{tab:ontology_mapping} The semantics description for the different columns of the MWS data containing physical measured values. [OM2] is used for the namespace of the "Ontology for units of Measurements", [CF] is used for the namespace of the CF Convention vocabulary. The header of the most right column ([OM2]:hAF) is a shortcut for "[OM2]:hasAggregateFunction". } 
\end{table*}

The following general steps were performed in order to reach the goal of FAIR and machine actionable
data by using semantic technologies: 
\begin{enumerate}
    \item Select an implementation for FAIR digital objects (FDO) using a general metadata schema.\\
    \item Describe general metadata like creator, contributors, license, etc. using the general metadata scheme. \\
    \item Select established schemes/vocabularies/ontologies to describe the domain-specific content of the dataset. \\ 
    \item Use the selected vocabularies to add the domain specific description to the metadata. \\
    \item Package the data and its metadata as a FDO. \\
    \item Publish the FDO in a common and sustainable FAIR repository.   
\end{enumerate}
All these steps are now described in more detail.

\subsubsection{FDO implementation}

RO-Crate~\citep{soiland2022creating} is an implementation of FAIR digital objects based on the \textit{schema.org}\footnote{see \url{https://schema.org}} vocabulary~\citep{mika2015schema}.
This vocabulary is a collaborative community-based activity with the goal to provide schemes for structured 
data. RO-Crate can be seen as a container for reusable datasets, files, etc. and for their metadata file
called \textit{ro-crate-metadata.json}. 
This metadata file contains general and domain-specific terms to
describe the dataset. 

\subsubsection{General metadata}

In the next step, we describe the general metadata of the data set using  \textit{schema.org} terms. As license\footnote{see \url{https://schema.org/license}} we choose the creative common
license BY-SA\footnote{see \url{https://creativecommons.org/licenses/by-sa/4.0/}} as it allows the 
usage and distribution of the dataset, but forces the consumer to give appropriate credit to the data providers, to indicate changes and to use the same license for redistribution. Furthermore, the creators\footnote{see \url{https://schema.org/creator}} and the contributors to the data set were identified using their ORCID identities~\citep{haak2012orcid}. 

\subsubsection{Domain specific vocabularies}

Due to the large number of existing vocabularies in semantic research, selecting domain-specific vocabularies is not easy.
The content of the WS data set can be assigned to earth sciences, and therefore the measured quantities should be described using a vocabulary from this domain. However, even within that domain many schemes, vocabularies, and ontologies exist. 
 Following a best-practice approach discussed within the
earth science research data project NFDI4EARTH\footnote{see \url{https://www.nfdi4earth.de/}}, 
we studied two candidates, the sweet ontology~\citep{buttigieg2018sweet} and the CF-Convention vocabulary~\citep{rew2007cf}. The CF-convention was finally chosen as it contains term descriptions for all weather parameters in our dataset (see Table~\ref{tab:parameters}) and is used in many data repositories of the earth science community. Although the  original CF-Convention does not fully support the Linked Open Data approach using URL endpoints for terms, we can use the service of the \textit{MMI Ontology Registry and Repository}\footnote{\url{https://mmisw.org/ont/cf/parameter}} based on 
a transformation of the original scheme.  
The established \textit{OM2: Ontology of units of Measure} (OM)~\citep{rijgersberg2013ontology} \footnote{\url{http://www.ontology-of-units-of-measure.org/page/om-2}} provides the terms for the semantic descriptions of our parameters' units (see Table~\ref{tab:parameters}).

As shown in Table~\ref{tab:ontology_mapping}, we combine both ontologies/vocabularies 
to describe the domain-specific content of our measurements in our mapping in a
machine-actionable approach. 
The description of the windSpeedAverage parameter is shown in the following code segment as an example: 

%\begin{tiny}
%\begin{verbatim}
%... 
%{
%  "@type": ":wind_from_direction [CF-Standard Name]",
%  "OM2:hasAggregateFunction": "[OM2]:average",
%  "OM2:hasUnit": "OM2:degree",
%  "dataType": "float64",
%   "name": "windDirectionAverage"  
%} ... 
%\end{verbatim}
%\end{tiny}

\begin{tiny}
\begin{verbatim}
{"@type": "PropertyValue",
 "name":"windDirectionAverage",
 "@id" : 
    "https://mmisw.org/ont/cf/parameter/wind_from_direction",
 "http://www.ontology-of-units-of-measure.org/\_
   resource/om-2/hasUnit":
    "http://www.ontology-of-units-of-measure.org/\_
      resource/om-2/degree", 
 "hasDataType" : "Float" , 
 "http://www.ontology-of-units-of-measure.org/\_
   resource/om-2/hasAggregateFunction" : 
     "http://www.ontology-of-units-of-measure.org/\_
       resource/om-2/average"  }
\end{verbatim}
\end{tiny}
In this case, the entity for the column with name \textit{windDirectionAverage} is connected with a vocabulary term from the CF-convention standard. 
The measurement properties values are described with the OM2 ontology with units of degrees, and an average measurement as aggregate function. The \textit{DataType} \textit{Float} is part of the \textit{schema.org} vocabulary.  For each measurement column of the dataset such a definition of metadata is provided. 

\subsubsection{Packaging of the FAIR Digital Object}

The RO-Crate tool for \textit{python} offers functionality to package the FAIR digital object as a zip file. 
The RO-Crate for the WS data contains only two artefacts. The data set itself and
its metadata file \textit{ro-crate-metadata.json}. As it is good practise to connect data, analysis software and scientific results, we add an additional entity in the metadata file containing the 
information of the software repository
% can be added if the final software repo is agreed on!! hk
%\footnote{see \url{https://gitlab.pic.es/magic/wsd}} 
used for this publication as an external link. 

\subsubsection{Publishing the FAIR data}

Findability and accessibility of scientific datasets are crucial for their reusage 
beyond the provision of metadata. The Zenodo repository\footnote{see \url{https://zenodo.org}} 
offers both plus the possibility to collect the datasets of a community like the MAGIC collaboration\footnote{see \url{https://zenodo.org/communities/magictelescopes}}. The authors publish the WS data within this community under the URL \url{https://dx.doi.org/10.5281/zenodo.11279074}. The analysis code is published under the URL \url{https://github.com/mgaug/WS-Analysis}.

\section{Conclusions}
\label{sec:conclusions}

We have analysed more than 20~years of weather data taken with the MAGIC weather station, acquired at a rate of once every 2~s, but stored and used for this analysis at a rate of once every 2~min. 

\begin{table*}
\centering
\begin{tabular}{cccccc}
\toprule
Parameter & MAGIC site & \multicolumn{4}{c}{Mountain rim}  \\
& this work & \multicolumn{3}{c}{\protect\citet{lombardi2006} and \protect\citet{lombardi2007}} & \protect\citet{Varela:2014} \\ %\protect\citet{Hidalgo:2021}  \\
& 2188~m & (CMT site, 2336~m) & (TNG site, 2397~m) & (NOT site, 2385~m) & (ELT test site, 2356~m) \\ % & (UERRA 800 and 750 hPa), \\
& 2003-2023 & 1985-2004 & 1998-2005 & 1998-2005 & 2008-2009  \\
\midrule
\multicolumn{6}{l}{Temperature} \\
%\midrule
Annual mean & 11.2\degC  &  8.8\degC  &  9.8\degC  &  & 9.5\degC \\
\midrule
\multicolumn{6}{l}{Relative humidity} \\
Annual mean   & 38\% & 36\%  &  41\%  & 43\% & 35\% \\ 
Annual median & 30\% &    &   &  & 22\% \\
\midrule
\multicolumn{6}{l}{Atmospheric pressure} \\
Annual mean & 788 hPa & 775 hPa & 772 hPa & 772 hPa & 772 hPa \\
\midrule
\multicolumn{6}{l}{Wind} \\
Prevailing dir. (night-time) &  E  & N & NE & E & NE \\
Next-to-prevail. dir. (night-time) & ESE & SW & S & W & NNE \\  
Mean wind speed &  12.8 km/h  & 7.9 km/h  &  16.6 km/h  & 25.9 km/h &  29.5 km/h \\
Median wind speed & 11.5 km/h &   &    &  &  25.6 km/h \\
Maximum wind speed average &   102 km/h  &  65.9 km/h &  96.8 km/h  & 107 km/h &    \\ 
Maximum wind gust  & 159 km/h  &  & & & 159 km/h \\
\toprule
\multicolumn{1}{l}{Weather downtime} & this work & \multicolumn{4}{c}{\protect\citet{garciagil2010}} \\
& (MAGIC, 2200~m) & (CMT, 2326~m) & (TNG, 2387~m) & (NOT, 2382~m) & (WHT, 2333~m) \\
& 2003-2023 & 1999-2003 & 2000-2005 & 2006-2008 & 1990-2007 \\
%& 189 months & 56 months & 72 months & 26 months & 216 months \\
\midrule
Average annual weather downtime & 19.5\%  & 20.7\% & 30.2\%  & 26.1\% & 26.3\% \\
\bottomrule
\end{tabular}
\caption{\label{tab:comparisonothersites}\ccol{Comparison of weather conditions at the MAGIC site with those published from telescope sites at the mountain rim. The fields left blank mean that the corresponding parameter is not available in the cited study. The abbreviations refer to the Carlsberg Meridian Telescope (CMT), Telescopio Nazionale Galileo (TNG), Nordic Optical Telescope (NOT), Extremely Large Telescope (ELT) and the William Herschel Telescope (WHT). Altitudes refer to meters a.s.l., in the upper part those of the automated weather stations, below the ones of the telescopes. Note the different time coverage of the cited studies. }}
\end{table*}

The MAGIC Telescopes site is characterized by median temperatures of $\sim$18\degC in summer and 6--7\degC in winter, about 2\degC higher than on the mountain rim, with absolute extremes found at $-5.0$\degC and $+30.4$\degC. Nighttime temperatures are about 2\degC lower on average. We find a significant temperature increase over time of 
$0.55\pm0.07\mathrm{(stat.)}\pm 0.07\mathrm{(syst.)}$\degC/decade,
statistically compatible, but at the upper end of the
predictions for the period 2015-2050 obtained from simulations by~\citet{Haslebacher:2022}. 
Furthermore, we find a significant increase in the diurnal temperature range (DTR) of
$0.13\pm0.04\mathrm{(stat.)}\pm0.02\mathrm{(syst.)}$\degC/decade, 
accompanied by an increase in its seasonal oscillation amplitude of $0.29\pm0.10\mathrm{(stat.)}\pm0.04\mathrm{(syst.)}$\degC/decade, indicating that the increase in DTR is stronger for summers than for winters.
This finding may be explained by a slight reduction in cloud cover over time and a decrease of soil moisture, as predicted by~\citet{Exposito:2015} for the Canary Islands. 

Temperature change rates over time are always found above $-0.4$\degC/min and below $+0.4$\degC/min. The site therefore satisfies the CTAO requirement of air temperature gradients of $<$0.5\degC/min 
for 20 minutes. We also find a significant correlation of the temperature change rate with the relative humidity (\textit{RH}) change rate, suggesting that the main cause of fast temperature changes are local clouds moving in and out of the site.

The \textit{RH} median is found to be $\sim$20\% during the two summer months July and August and $\sim$40\% during the winter months, with a general median of 29\% (mean: 38\%). Contrary to previous claims from data taken at the caldera rim~\citep{lombardi2007}, we cannot reproduce the claimed higher \textit{RH} during night, but instead find even a slight preference for higher day-time \textit{RH}, correlating with solar altitude. 
More surprisingly, we find that the daily median \textit{RH} increases by
$4.0\pm0.4\mathrm{(stat.)}\pm1.1\mathrm{(syst.)}$\%/decade
 when periods of precipitation are excluded. This finding is consistent with \citet{Haslebacher:2022}'s observation that humidity decreases over time for all astronomical sites, except for the two island observatories on Mauna Kea and La~Palma, where specific humidity increases instead. A possible explanation for this effect is the enhanced evaporation of sea water as a result of global warming.

Periods of precipitation were always shorter than 6.5~days, except on one occasion in Jan.~2021, when an unusually long period of 9~days of precipitation including snowfall was recorded. Long periods of precipitation lasting longer than 3~hours and occurring during the winter months are responsible for the largest fraction of telescope downtime. However, the median duration of a precipitation period is only about half an hour; snowfall lasts even shorter on average. We find a mean annual occurrence of 230 precipitations that last less than 3~hours, distributed almost uniformly throughout the day with a slight preference around noon for winter and fall, while the less frequent spring and summer short precipitations present a slight preference for night time. In the past 20 years, no significant increase or decrease in the probability of occurrence of short precipitations could be found, with limits of  $>$\,--1.9/year and $<$\,0.6/year (95\%~CL). For precipitations lasting longer than 3~hours, we studied mean yearly occurrence and occurrence probability increases or decreases over time, but found no significant trends. So far, the amount and impact of precipitation at the site have not shown significant variation as a consequence of climate change.

We observe that the MAGIC Telescopes site shows a predominant trade wind direction from east, followed by ENE during the day and ESE during astronomical night. Strong storms seem to blow from arbitrary directions through, with the exception of the fourth quadrant comprised within North and West. Median wind speeds at the MAGIC site are only about 30\%-50\% of those observed at the mountain rim~\citep{lombardi2007} suggesting that the lower altitude protects the site from breeze, but not from the less frequent, but potentially harmful, strong storms. The WS detected wind speeds up to its detection limit of 157~km/h, however, from the distribution of wind gust speeds during strong storms, we could estimate a maximum wind gust speed of 201$^{+4\mathrm{\,(stat)}+2\mathrm{\,(syst)}}_{-7\mathrm{\,(stat.)}-7\mathrm{\,(syst.)}}$\,km/h for a return period of 475~years. Contrary to other findings on the neighbouring island of Tenerife, we find a significant \textit{decrease} of average trade wind speeds over time, of
0.85$\pm$0.12(stat.)$\pm$0.07(syst.) (km/h)/decade.
In turn, no changes in the occurrence of strong storms have been detected over time. For example, an increase or decrease in storm occurrence with wind gusts stronger than 100~km/h (with a current expectation value of approximately once per year) can be excluded at 95\%~CL for more than $\pm$1 additional event per decade. 

\ccol{Due to the relatively frequent replacement of broken weather stations by new or repaired (and hence re-calibrated) stations, and a series of robustness tests, we have been able to estimate the effect of possible sensor drifts and incorporate them into the systematic uncertainties. Given the outcome of these tests, we are unable to explain the observed long-term temperature increase, the increase in DTR, \textit{RH} and the decrease in trade wind speeds other than due to the evolution of the local climate itself.  }

\ccol{The observed absence of long-term evolution of $>$3-hours-long precipitation periods and storms is less prone to systematic effects, as long as the counting algorithm is correct and the data gaps are taken into account correctly. For this reason, we have developed the new statistical algorithm described in Appendix~\ref{sec:likelihood}.}

We find weather-related downtime ranging from 18\%-20\%, depending on the method applied to count downtime, compatible with the one obtained from the automatic CMT telescope, but considerably lower than the weather-induced downtime of the optical telescopes on the mountain rim~\citep{garciagil2010}. The differences are due to less strict observation criteria for Imaging Atmospheric Cherenkov Telescopes, but also to the calmer winds at the MAGIC site. Relaxed waiting times after each wind gust can even lead to an additional gain of 3\% in weather downtime for the CTAO. 

\ccol{%
Finally, Table~\ref{tab:comparisonothersites} provides a summary of the main weather statistics found in this study with those available from telescopes sites at the mountain rim of the observatory. }

\ccol{
%In this sense and beyond this finding, 
Characterizing the scaling properties of meteorological data by means of fractal and multifractal analysis allows us to better understand how locally measured fluctuations can be linked to variations at larger scales, and vice versa \citep{baranowski2015multifractal}. }

%In this article, the time series of meteorological parameters acquired from the weather station at the MAGIC telescope facility was analyzed.
%Meteorological data are acquired in real time both for MAGIC data reconstruction and for safe operation of the detector.
For the fractal analysis, the data set was downsampled to a one hour sampling rate. The DFA algorithm used revealed the presence of long-range correlations on a wide range of scales. It allowed us to evaluate data persistence in terms of the Hurst exponent $H$. The data analysed show values of $H>0.5$, and are therefore persistent. 
\ccol{The relatively large Hurst exponent, obtained by means of DFA, of the temperature anomalies is consistent with the near-ocean location of the MAGIC telescopes. }
To evaluate whether the meteorological time series data are multifractal, the MFDFA algorithm was used. First, a generalised Hurst exponent $h(q)$ was computed for different values of the scaling order $q$. A slope different from zero was found for all parameters, indicative of multifractality. To further verify this finding, the multifractal spectra were evaluated. 
All meteorological parameters acquired by the WS have revealed the presence of multifractality in the data, as a consequence of the width of the multifractal spectrum. 
The strongest multifractality is exhibited by atmospheric pressure and \textit{RH}. 

The asymmetry of the spectrum can be linked to changes from high positive to low negative values. A more pronounced left tail is indicative of the fact that extreme events dominate in the data. Instead, a more pronounced right tail is indicative of a slower variation of the data, which was found in the atmospheric pressure, wind direction and \textit{RH} time series.
%The adopted methodology proved to be useful for the characterization of weather data monitoring the MAGIC telescope, which are relevant both for data reconstruction and for safe operation of the detector. 

Characterisation of the temporal scaling properties of meteorological time series by fractal analysis is of relevance for linking changes in local fluctuations to fluctuations on larger scales, and vice versa. \citet{baranowski2015multifractal} have shown that MFDFA allows us to relate differences in the dynamics of meteorological processes to areas within different climatic zones. This suggests that the analysis presented here could be extended to data sampled at other weather stations of the ORM.  
MFDFA can then be used to characterise the time and space dynamics of the analysed data. Such dynamics, in turn, can be attributed to climatic conditions.

The characterisation of the long-term stability of weather conditions and climate at the location of the northern site of the Cherenkov Telescope Array Observatory (CTAO) has a direct impact on its future duty cycle and data quality. We find a minimum weather downtime of 16.5\%, achievable if the CTAO is allowed to operate up to wind gust speeds of 60~km/h, almost independently of the waiting time required after each gust. In contrast, if a wind gust limit $<$40~km/h is established, only 18.5\%-20.5\% weather downtime is possible, depending on the waiting time required. 

For the moment, and after the detailed analysis of long-term evolution of our weather data carried out in this article, we cannot find any hints that the weather downtime may considerably increase in the near future, unless strong non-linear changes of weather behaviour accrue from climate change. 

%\begin{comment}
\section*{Acknowledgements}

This work would have been impossible without the support of our colleagues from the MAGIC collaboration. We are especially grateful to the many shifters and local technicians who helped maintain, debug, and improve the MWS. We also thank the Instituto de Astrof\'{\i}sica de Canarias for the excellent working conditions at the Observatorio del Roque de los Muchachos on La Palma. 
The financial support of the Spanish grants PID2022-139117NB-C41 and PID2022-139117NB-C43, funded by MCIN/AEI/10.13039/501100011033/FEDER, UE, the German BMBF and MPG, the Departament de Recerca i Universitats de la Generalitat de Catalunya (grant SGR2021 00607) are gratefully acknowledged.

\section*{Data availability}

The data underlying this article are available on Zenodo, at \url{https://dx.doi.org/10.5281/zenodo.11279074}
The analysis code is published under the URL \url{https://github.com/mgaug/WS-Analysis}.  
%\end{comment}

\bibliographystyle{mnras}
\interlinepenalty=10000
\bibliography{MagicWeather}

\appendix

\section{Likelihood model for occurrence of extreme events}
\label{sec:likelihood}

We numerate the $N=20$ years with index $i=0$ for the year 2004, ending with $i=N-1$
for the year 2023. We will test a linear model, with a mean occurrence  $p_i$ of storms or rains being:  
\begin{align}
   p_i &= p_A + \alpha \cdot \left(i - \frac{N+1}{2} \right) \quad,
\end{align}
\noindent
where $p_A$ denotes the average storm or rain occurrence for the full sample and $\alpha$ the yearly increase or decrease of occurrence. 

In the following, we assume a Poissonian probability mass function for each year, compared with the actually counted number of events, $k_i$, every year. Large data gaps are taken into account through a weight $w_i$ describing the probability not to miss a storm or rain due to missing data. The likelihood $\mathcal{L}$ for such a process can then be written as: 

\begin{align}
  \ln(\mathcal{L})  &= \sum_{i=0}^{N-1} \left(-p_A -\alpha\cdot i + \alpha \cdot  \frac{N+1}{2}\right)\cdot w_i + {} \nonumber\\
  & {}  + k_i \cdot \ln\left[ \left(p_A + \alpha \cdot i - \alpha\cdot \frac{N+1}{2} \right) \cdot w_i  \right] \quad.  \label{eq:l}
\end{align}
%where $k_i$ denotes the actually counted number of storms in year $i$ since 2003. 

\begin{figure}
\centering
\includegraphics[width=0.99\linewidth, clip, trim={0cm 0 0cm 0}]{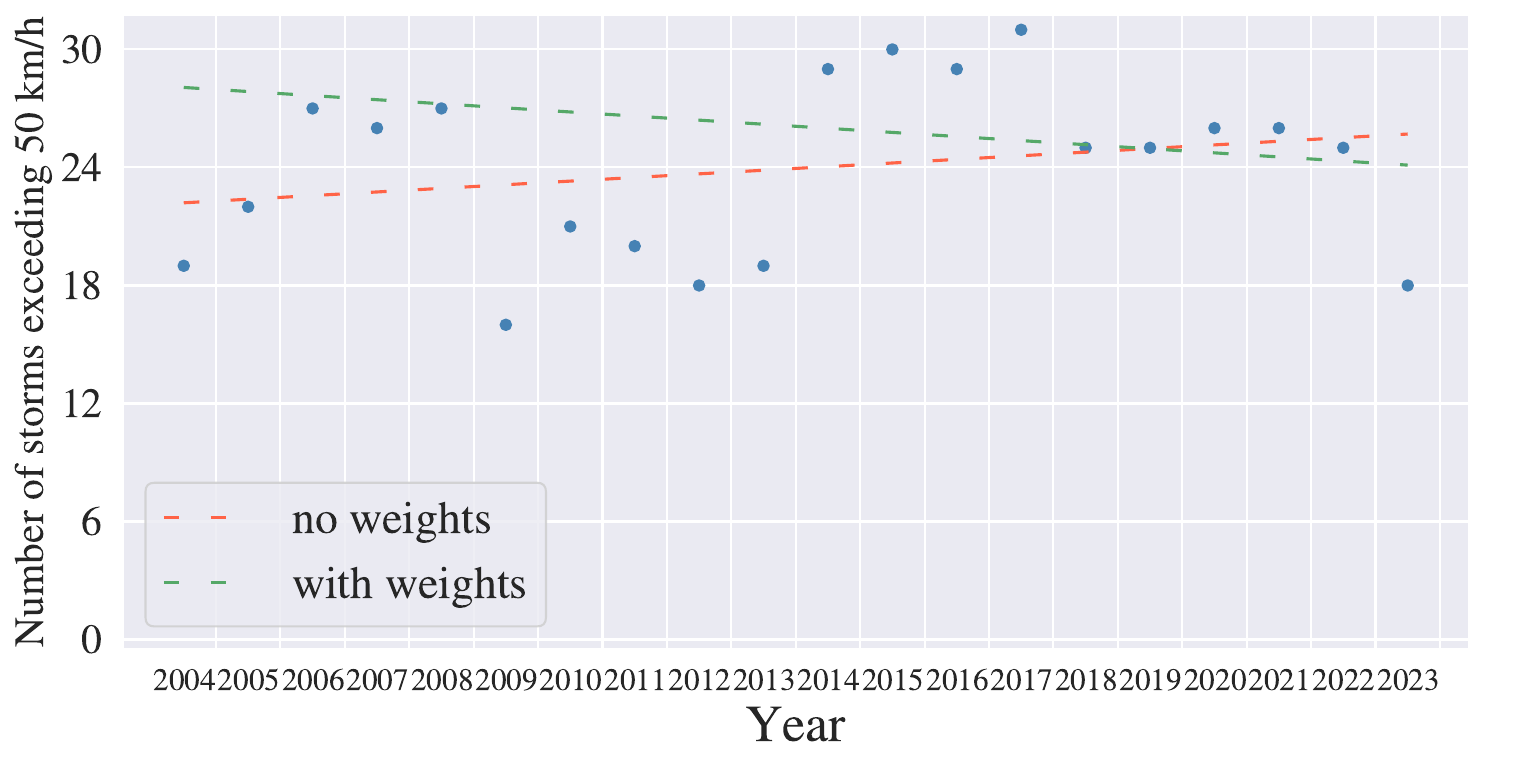}
\includegraphics[width=0.99\linewidth, clip, trim={0cm 0 0cm 0}]{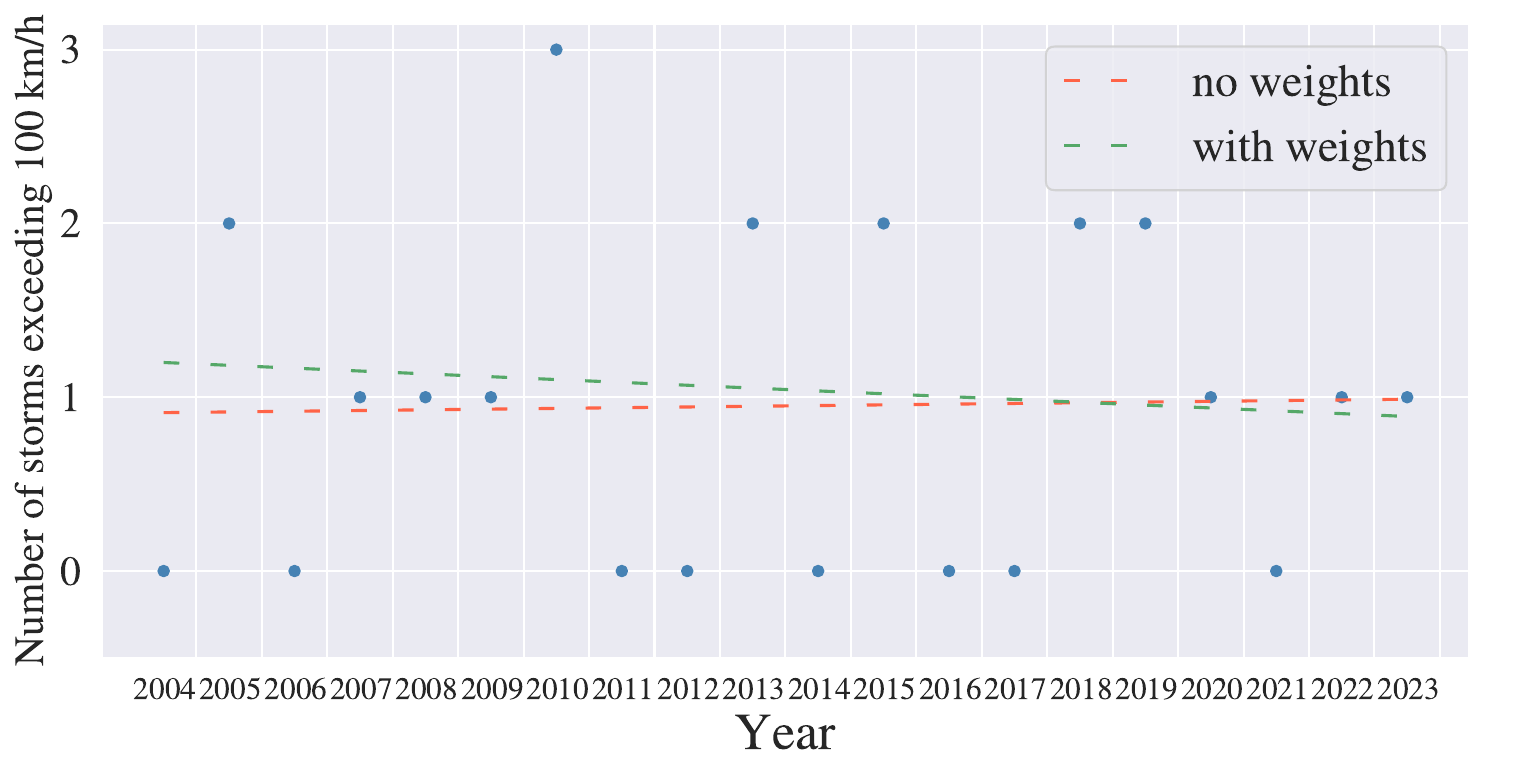}
\caption{
  \label{fig:huracans}
  Two examples of the yearly occurrence of storms: On top, for wind speeds larger than 50\,km/h, below for very strong storms with wind speeds larger than 100\,km/h. The red dashed lines show the results of the most probable expectation values following a linear model without weights, the green line is obtained after taking into account the weights due to data gaps. }
\end{figure}

Minimization of the log-likelihood with respect to the parameters of interest, $p_A$ and $\alpha$ leads to a set of two equations, which can be easily solved numerically yielding expectation values $\hat{p}_A$ and $\hat{\alpha}$:
\begin{align}
  0 &= \sum_{i=0}^{N-1} -w_i + \frac{k_i}{\hat{p}_A + \hat{\alpha}\cdot (i - \frac{N+1}{2})}   \\
  0 &= \sum_{i=0}^{N-1} -(i-\frac{N+1}{2})\cdot w_i + \frac{k_i \cdot (i-\frac{N+1}{2})}{\hat{p}_A + \hat{\alpha}\cdot (i - \frac{N+1}{2})} \quad.  
\end{align}
Two examples of such solutions (with and without application of weights) are shown in Fig.~\ref{fig:huracans}. 

With the resulting solutions for $\hat{p}_A$ and $\hat{\alpha}$, we can calculate a test statistic for the hypothesis $\mathcal{L}(\hat{p}_A,\hat{\alpha})$ with respect to the null hypothesis $\mathcal{L}_0(\hat{p}_0^{\alpha=0},\alpha=0)$: 

\begin{align}
  \textit{TS} &=  2 \cdot \ln\left(\frac{\mathcal{L}}{\mathcal{L}_0}\right) \\
 &=  2 \cdot \Bigg\{  \sum_{i=0}^{N-1} \Big( -\hat{p}_A -\hat{\alpha}\cdot (i - \frac{N+1}{2}) \Big)\cdot w_i + {} \nonumber\\
   & {} + \sum_{i=0}^{N-1} k_i \cdot \ln\left[\left(\hat{p}_A +\hat{\alpha}\cdot (i - \frac{N+1}{2})\right)\cdot w_i \right]   + {}\nonumber\\  
   & {} +  \sum_{i=0}^{N-1}k_i  - \sum_{i=0}^{N-1} k_i \cdot \ln \left(\frac{\sum_{j=0}^{N-1} k_j}{\sum_{j=0}^{N-1} w_j} \cdot w_i\right)  \Bigg\} \quad  \nonumber\\
  &= -2 \sum_{i=0}^{N-1} \Big( \hat{p}_A +\hat{\alpha}\cdot (i - \frac{N+1}{2}) \Big)\cdot w_i  + {}\nonumber\\
  & {} + 2\cdot \sum_{i=0}^{N-1} k_i \cdot \left\{1 +\ln\left(\sum_{j=0}^{N-1} w_j \cdot \frac{\hat{p}_A +\hat{\alpha}\cdot (i - \frac{N+1}{2})}{\sum_{j=0}^{N-1} k_j}    \right) \right\}  \quad . \\
   &= 2\cdot \Bigg( \sum_{i=0}^{N-1}  k_i \cdot \left\{1 +\ln\left(\frac{\sum_{j=0}^{N-1} w_j \cdot \left(\hat{p}_A +\hat{\alpha}\cdot (i - \frac{N+1}{2})\right)}{\sum_{j=0}^{N-1} k_j}    \right) \right\} -  {}\nonumber\\
    & {}\qquad\qquad  - \left(\hat{p}_A +\hat{\alpha}\cdot (i - \frac{N+1}{2}) \right)\cdot w_i \Bigg) \quad . 
\end{align}
\noindent
In order to calculate confidence intervals for the estimated slope parameter $\hat{\alpha}$, we re-formulate the likelihood Eq.~\ref{eq:l}  in the form of a profile likelihood~\citep{MurphyVanderWaart:2000} ratio test, with $p_A$ treated as nuisance parameter: 

\begin{align}
D(\alpha) = -2 \ln \left(\frac{\;\mathcal{L}(\alpha;{\widehat{\widehat{p}}_A}(\alpha)}
                                                  {  \mathcal{L}(\widehat{\alpha}; \widehat{p}_A)} \right) \quad, \label{eq.lr1}
\end{align}
\noindent
where $\mathcal{L}(\widehat{\alpha}; \widehat{p}_A)$ denotes the likelihood Eq.~\ref{eq:l} evaluated its the global maximum (located at $\widehat{\alpha}; \widehat{p}_A$) and $\widehat{\widehat{p}}_A(\alpha)$ denote those values of $p_A$, which maximize the likelihood  at a given test value of $\alpha$. It can be shown~\citep{MurphyVanderWaart:2000} that in the large sample limit, the new test statistic $D(\alpha)$ can be used as a likelihood ratio test statistic for $\alpha$, behaved asymptotically normal and efficient, if $\ln(\mathcal{L}(\alpha;{\widehat{\widehat{p}}_A}))$ is continuously differentiable twice for all $\alpha$.

\begin{figure}
    \centering
    \includegraphics[width=0.88\linewidth]{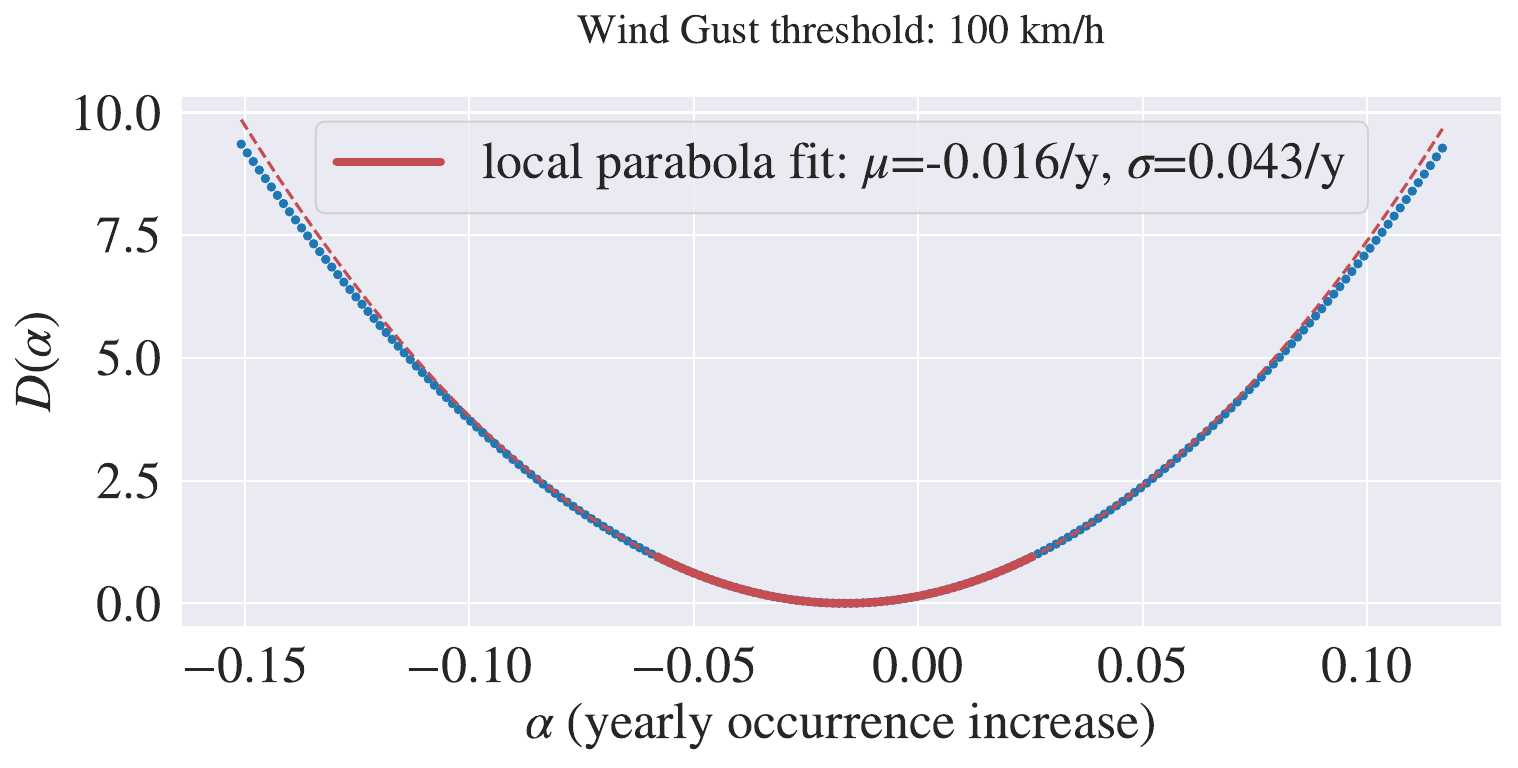}
      
      \vspace{0.3cm}
      
 \includegraphics[width=0.88\linewidth]{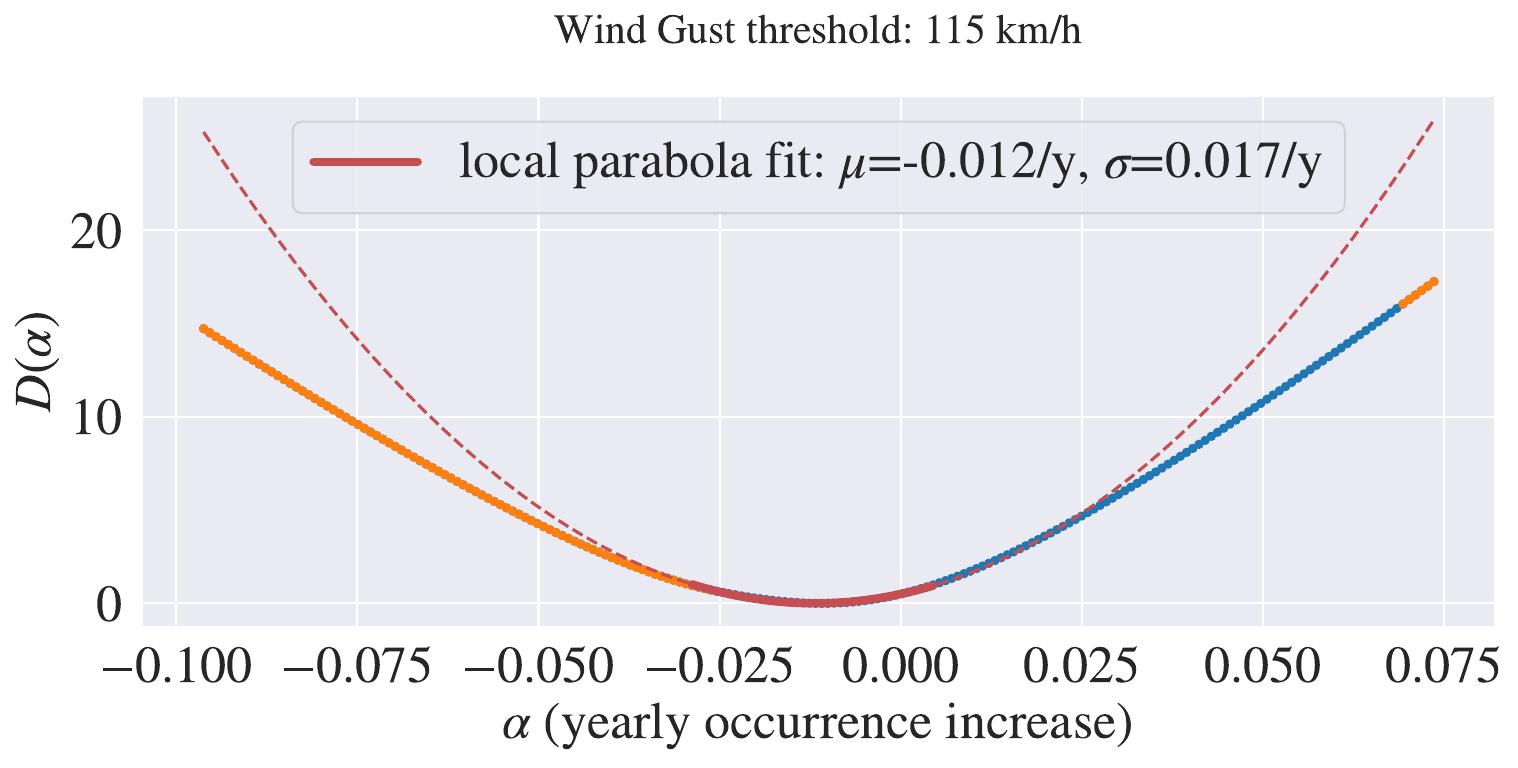}
    \caption{Two examples of a profile likelihood $D(\alpha)$ (blue dots). A parabola fit has been made around the minimum (red line) and extrapolated to higher values of $\alpha$ (dashed red line). On top, no unphysical values for $\widehat{\widehat{p_A}}(\alpha)$ are probed and correct coverage of confidence intervals can be trusted. Below, the profiling procedure has led to unphysical values of $\widehat{\widehat{p_A}}(\alpha)$ (orange dots) and $\widehat{\widehat{p_A}}$ needed to be adjusted according to Eq.~\ref{eq:p0adjust}. The extrapolated parabola fit does soon not agree with $D(\alpha)$ anymore, and some over-coverage needs to be assumed. \label{fig:non-diff}}
\end{figure}

The differentiability condition is sometimes not fulfilled, when at least one $p_i(\widehat{\widehat{p}}_A,\alpha)$ gains unphysical negative values (normally far from $\hat{\alpha}$). In that case, $\widehat{\widehat{p}}_A(\alpha)$ needs to get re-defined in order to guarantee physically correct positive occurrence probabilities always: 
\begin{align}
    \widehat{\widehat{p}}_A(\alpha) \rightarrow \left\{ 
\begin{array}{cl}
    -\dfrac{\alpha \cdot (N-1)/2}{w_{N-1}} & \mathrm{if}~  \alpha < 0\\[0.4cm]
     \dfrac{\alpha \cdot (N-1)/2}{w_0} & \mathrm{if}~ \alpha > 0
\end{array}
    \right.  
    \label{eq:p0adjust}
\end{align}\noindent
An example of its effect on $D(\alpha)$ is shown in Fig.~\ref{fig:non-diff}. Such exceptional cases lead to some over-coverage of the retrieved confidence intervals and have occurred at very high wind gust thresholds $>100$\,km/h.

\begin{figure*}
    \centering
    \includegraphics[width=0.485\textwidth]{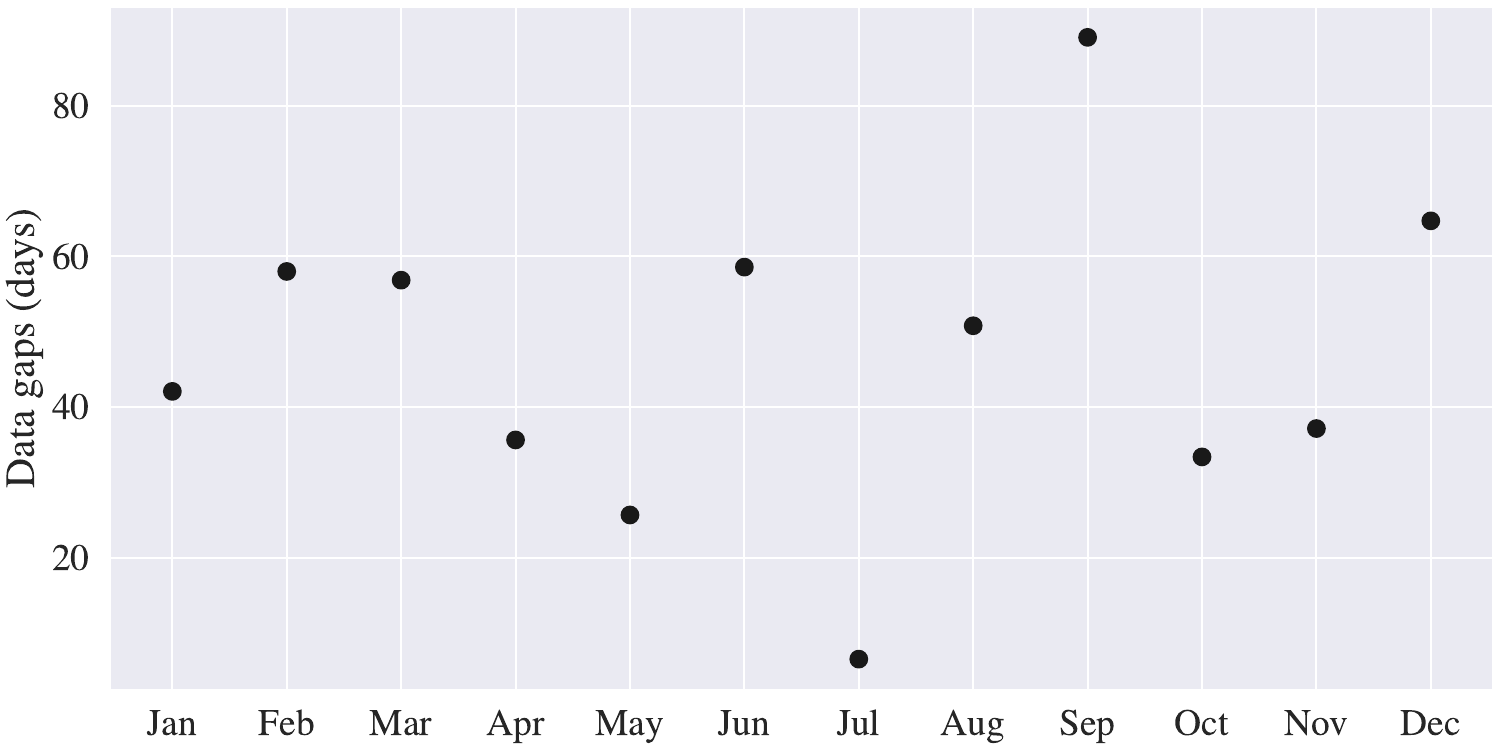}
    \caption{Seasonal cycle of large data gaps of the MAGIC wind data.}
    \label{fig:data_gaps}
\end{figure*}

The weights for each year, $w_i$, have been determined from the number of days lost in each month $m$, $N_\mathrm{lost}(m)_i$, due to data gaps greater than one day. Smaller gaps have been considered not to impede detection of an event. However, since storm or rain occurrence shows a strong seasonal dependency (see Fig.~\ref{fig:data_gaps}), each monthly observation duty cycle $(N_\mathrm{days}(m)-N_\mathrm{lost}(m)_i)/N_\mathrm{days}(m)$ has been reweighted according to the (normalized) occurrence probability $P(m)$ of a storm during the corresponding month $m$ (obtained, for example, from the normalized values of Fig.~\ref{fig:storm_count}): 

\begin{align}
    w_i &= \sum_{m} \dfrac{N_\mathrm{days}(m)-N_\mathrm{lost}(m)_i}{N_\mathrm{days}(m)} \cdot P_m \quad, \label{eq:wi}
\end{align}
\noindent
where $N_\mathrm{days}(m)$ denote the number of days of a given month $m$. Note that for a hypothetical full duty cycle during a given year $i$, $w_i$ sums to one exactly by construction.  

%%%%% Fractal analysis formula

\section{Detrended Fluctuation Analysis}\label{sec:DFA}
The output of Detrended Fluctuation Analysis (DFA) is the Hurst exponent $H$ of the analysed time series. DFA has several applications in atmospheric physics and astrophysics \citep{ivanova,varostos3,kosc,dos2021mixture,talkner,eichner,fraedrich,patta,matsoukas,moret,kavasseri,ausloos2}. 
DFA is applied to random walk-like time series $Y(t)$, where $t=1\dots N$ and $N$ is the length of the data. If the time series to be analysed is instead a noise-like time series $x(t)$, the mean subtraction and integration is carried out first and a random walk-like time series is obtained, according to
\begin{equation}
Y(t)=\sum_{t'=1}^{t}\bigl(x(t')-\bar{x}\bigr) \quad.
\label{Eq:Y}
\end{equation}
$Y(t)$ is then divided into $N_n = \lfloor N / n \rfloor$ non-overlapping time intervals of length $n$. If the length of the time series is not
a multiple of $n$, the last parts of the data from $Y(t)$ should be excluded. To avoid this, the same procedure can be repeated starting from the end of the integrated time series \citep{kant}, and in this case, the total number of time intervals considered is $N_{tot} = 2N_n$. If this is not feasible, the total number of time intervals will be chosen as $N_{tot} = N_n$. After this, the data in each interval $s$ are fitted with a least squares line $Y_s^\mathrm{fit}$. The standard deviation $\textit{RMS}\,( n,s )$ is then computed (locally) in each window,
\begin{equation}
\textit{RMS} \,( n,s ) = \sqrt{\frac{1}{n} \sum_{i = 1}^n \ [ Y ( ( s - 1 ) n + i ) - Y_s^\mathrm{fit} (i) ]^2}
\label{eq:DFA_var1}
\end{equation}
for $s = 1, \dots , N_n$ and, if DFA is also applied backwards,
\begin{equation}
	\textit{RMS} \,( n,s ) = \sqrt{\frac{1}{n} \sum_{i = 1}^n \  [ Y ( N - ( s - N_n ) n + i ) - Y_s^\mathrm{fit} (i) ]^2}
\label{eq:DFA_var2}
\end{equation}
for $s = N_n + 1, \dots , 2N_n$.
The overall root mean square fluctuation $F ( n )$, also known as the scaling function, is obtained averaging the squared \textit{RMS} over all the time intervals with same length $n$,
\begin{equation}
	F ( n ) = \biggl[ \frac{1}{N_\mathrm{tot}} \sum_{s = 1}^{N_\mathrm{tot}} \textit{RMS}^2 ( n,s ) \biggr]^{1 / 2}
	\label{eq:DFA}
\end{equation}
This is repeated for different windows length, i.e.\ for different $n$. The scaling function $F(n)$ scales as a power law with window size
\begin{equation}
F(n)\sim n^{H}
\label{Eq:DFA}
\end{equation}
where $H$ is the Hurst exponent characterising the data. It can be estimated using Eq.~\ref{Eq:DFA} from the slope of a straight line fitting $F(n)$ to $n$ in a double logarithmic plot.
%%%%%%%%%%%%%%LOCAL HURST
\section{Multifractal Detrended Fluctuation Analysis}\label{sec:H}
Since a multifractal time series is better described by a set of Hurst exponents, DFA needs to be generalized to higher orders using MFDFA \citep{kant,Ihlen_2012}. $F(n)$ is computed at various orders $q$ and different Hurst exponents, $h(q)$, are obtained,
\begin{equation}
F_q ( n ) = \biggl[ \frac{1}{N_\mathrm{tot}} \sum_{s = 1}^{N_\mathrm{tot}} [\textit{RMS\,}^2 ( n,s )]^{q/2} \biggr]^{1 / q} \sim n^{h(q)} \quad,
\label{Eq:qFn}
\end{equation}
where $q$ can take any real value. For $q>0$, intervals of data characterized by a large variance will dominate the average when computing Eq.~\ref{Eq:qFn}. Thus, for $q>0$, $h(q)$ quantifies the scaling behavior of intervals that have large fluctuations. Instead, for $q<0$, time intervals with small variance will dominate.
%in Eq.~\ref{Eq:qFn}. 
Hence, for negative values of $q$, the $q$-order Hurst exponent $h(q)$ describes the scaling behaviour of time intervals that have small fluctuations. For $q = 0$ $F_{q}(n)$, diverges and can be replaced by an exponential of a logarithmic sum
\begin{equation}
F_0(n)=\exp \biggl[ \frac{1}{2N_\mathrm{tot}} \sum_{s = 1}^{N_\mathrm{tot}} \ln(\textit{RMS\,}^2 ( n,s )) \biggr] \sim n^{h(0)}\quad. 
\label{Eq:zeroFn}
\end{equation}
When $n$ is equal to the length of the time series, and $n=N$, Eq.~\ref{Eq:qFn} becomes independent of $q$. The sum in Eq.~\ref{Eq:qFn} runs over only one or two segments, depending on whether MFDFA is also computed backwards or not (like DFA).

The multifractal spectrum $f(\alpha)$ is related to $h(q)$ via a Legendre transform \citep{kant, feder1988random}:
\begin{equation}
\alpha = h(q) + q h'(q) \quad f(\alpha) = q(\alpha - h(q)) + 1
\label{mf_spectrum}
\end{equation}
and is used to quantify multifractality based on its width. The shape of the spectrum is concave down. The most important parameters of the multifractal spectrum are $\alpha_\mathrm{min}$ (the minimum value of $\alpha$), $\alpha_\mathrm{max}$ (the maximum value of $\alpha$), and $\alpha_0$ (the value of $\alpha$ where the spectrum has its maximum value). $\alpha_\mathrm{min}$ and $\alpha_\mathrm{max}$ indicate the most extreme and smoothest events in the time series, respectively. The wider the spectrum, the more multifractal the time series. The spectrum can also be asymmetric: a more pronounced left tail corresponds to a strong presence of extreme events, while a more pronounced right tail corresponds to the dominance of periods with a small variance.

%\clearpage

\section{Detrended Cross Correlation Analysis}\label{sec:DCCA}

%\mg{If the methodology section moves to the appendix, then the next paragraph would need to stay here. }

%To investigate if changes in the North Atlantic Oscillation Index (NAOI) affect the temperature recorded at the MAGIC telescope location, temperature anomalies were correlated with the NAOI. Detrended Cross Correlation Analysis (DCCA) was used to this end. The temperature anomalies were obtained subtracting the main seasonality of the data. 
%\mg{Suggest to move all explanations of NAO to either where they appear first, or the discussion}
% As described in \citet{tatli2019detrended}, during the positive phase of NAO, in the south of Europe, a cold and dry period prevails. In the negative phase of NAO, warm and rainy conditions prevail over the Mediterranean basin. 
%\mg{until HERE}
%A negative correlation is expected between the NAOI and the MAGIC temperature residuals. 

The DCCA method is described below on the basis of~\citet{vassoler2012dcca} and references therein. % \mg{From here on to the appendix again. }
DCCA is a generalisation of the DFA algorithm, which makes use of the detrended covariance. It looks for power-law cross-correlations between two nonstationary time series $y(t)$ and $y'(t)$ of equal length $N$. First, the two integrated signals $R_{k}=\sum_{i=1}^{k}y_{i}$ and $R'_{k}=\sum_{i=1}^{k}y'_{i}$ are computed, where $k = 1,...,N$. Then the entire time series is divided into $N-n$ overlapping boxes, each with $n + 1$ values. In each box, starting at $i$ and ending at $i + n$, the local trend is defined by $\hat{R}_{k,i}$ and $\hat{R'}_{k,i}$ with $(i \leq k \leq i + n)$ and is the ordinate of a linear least-squares fit. The detrended walk is defined as the difference between the original walk and the local trend. The covariance of the residuals in each box is
 \begin{equation}
     f_\mathrm{DCCA}^{2}(n,i)=\frac{1}{n+1}\sum_{k=i}^{i+n}(R_{k}-\hat{R}_{k,i})(R'_{k}-\hat{R'}_{k,i}) \quad.
 \end{equation} 
 The detrended covariance function is obtained by summing over all overlapping $N-n$ boxes of size $n$:
 \begin{equation}
 F^{2}_\mathrm{DCCA}=\frac{1}{N-n}\sum_{i=1}^{N-n}f^{2}_\mathrm{DCCA}(n,i) \quad.
\end{equation}
If one random walk is analysed, ($R_{k} =R'_{k}$), the detrended covariance $F^{2}_\mathrm{DCCA}(n)$ is simply the detrended variance $F^{2}_\mathrm{DFA}(n)$ of the DFA. If self-affinity is present, $F_\mathrm{DCCA}^{2}(n) \sim n^{2\lambda}$.  
The $\lambda$ exponent quantifies long-range power-law cross-correlations. It also identifies seasonality~\citep{zebende2009cross}.
Some applications of the DCCA algorithm are found in \citet{podobnik2009quantifying,podobnik2009cross,yuan2014different}. 
As $\lambda$ does not quantify the level of cross-correlations, the DCCA cross-correlation coefficient can be used~\citep{zebende2011dcca}. It is defined as the ratio between the detrended covariance function $F^{2}_\mathrm{DCCA}$ and the detrended variance function $F_\mathrm{DFA}$, i.e.,
\begin{equation}
\rho_\mathrm{DCCA}=\frac{F^{2}_\mathrm{DCCA}}{F_{\mathrm{DFA}\{y_{i}\}}F_{\mathrm{DFA}\{y'_{i}\}}}
\label{eq:dcca}
\end{equation}
The value of $\rho_\mathrm{DCCA}$ ranges between $-1 \leq \rho_\mathrm{DCCA} \leq 1$. A value of $\rho_\mathrm{DCCA} = 0$ means that there is no cross-correlation. The $\rho_\mathrm{DCCA}$ coefficient has been tested and proven to be quite robust \citep{vassoler2012dcca}.

\section{Turbulence Intensity tests}
\label{sec:TI}

The turbulence intensity (TI), which is defined as the ratio of the standard deviation of fluctuating wind velocity to the mean wind speed, has been calculated for the entire data sample. Its medians as a function of wind velocity have been fitted to a linear function and residuals with respect to the fitted median calculated, the \textit{normalized TI}. 

\begin{figure}
\centering
\includegraphics[width=0.99\linewidth]{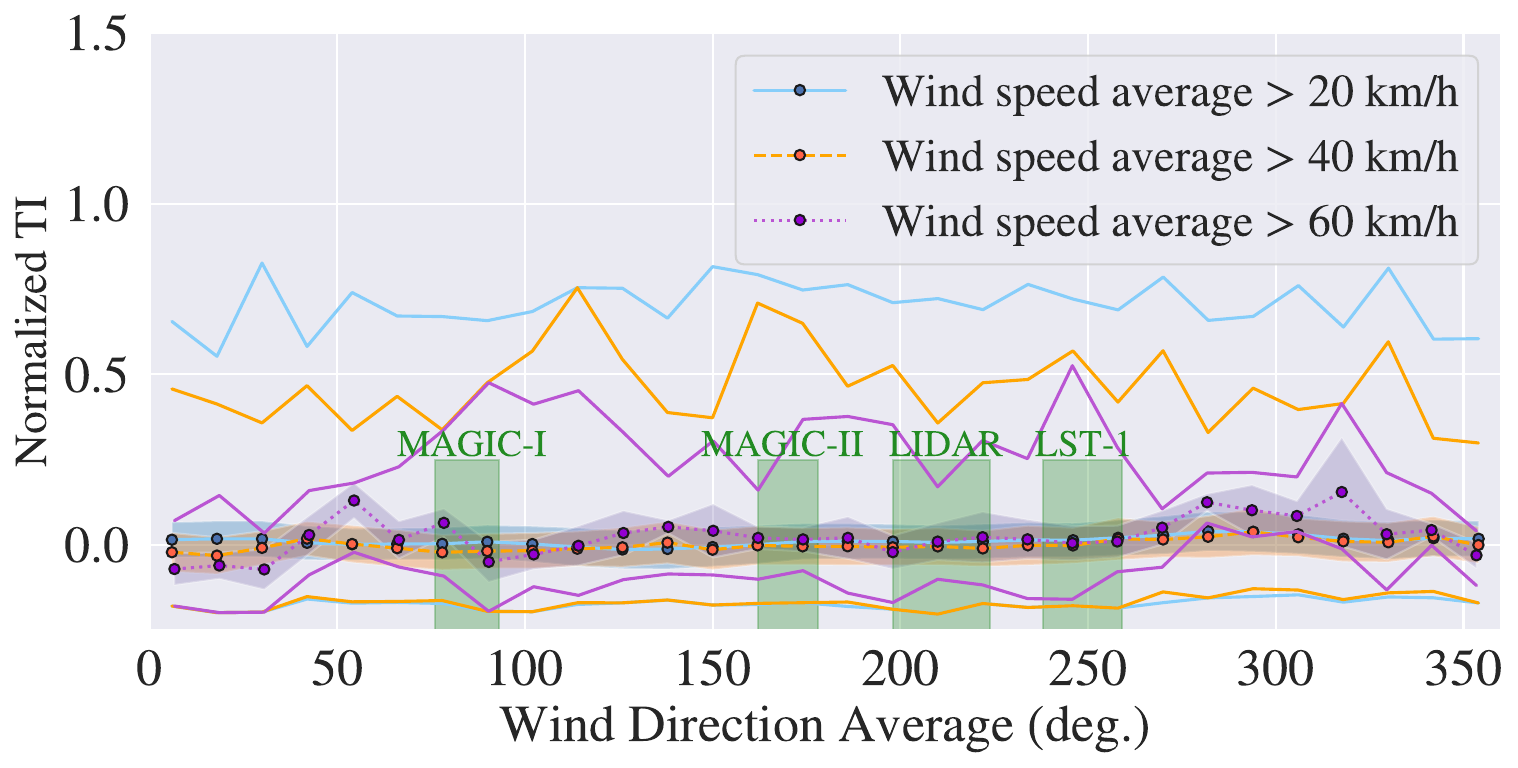}
\caption{\label{fig:windTI}Wind-speed normalized turbulence intensity (TI) as a function of wind direction. Circles display the bin-wise medians, the shadow fills the IQR, and the lines display the  maxima and minima observed. Three wind speed thresholds have been applied to the wind speed average: $>20$~km/h (blue), $>40$~km/h (orange) and $>60$~km/h (violet). For convenience, the four main possible wind shadows are drawn as green rectangles: the two MAGIC Telescopes, the LIDAR tower, and the LST1 telescope. }
\end{figure}

Figure~\ref{fig:windTI} shows the statistical distributions of the normalized TI as a function of wind direction, for three different threshold wind velocities. One can observe that above a threshold of 60~km/h, turbulence is observed from a few directions, which do not correlate, however, with the large obstancles found: the three telescopes MAGIC-I, MAGIC-II and the LST-1, and nearby the LIDAR tower. 

Based on this test, we conclude that these obstacles do not generate an excess of wind turbulence. 

% Don't change these lines
\bsp	% typesetting comment
\label{lastpage}
\end{document}